%% file: topmass.tex
\documentstyle[twoside,epsf]{topmassprd}
\input{prd_definitions}
\begin{document}
\markboth{ }{ }
\pagestyle{myheadings}
\bibliographystyle{unsrt}
\input title

\pagenumbering{roman}
\input author_list

\newpage
\input abstract             % abstract
%\begin{bf}
%\tableofcontents
%\end{bf}
%
\input {introduction}

\include{detector}

\include{data_samples}

\include{sim_and_back}

\include{chapter_5_mod3}

\include{event_fitting}

\include{likelihood}

\include{results}

\include{chapter_9}

\include{ch10}

\include{conclusion2}

\include{appendix}

\include{other}

\include{dilep-appendix}

\include{bibliography}
\end{document}

%% file: prd_definitions.tex
\def \Et {{\rm E}_{\rm T}}
\def \Pt {{\rm P}_{\rm T}}
\def \Pz {{\rm P}_{\rm Z}}

\newcommand{\MET}{\mbox{$\protect \raisebox{.3ex}{$\not$}\et$}}

\def\qbar{{\bar q}}

\def\Z0{${\em Z^0\/}$}

\def\r#1 {$^{#1}$}

\newcommand{\et}{{\rm E}_{\scriptscriptstyle\rm T}}

\newcommand{\met}{\mbox{$\protect \raisebox{.3ex}{$\not$}\et \ $}}

\newcommand{\ppbar}{p\bar{p}}

\newcommand{\ttbar}{t\bar{t}}
\newcommand{\bbbar}{b\bar{b}}
\newcommand{\ccbar}{c\bar{c}}

\newcommand{\bbar} {\bar{b}}   
\newcommand{\tbar} { \bar{{t}}  }                                
\newcommand{\Lik}{\mbox{$\mathcal{L}$ }}
\newcommand{\Ki}{\mbox{$\chi^{2}$ }}
\newcommand{\mPr}{\mathcal{P}}

\newcommand{\mLik}{\mathcal{L}}
\newcommand{\mCPr}{\mathcal{C}}
\def \pb    {pb$^{-1} $}
 
\def \mtop {$M_{top} \ $}

\def \Mt {M_{top}}

\def \tljx {\ttbar \rightarrow \ell \nu q \bar{q}^{\prime} b \bar{b} X}
\def \tllx {\ttbar \rightarrow \ell^+ \bar{\nu} \ell^- \nu b \bar{b} X}

\newcommand{\gev}  { \rm GeV }
\newcommand{\gevc} { {\rm GeV}/c }
\newcommand{\gevcc}{ {\rm GeV}/c^{2} }

\def\gepsfcentered#1{
  \def\testit{#1}
  \def\lbracket{[}
  \ifx\testit\lbracket
    \let\dofilecmd=\gepsfwithopt
  \else
    \let\dofilecmd=\gepsfnoopt
  \fi
  \dofilecmd}

\def\gepsfnoopt#1{
  \begin{center}
  \leavevmode
  \epsffile{#1}
  \end{center}}

\def\gepsfwithopt#1 #2 #3 #4]#5{
  \begin{center}
  \leavevmode
  \gepsfmaxx=0.94\textwidth
  \epsffile[#1 #2 #3 #4]{#5}
  \end{center}}

\newdimen\gepsfmaxx
\def\epsfsize#1#2{
  \ifnum \epsfxsize=0
    \ifnum \epsfysize=0
      \ifnum #1 > \gepsfmaxx
        \gepsfmaxx
      \else
        #1
      \fi
    \else
      \epsfxsize
    \fi
  \else
    \epsfxsize
  \fi
}

%% file: title.tex
%\begin{flushright}
%{\tt CDF/TOP/PUB/4953}\\
%Version 2.0 \\
%\today
%\end{flushright}
%
\begin{center}
\begin{bf}
Measurement of the Top Quark Mass with the Collider Detector at Fermilab\\
\end{bf}
\end{center}

%\vskip0.35in

\dsp

%% file: author_list.tex
\font\eightit=cmti8
\def\r#1{\ignorespaces $^{#1}$}
\hfilneg
\begin{sloppypar}
\noindent
T.~Affolder,\r {21} H.~Akimoto,\r {43}
A.~Akopian,\r {36} M.~G.~Albrow,\r {10} P.~Amaral,\r 7 S.~R.~Amendolia,\r {32} 
D.~Amidei,\r {24} K.~Anikeev,\r {22} J.~Antos,\r 1 
G.~Apollinari,\r {10}
% ...
\end{sloppypar}
\vskip .026in
\begin{center}
(CDF Collaboration)
\end{center}

\vskip .026in
\begin{center}
\r 1  {\eightit Institute of Physics, Academia Sinica, Taipei, Taiwan 11529, 
Republic of China} \\
\r 2  {\eightit Argonne National Laboratory, Argonne, Illinois 60439} \\
\r 3  {\eightit Istituto Nazionale di Fisica Nucleare, University of Bologna,
I-40127 Bologna, Italy} \\
\r 4  {\eightit Brandeis University, Waltham, Massachusetts 02254} \\
\r 5  {\eightit University of California at Los Angeles, Los 
Angeles, California  90024} \\  
\r 6  {\eightit Instituto de Fisica de Cantabria, CSIC-University of Cantabria, 
39005 Santander, Spain} \\
\r 7  {\eightit Enrico Fermi Institute, University of Chicago, Chicago, 
Illinois 60637} \\
\r 8  {\eightit Joint Institute for Nuclear Research, RU-141980 Dubna, Russia}
\\
\r 9  {\eightit Duke University, Durham, North Carolina  27708} \\
\r {10}  {\eightit Fermi National Accelerator Laboratory, Batavia, Illinois 
60510} \\
\r {11} {\eightit University of Florida, Gainesville, Florida  32611} \\
\r {12} {\eightit Laboratori Nazionali di Frascati, Istituto Nazionale di Fisica
               Nucleare, I-00044 Frascati, Italy} \\
\r {13} {\eightit University of Geneva, CH-1211 Geneva 4, Switzerland} \\
\r {14} {\eightit Harvard University, Cambridge, Massachusetts 02138} \\
\r {15} {\eightit Hiroshima University, Higashi-Hiroshima 724, Japan} \\
\r {16} {\eightit University of Illinois, Urbana, Illinois 61801} \\
\r {17} {\eightit The Johns Hopkins University, Baltimore, Maryland 21218} \\
\r {18} {\eightit Institut f\"{u}r Experimentelle Kernphysik, 
Universit\"{a}t Karlsruhe, 76128 Karlsruhe, Germany} \\
\r {19} {\eightit Korean Hadron Collider Laboratory: Kyungpook National
University, Taegu 702-701; Seoul National University, Seoul 151-742; and
SungKyunKwan University, Suwon 440-746; Korea} \\
\r {20} {\eightit High Energy Accelerator Research Organization (KEK), Tsukuba, 
Ibaraki 305, Japan} \\
\r {21} {\eightit Ernest Orlando Lawrence Berkeley National Laboratory, 
Berkeley, California 94720} \\
\r {22} {\eightit Massachusetts Institute of Technology, Cambridge,
Massachusetts  02139} \\   
\r {23} {\eightit Institute of Particle Physics: McGill University, Montreal 
H3A 2T8; and University of Toronto, Toronto M5S 1A7; Canada} \\
\r {24} {\eightit University of Michigan, Ann Arbor, Michigan 48109} \\
\r {25} {\eightit Michigan State University, East Lansing, Michigan  48824} \\
\r {26} {\eightit University of New Mexico, Albuquerque, New Mexico 87131} \\
\r {27} {\eightit The Ohio State University, Columbus, Ohio  43210} \\
\r {28} {\eightit Osaka City University, Osaka 588, Japan} \\
\r {29} {\eightit University of Oxford, Oxford OX1 3RH, United Kingdom} \\
\r {30} {\eightit Universita di Padova, Istituto Nazionale di Fisica 
          Nucleare, Sezione di Padova, I-35131 Padova, Italy} \\
\r {31} {\eightit University of Pennsylvania, Philadelphia, 
        Pennsylvania 19104} \\   
\r {32} {\eightit Istituto Nazionale di Fisica Nucleare, University and Scuola
               Normale Superiore of Pisa, I-56100 Pisa, Italy} \\
\r {33} {\eightit University of Pittsburgh, Pittsburgh, Pennsylvania 15260} \\
\r {34} {\eightit Purdue University, West Lafayette, Indiana 47907} \\
\r {35} {\eightit University of Rochester, Rochester, New York 14627} \\
\r {36} {\eightit Rockefeller University, New York, New York 10021} \\
\r {37} {\eightit Rutgers University, Piscataway, New Jersey 08855} \\
\r {38} {\eightit Texas A\&M University, College Station, Texas 77843} \\
\r {39} {\eightit Texas Tech University, Lubbock, Texas 79409} \\
\r {40} {\eightit Istituto Nazionale di Fisica Nucleare, University of Trieste/
Udine, Italy} \\
\r {41} {\eightit University of Tsukuba, Tsukuba, Ibaraki 305, Japan} \\
\r {42} {\eightit Tufts University, Medford, Massachusetts 02155} \\
\r {43} {\eightit Waseda University, Tokyo 169, Japan} \\
\r {44} {\eightit University of Wisconsin, Madison, Wisconsin 53706} \\
\r {45} {\eightit Yale University, New Haven, Connecticut 06520} \\
\end{center}

%% file: abstract.tex
%\begin{abstract}
This report describes a measurement of the top quark mass in $\ppbar$ 
collisions at a center of mass energy of 1.8 TeV. The data sample was 
collected with the CDF detector during the 1992--95 collider run at the 
Fermilab Tevatron, and corresponds to an integrated luminosity of 106 \pb. 
Candidate $t\bar{t}$ events in the ``lepton+jets'' decay channel provide
our most precise measurement of the top quark mass. For each event
a top mass is determined by using energy and momentum
constraints on the production of the $\ttbar$ pair and its
subsequent decay.
%the constraints on the energy and
%momentum of the top and anti-top quarks decay products.
A likelihood fit to the distribution of reconstructed masses
in the data sample gives a top mass in the lepton+jets
channel of $176.1\pm 5.1 (stat.)\pm 5.3 (syst.) \ \gevcc$.
Combining this result with measurements from the ``all-hadronic'' and 
``dilepton'' decay topologies yields a top mass of $176.1\pm 6.6\ \gevcc$.

\noindent PACS numbers 14.65.Ha, 13.85.Qk, 13.85.Ni

%\end{abstract}
%\newpage

%% file: introduction.tex
\chapter {INTRODUCTION}
\label{s-intro}
\pagenumbering{arabic}
\setcounter{secnumdepth}{2}
\setcounter{section}{0}
%\pagenumbering{arabic}
\setcounter{page}{1}
This paper describes a measurement of the top quark
mass using events produced 
in proton-antiproton ($\ppbar$) collisions at the Fermilab Tevatron
with a center-of-mass energy of  1.8~TeV and reconstructed through
the decay mode
 $t \bar {t} \rightarrow W^{+} b + W^{-} \bar{b}
             \rightarrow \ell^{+}\nu b + q\bar{q}^{\prime}\bar{b} \ $ 
(and charge conjugate mode). Throughout this paper the symbol $\ell$ will be 
used to denote either an electron or a muon exclusively.
  We present results from two data samples with integrated 
luminosities of 19.7~pb$^{-1}$ (Run 1a) and 86.3~pb$^{-1}$ 
(Run 1b) collected with the Collider Detector at
Fermilab (CDF) from September 1992 to June 1993 and from 
February 1994 to July 1995, respectively.

 The existence of the top quark was established by
direct experimental observation at the
Fermilab Tevatron by the CDF~\cite{cdf-evidence,cdf-discovery}
and D\O\  collaborations~\cite{d0-discovery}. These analyses led to
$\ttbar$ cross section and top quark mass measurements. 
Additional analyses showed that the kinematics of the observed events 
were inconsistent with being solely from background sources
and were consistent with standard model $\ttbar$~\cite{htanal}.
With substantially larger data samples and improved
understanding of systematic uncertainties, more precise measurements 
of the top quark mass~\cite{d0-mass-prl,cdf-mass-prl}
and $\ttbar$ production cross section~\cite{xsec,d0xsec} in $\ppbar$ 
collisions were recently reported.  The larger data samples were 
used to perform detailed comparisons of kinematic variables
between $\ttbar$ candidate events and simulated standard model $\ttbar$ 
and background events~\cite{andy}. The data samples were
also used in the identification and
analysis of $\ttbar$ production into fully hadronic final 
states~\cite{all-hadron,d0-all-hadron} and final states involving two 
leptons, $\ell \overline{\ell}$~\cite{cdf-dilepton,d0-dilepton} or 
$\ell\tau$~\cite{tau-dil}.

The top quark is defined as the $I_{3}=+1/2$ member of a
weak SU(2) isodoublet that also contains the $b$ quark. In 
$\ppbar$ collisions, top quarks are expected to be
produced primarily in $\ttbar$ pairs via quark-antiquark annihilation ($\approx$90\%)
or gluon fusion ($\approx$10\%) and decay through the electroweak interaction to a final state 
consisting of a $W$ boson and $b$ quark.  In the standard 
model, the branching fraction for $t\to Wb$ is expected to be nearly 
100\%. The decay width is calculated to
be $1.6-1.7 \ \gev$ for masses between 150 and $180 \ \gevcc$~\cite{twidth}.
The top quark mass is sufficiently large that top-flavored hadrons are
not expected to form~\cite{notophad}. 

The mass of the top quark, $M_{top}$, is an important parameter in 
calculations of electroweak processes since it is approximately 35 times 
larger than that of 
the next heaviest fermion.  Like other fermion masses,
$M_{top}$ is not predicted in the standard model~\cite{SM}.
%, while extensions to the theory relate it to an underlying smaller set of 
%fundamental parameters~\cite{SM-extensions}
On the other hand, the standard model relates the masses of the top quark
and $W$ boson to that of the Higgs boson, so that
precise measurements of the former imply bounds on the latter.
With the assumption of the validity of the standard model,
experimental studies of the electroweak interaction can
alternatively be used to estimate the value of \mtop.
For instance, a fit to LEP (including LEP-II) data, leaving
the top quark mass and the Higgs mass as free parameters, 
yields an inferred top quark mass of
%Measurements performed at LEP (including LEP-II)
%and SLC
%Studies of $Z$ boson production and decay at $e^+ e^- $ colliders, 
%for instance,  yield an inferred value of the top
%quark mass of 
%\mtop $=177^{+7 +18}_{-8 -12} \gevcc$~\cite{lep-topmass}.
%\mtop $=181.3^{+6.1 +15.7}_{-6.2 -17.3} \ \gevcc$~\cite{lep-topmass}.
$ 160^{+13}_{-9} \ \gevcc$
and a Higgs mass of $60^{+127}_{-35}~\gevcc$~\cite{lep2-topmass}.
%The uncertainty can be mostly attributed to the 
%uncertainty in the Higgs boson mass in the fitting procedure.

 The decay modes of the $W$ bosons into either 
lepton-neutrino ($\ell\nu$), ($\tau\nu$) or quark-antiquark 
($q\bar q^{\prime}$) final
states classify candidate $\ttbar$ events into four main
categories. All-hadronic final states, which comprise approximately
44\% of $\ttbar$ decays, correspond to those events in which both
$W$ bosons decay hadronically. Lepton+jet events are
those events in which only one of the two $W$ bosons decays
hadronically while the other decays into $\ell\nu$ and form 30\% 
of $\ttbar$ decays.
Dilepton events are defined as those in which the 
$W$ bosons decay into either $e\nu$ or $\mu\nu$ final states 
and occur only about 5\% of the time. Lastly, there is an additional 21\%
of events for which the final state includes one or more $\tau$ 
leptons. The $\tau$ events are particularly difficult to identify because
$\tau$'s decay into leptons or hadrons and are often indistinguishable from 
the other final states, thus contaminating the other samples. 
%~\cite{tau-dil}. While these $\tau$ events do contribute
%to the dilepton event acceptance, they
%are treated separately from dilepton events because 
%they are particularly difficult to identify in $\ppbar$ collisions.
Each $\ttbar$ decay mode is characterized by a final state
consisting of two $b$ hadrons and either zero, two, or four additional
jets, depending on the decay mode of the $W$'s in the event.
Additional jets beyond those from the $\ttbar$ decay may also
arise from initial and final state radiation of the incoming and 
outgoing partons.
%~\cite{lynorr1}. 

  The direct experimental determination of \mtop through
analysis of $\ttbar$ pairs produced in $\ppbar$
collisions can be obtained by comparing observed kinematic
features of top events to those predicted for different top 
quark masses~\cite{andy}. 
%A thorough comparison of the kinematics
%of standard model $\ttbar$ events and background events with the 
%data sample support their usage in modeling observed kinematic
%distributions~\cite{andy}. 
While any kinematic variable which exhibits 
sensitivity to the mass of the top quark may be used to measure \mtop,
%Candidate top events observed in the dilepton decay mode 
%have been used to measure \mtop by studying the transverse energy 
%distribution of associated jets~\cite{cdf-dilepton}. 
the lowest statistical uncertainty is achieved by explicitly reconstructing
the top mass  from the $\ttbar$ daughter decay products. 
In this paper, we discuss the complete reconstruction of
top events in the lepton+jets topology and report the measurement
of \mtop obtained using the distribution of the
reconstructed top quark masses from the data sample. 
%Because the technique also
%yields the reconstructed four-momenta of the top quarks, it
%has further importance for analyses of other kinematic
%features such as the $\ttbar$ invariant mass.

This paper is structured as follows.  Section~\ref{s-detector}
presents a description of the CDF detector, emphasizing the
subsystems most important to this analysis. Section~\ref{s-part} 
discusses the reconstruction of jets and leptons in the CDF detector
and defines the sample of events which are used in
the measurement of the top quark mass.
Section~\ref{sim_back} describes the simulations used and
discusses the details of the background calculation.
Section~\ref{s-recon} describes the corrections which are
applied to the raw calorimeter measurements. Section~\ref{s-algor}
presents the algorithm used to estimate the top quark mass
on an event-by-event basis and describes the results of 
the algorithm when applied to simulated samples of both $\ttbar$ and
background events. The description of the likelihood procedure
and the subsequent extraction of \mtop are the subjects of 
Sections~\ref{s-like} and ~\ref{like-data}.
Section~\ref{s-sys} describes the systematic uncertainties associated
with the top quark mass measurement. Combining the measurements from
the lepton+jets, dilepton, and all-hadronic analyses is the focus
of Section~\ref{s-combmass}.  Conclusions are given in 
Section~\ref{s-conclusion}.

%% file: detector.tex
\chapter{THE CDF DETECTOR}
\label{s-detector}
\begin{figure}
\epsfysize=1.0\textwidth
\epsfclipon
%%\epsfysize=6.0in
%%\epsffile[0 72 612 720]{cdf-detector.ps}
%%\hspace{-1.3truein}{\epsffile[0 72 612 720]{det.ps}}
%\epsffile[100 265 477 565]{det.ps}
\epsffile[90 265 477 665]{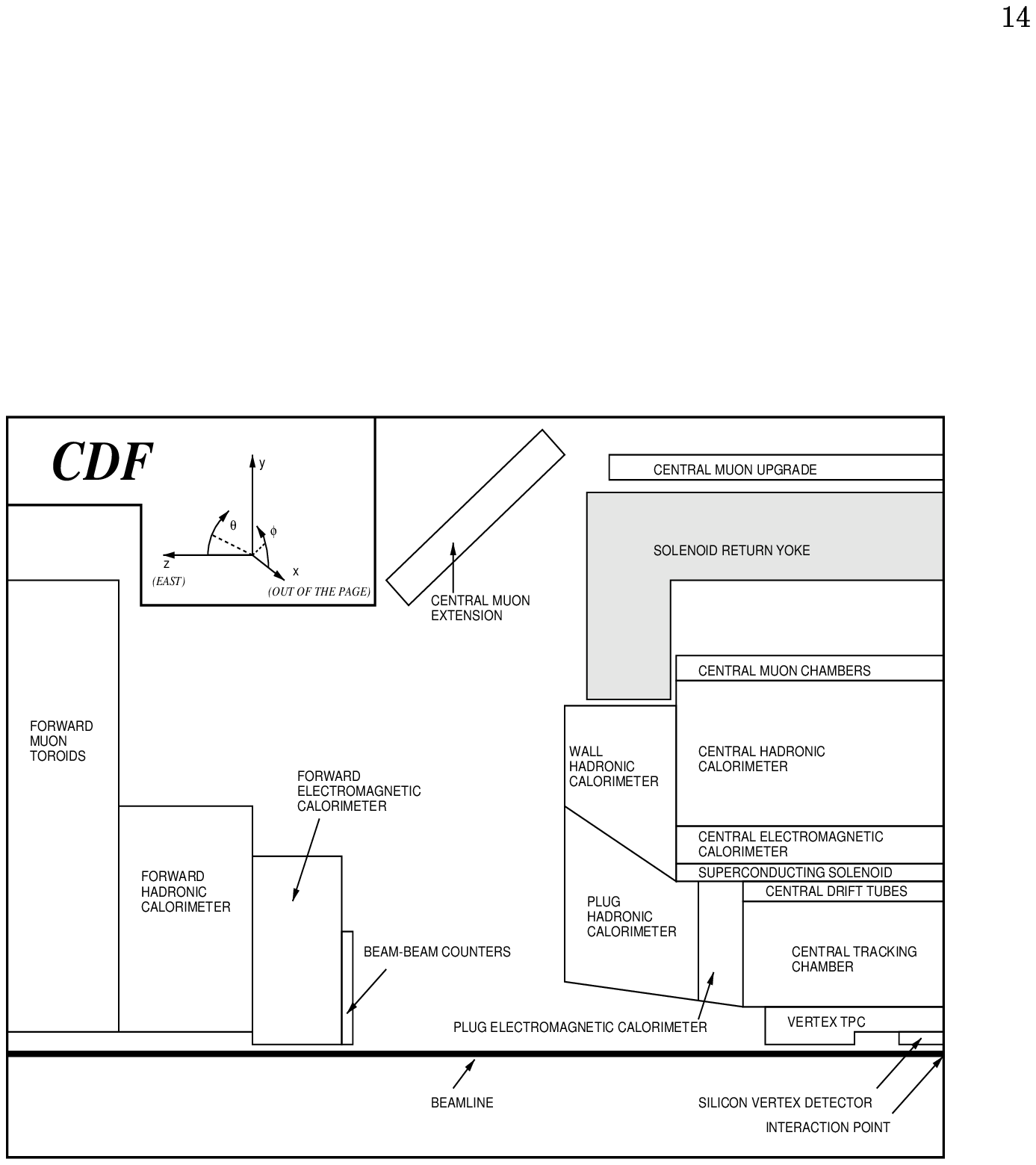}
\caption{Side view of one quadrant of the CDF detector for Run 1.
The detector is symmetric about the interaction point.}
\label{cdf-detector}
\end{figure}

The CDF detector is an azimuthally symmetric general purpose detector.
It consists of independent subsystems designed for
distinct tasks.  The three most relevant subsystems to $\ttbar$ 
detection are the tracking chambers, the calorimetry, and the muon chambers. 
In this section, we briefly describe these subsystems. The various
subsystems are shown in the side view of one quadrant of the detector 
in Fig.~\ref{cdf-detector}.
A more detailed description of each of these components can be
found in Refs.~\cite{cdf-evidence,cdf_det}

\section{Detector subsystems}

The tracking system consists of three subsystems that are all immersed in
a 1.4 T solenoidal magnetic field. The outermost system, the central tracking
chamber (CTC)~\cite{ctc}, is a wire drift chamber consisting of 84 concentric 
cylindrical layers 
of sense wires. The CTC has a length of 3.2 m and an outer radius of 1.32 m
which results in full acceptance for charged particles in the region 
$|\eta | < 1$~\cite{etadef}. The momentum transverse to the beamline ($\Pt$) 
is measured by the CTC with a precision given by 
$\delta (\Pt )/ \Pt=0.0011 \Pt$
($\Pt$ in $\gevc$), when the track is constrained to go through the beam 
position determined for each run.

 Inside the CTC is a set of time projection chambers (TPC)~\cite{vtx}, 
with tracking coverage in the region $|\eta | < 3.25$. This
detector, referred to as the VTX, is used to measure the position
of the $\ppbar$ interaction vertex along the $z$-axis with a resolution
of 1 mm. In events with more than one reconstructed vertex, the primary
vertex is taken to be the one with the largest number of VTX hits on 
its associated tracks. Primary collisions are spread with
an approximately Gaussian density along the {\it z}-axis 
with $\sigma \sim 30 \ {\rm cm}$. The primary vertex is required
to be within $\pm$60 cm of $z=0.0$. The efficiency of this requirement
is evaluated using the same techniques described in Ref.~\cite{vtxeff} and 
is estimated to be 95.6\%.
%The fraction of events 
%where two nearby and well measured vertices are misidentified
%as a single vertex is estimated to be $\sim 2-3\%$.  

  The innermost tracking system, the silicon vertex detector, SVX, consists 
of four layers of single-sided silicon detectors
(the Run 1a detector was replaced for Run 1b due to radiation
damage)~\cite{svx},
mounted inside two cylindrical barrels having a combined length of
51.0 cm. The four layers are located at radii of
approximately 3.0, 4.2, 6.8 and 7.9 cm from the beamline. The axial strips
of the three innermost layers have 60 $\mu$m pitch, and the
outermost layer has 55 $\mu$m pitch.
The silicon detector measures hits in the transverse plane with a
precision of 13 $\mu$m and the impact parameter of tracks relative to the 
primary vertex has a precision of $(13+40/\Pt)\ \mu$m ($\Pt$ in $\gevc$). 
Secondary vertices (from weak decays, for example) are
identified and reconstructed by augmenting reconstructed CTC tracks 
with hits found in the SVX. The precision of the SVX enables efficient
identification of secondary vertices from the decays of $b$ hadrons 
($c\tau \sim 400 \mu$m). The momentum resolution of a track reconstructed
using both the SVX and CTC detectors
is given by $\delta \Pt / \Pt = \sqrt{(0.0009\Pt)^2+(0.0066)^2}$ , where
$\Pt$ is in $\gevc$ and the second term is due to multiple scattering.
 
Muons are identified by the association of reconstructed track segments  
in the proportional wire chambers of either of the three muon systems, 
the central muon system (CMU)~\cite{cmu}, the central muon upgrade (CMP), 
or the central muon extension (CMX), with charged particle tracks observed 
in the CTC.  The CMU and CMP, separated by 0.6 m of steel, each cover the 
pseudorapidity region $|\eta |<0.6$.  In that region CMU covers 
$\approx 84\%$ of the solid angle, CMP $\approx 63\%$, and both combined 
$\approx 53\%$.  At larger pseudorapidities the CMX provides $\approx 71\%$ 
coverage of the solid angle for $0.6<|\eta |<1$.

The CDF calorimeters are segmented into projective towers. The towers
are further divided into compartments designed to separately measure 
electromagnetic and hadronic energy.  Three separate regions
of calorimetry provide coverage in $\eta$ from $-$4.2 to 4.2. 
All of the electromagnetic calorimeters use lead as the absorber,
while the hadronic calorimeters use iron.  In the central region, 
coverage is provided
%by scintillator electromagnetic and hadronic calorimeters, 
%CEM~\cite{cem}($|\eta |<1.1$)
%and CHA~\cite{cha}/WHA ($| \eta |<1.3$), respectively. Coverage at larger 
%pseudorapidities is provided by gas calorimeters PEM 
%and PHA in the region $1.1<|\eta |<2.4$, and FEM and FHA in
%the range from $2.4<|\eta |<4.2$. 
%The calorimeters provide identification of, and energy measurement for
%jets, electrons, photons, unclustered energy~\cite{unclus}, and
%missing transverse energy (\met)~\cite{metdef}.  The energy 
%resolution for the the CEM, PEM, and FEM have sampling terms 
%of 13.7\%, 22\% and 26\%, respectively, and constant terms of 2\%.
%The sampling terms for the hadronic calorimeters 
%increase from $50\%$ for the CHA to $137\%$ for the FHA,
%and have constant terms of $3-6\%$.
by electromagnetic and hadronic calorimeters, CEM~\cite{cem}
and CHA/WHA~\cite{cha}, respectively. The CEM is composed of
alternating layers of lead and scintillator, whereas the CHA and WHA
are composed of alternating layers of iron and scintillator.
Coverage at larger pseudorapidities is provided by calorimeters PEM 
and PHA, and in the far forward regions by the FEM and FHA. These
calorimeters use gas proportional chambers, instead of scintillators,
as the active sampling
medium. The calorimeters provide identification of, and energy measurement for
jets, electrons, photons, unclustered energy~\cite{unclus}, and
missing transverse energy (\met)~\cite{metdef}. The coverage
in pseudorapidity and the energy resolution for the
calorimeters are given in Table~\ref{cal-det}.

\begin{table}[ht]
\begin{center}
\begin{tabular}{crr}
\hline\hline
Detector & \multicolumn{1}{c}{$\eta$ range} & 
\multicolumn{1}{c}{Energy resolution} \\
\hline
\rule{0mm}{5mm} CEM & $    |\eta|<1.1$ & $13.5\%/\sqrt{\Et}\oplus 2\%$\\
\rule{0mm}{5mm} PEM & $1.1<|\eta|<2.4$ & $22\%/\sqrt{\Et}\oplus 2\%$  \\
\rule{0mm}{5mm} FEM & $2.2<|\eta|<4.2$ & $26\%/\sqrt{\Et}\oplus 2\%$  \\
\rule{0mm}{5mm} CHA & $    |\eta|<0.9$ & $50\%/\sqrt{\Et}\oplus 3\%$  \\
\rule{0mm}{5mm} WHA & $0.7<|\eta|<1.3$ & $75\%/\sqrt{\Et}\oplus 4\%$  \\
\rule{0mm}{5mm} PHA & $1.3<|\eta|<2.4$ & $106\%/\sqrt{\Et}\oplus 6\%$ \\
\rule{0mm}{5mm} FHA & $2.4<|\eta|<4.2$ & $137\%/\sqrt{\Et}\oplus 3\%$ \\
\hline\hline
\end{tabular}
\end{center}
\caption{Coverage in pseudorapidity and energy resolution for the various 
calorimeters. The symbol $\oplus$ signifies that the constant term is added
in quadrature with the sampling (first) term. Energy resolutions for the
electromagnetic calorimeters are for incident electrons and photons. For the
hadronic calorimeters, they are for incident pions. $\Et$ should be expressed
in $\gev$.}
\label{cal-det}
\end{table}

%
%The Central calorimeter is segmented into separate electromagnetic (EM)
%and hadronic (HAD) compartments, each providing azimuthally symmetric
%coverage over the pseudorapidity ($\eta = -log(tan(\theta/2))$)
%range $|\eta|<1.1$.  The EM calorimeter is comprised of 10 alternating
%layers of lead and plastic scintillator and has a measured resolution
%of $\sigma (E)/E=13.6\%/\sqrt(E) \oplus 1.\%$.  The hadronic calorimeter
%is made of 16 layers of steel and plastic scintillator with a total
%depth corresponding to 12 interaction lengths.
%
%The energy scale of the EM calorimeter is determined using electrons
%from leptonic W decays and is known to a precision of $0.2\%$
%~\cite{emscale}.
%The hadronic calorimeter is calibrated with single charged particles
%from testbeam facilities as well as from minimum bias data~\cite{hadscale}. 
%

\section{Luminosity and triggers}

  The events used in this analysis are extracted from two data samples
with integrated luminosities of 19.7 \pb (Run 1a) and 86.3 \pb (Run 1b)
collected during the period from September 1992 to June 1993, and
from February 1994 to July 1995 respectively.
Instantaneous luminosities varied between 
$1\times 10^{30} \ {\rm to } \ 2\times 10^{31} {\rm cm^{-2} sec^{-1}}$ 
during the data taking period, with averages that increased from 
$\approx 3.3\times 10^{30} \ {\rm cm^{-2} sec^{-1}}$ during Run 1a to
$\approx 1\times 10^{31} \ {\rm cm^{-2}s^{-1}}$  for Run 1b. 
The corresponding average number of interactions per 
crossing increased from 0.6 to 1.8. Since the measured
jet energies increase in the presence of additional interactions, the
corrections to the jet energies differ between Run 1a and Run 1b 
(see Section~\ref{s-fijetcor}).

   A multi-level trigger is used to select events containing 
high-$\Pt$ leptons~\cite{cdf-evidence,xsec}. To increase the $\ttbar$ 
acceptance in the muon channel, a trigger based on the missing 
transverse energy ($\met$) 
was added for Run 1b~\cite{xsec}. For the high-$\Pt$ inclusive 
lepton sample, only triggers from the central region are used in this analysis.
The CEM trigger efficiency for fiducial~\cite{deffid} electrons from $\ttbar$ 
events with $\Et>20\ \gev$ and $|\eta|<1$  is essentially 100\%.
The muon trigger is measured to be 85.4\% efficient for fiducial muons
from $\ttbar$ events that have $\Pt>20\ \gevc$.

%% file: data_samples.tex
\chapter {DATA SAMPLES}
\label{s-part}

  The data sample selection for this analysis is based on
standard model decay of top quark pairs through the
$\tljx$ channel. 
%where $\ell$ is either an electron or muon. 
The final state should therefore include a high-$\Et$
($\Pt$) electron (muon), significant missing transverse energy 
and four jets. The
momenta of these objects are measured from data recorded with each detector
subsystem, sometimes in combination. The four-momenta of electrons 
are expressed in terms of ($\Et$, $\phi$, $\eta$, $m$)  where
$\Et$ is the transverse energy ($\Et\equiv E\sin\theta$),
$\phi$ is the azimuthal angle, $\eta$ is the pseudorapidity and $m$ is
the mass.
For muons and jets 
%are measured in the tracking system, and are thus characterized by 
%$\Pt$, $\phi$ and $\eta$ from their reconstructed tracks. 
$\Pt$ is used rather than $\Et$. 
%Jet energies are calculated assuming a mass of $0.5 \ \gevcc$,
%unless the jet is assumed to be a $b$ quark, in which case its
%mass is taken to be $5 \ \gevcc$. 
In all cases, the direction of these objects
is measured with much greater precision than their energies.
In this section, we first describe the identification and reconstruction
of leptons and jets, and then we define the data samples.

\section{High-$\Pt$ leptons}
\label{lepid}

  We are most interested in identifying charged leptons
which are produced from the decay of a $W$ boson.
%, or through the semileptonic decay of $b$ or $c$-flavored hadrons. 
These leptons are distinguished from those
produced in semileptonic decay of $b$ or $c$ quarks because leptons 
from $W$-boson decay are not part of a jet and have typically much higher 
$\Pt$. A sample of high-$\Pt$ leptons is used to select leptons
which are consistent with having come from $W$-boson decay. 
%
%The
%second sample, the low $\Pt$ lepton sample, which has a substantially
%lower $\Pt$ requirement, is used to select leptons which are
%consistent with having come from semileptonic decay of bottom
%or charm hadrons.
%
%\subsection{High $\Pt$ leptons}
  
  A sample of events which contain high-$\Et$ electrons are selected 
from the Run 1 data sample by requiring the electron to have
$\Et>20\ \gevc$ and be in the central region of the
detector ($|\eta |<1$). Backgrounds from photon conversions and charged
hadrons are rejected by cutting on several
variables. Here we describe those cuts which provide the largest
discrimination against background.
A detailed discussion of other selection criteria can 
be found in Ref.~\cite{cdf-evidence}. 
Electrons are required to have a CTC track pointing to the electron shower 
in the CEM. The energy in the hadronic calorimeter divided by the energy 
detected in the electromagnetic calorimeter ($HAD/EM$) is required to be
less than 5\%. We also require that the energy of the shower divided by 
the momentum of the associated track is less than 1.8.
Electrons from photon conversions are removed with an efficiency of 
88\%~\cite{cdf-evidence} by requiring each electron to have a matching track 
in the VTX, and the invariant mass of this track with any other CTC track to 
be greater than 0.5 $\gevcc$.
The energy of high-$\Et$ electrons is measured using the calorimeter
energy in the tower to which the CTC track points plus the adjacent 
towers~\cite{wmass}. High-$\Et$ electrons are measured with a resolution 
of $\sigma(\Et)/\Et=13.5\%/\sqrt{\Et}\oplus 2\%$, where $\Et$ is in 
$\gev$.

%A number of additional requirement are made to remove the background
%from charged hadrons. 

%   Candidate muons from $W$ decays used in this analysis are also

  The high-$\Pt$ muon sample is selected by requiring that each event 
contain at least one muon candidate which has $\Pt>20\ \gevc$ and 
is in the central region of the
detector ($|\eta |<1$). Muon candidates are identified by a 
match between a track segment in CMU, CMP, or CMX and the CTC.
The primary backgrounds are from secondary
particles in charged hadron showers which ``punch through'' the
calorimeter and produce tracks in the muon chambers, and cosmic rays.
To reject the charged hadron background, the muon is required to
have an energy deposition in the calorimeters which is characteristic of a
minimum ionizing particle. Backgrounds from cosmic rays are rejected
by requiring that the track extrapolates back (in $r-\phi$) to within 3 mm 
of the beamline and that in the $r-z$ plane it is within 5 cm (at r=0) 
of the primary 
vertex. A number of other selection requirements are made which are
described in Ref.~\cite{cdf-evidence}. The momentum of high-$\Pt$ muons 
is measured by constraining the CTC track to the average beam position.
Its transverse momentum is measured with a
resolution of $\sigma(\Pt)/\Pt=0.11\% \Pt$, where $\Pt$ is in $\gevc$.

From these high-$\Pt$ lepton samples, we further select those events
in which the high-$\Pt$ lepton is isolated~\cite{defiso} from 
jet activity. For the lepton+jets analysis, we require
that there is only one $W\to\ell\nu$ candidate in the event.
The lepton $\ell$ is referred to as the primary lepton in the event.

\section{Jet reconstruction}
\label{s-jetrecon}

Jets are constructed from calorimeter tower information using a cone algorithm 
with cone radius $\Delta R \equiv\sqrt{\Delta \eta^2 +\Delta \phi^2 } = 0.4$.
The jet transverse energy is defined as the sum of the energy
deposited in calorimeter towers within the cone, multiplied by
$\sin{\theta}$, where $\theta$ is the polar angle of the $\et$-weighted 
centroid 
of the clustered towers. After correcting for 
the various energy losses (see Section~\ref{s-recon}), jets which do
not contain heavy flavor, and have
$\Pt>80~\gev$, have a transverse momentum
resolution of $\delta\Pt/\Pt \approx 12\% $.
A discussion of the jet reconstruction algorithm can be 
found in Ref.~\cite{three_jet,scale}.

%  Jets are reconstructed according to a fixed cone algorithm~\cite{cdfjetid}
%from the detected energy clusters in the event. Their transverse energy is taken to 
%be the scalar sum of the $\et$ measured in each calorimeter tower
%that lies within a fixed $\Delta R \equiv\sqrt{\Delta \eta^2 +\Delta \phi^2 }$,
%or cone size, relative to the $\et$ weighted center of the jet.  
%Unless stated otherwise, the cone size used in this analysis is 
%$0.4$.
% and the mass, m, is taken to be $5 \ \gevcc$
%if the parent parton is assumed to be a {\it b} quark and $0.5 \ \gevcc$ 
%otherwise. 

\subsection{Identification of $b$-quark jets}

   The identification of jets that arise from $b$ quarks ($b$-quark jets
or simply $b$ jets) plays an important role in
the analysis described in this report.
The identification relies on finding evidence for a $B$-hadron decay,
using two separate tagging algorithms.
% the secondary vertex tagger (SVX) and the soft lepton tagger (SLT). 

The silicon vertex (SVX) tag  algorithm~\cite{cdf-evidence,xsec} 
searches within a jet for displaced vertices due to $B$-hadron decays.
It is applied to jets that have 
raw $\Et>15 \ \gev$  and uses tracks which 
 are within $\Delta R<0.4$ of the jet axis and have hits in the silicon vertex 
detector. 
The algorithm allows for two passes. In the first pass, a secondary vertex
is required to have at least three tracks with $P_T>0.5 \ \gevc$, at least
one of which has $P_T>2.0 \ \gevc$. In the second pass,
tighter track quality cuts are applied, and a secondary vertex is required 
to have at least two tracks with $P_T>1.0 \ \gevc$, including at least one
with $P_T>2.0 \ \gevc$. 
This  algorithm has an efficiency of 
about 48\% for tagging at least one $b$ jet in a $\ttbar$ event.

  The soft lepton  tag (SLT) algorithm~\cite{cdf-evidence,kestenbaum} 
searches for additional leptons which are
consistent with having come from a semileptonic $B$-hadron decay.
The lepton is required to have  $\Pt>2\ \gevc$ and to be
within $\Delta R < 0.4$ of a jet with raw jet 
$\Et>8 \ \gev$. 
%A number of other 
%selection criteria are applied to improve the
%purity and reduce backgrounds~\cite{kestenbaum}. 
The efficiency for tagging at least one $b$ jet in a $\ttbar$ event 
with this algorithm  is about 15\%.  
%Because these leptons typically are of lower transverse momentum,
%we refer to these as ``soft lepton tags''. The

  The SVX algorithm obtains both  higher purity and higher efficiency 
then the SLT algorithm. However, the SLT algorithm is also employed
%has a significantly smaller efficiency 
%and a larger background than SVX, but 
for tagging $b$  jets because it uses nearly uncorrelated
information and adds significantly to the acceptance.

\section {Top mass candidate sample}
\label{s-data}

 Full reconstruction of candidate $\ttbar$ events 
is possible if the event has at least four jets and a
$W$ candidate decaying into either $e\nu$ or $\mu \nu$.
The majority of such events are not from $\ttbar$ production
but rather from the production of a $W$ boson in association with jets.
The fraction of these background
events containing at least one $b$ jet is of the order of
$1\%$~\cite{cdf-evidence}, while standard model $\ttbar$ decays
are expected to always have two $b$ jets.
Data samples with larger fractions of $\ttbar$ events can therefore
be formed by requiring evidence of $b$ hadrons in one
or more jets. 

%Of course this comes at the expense of some efficiency.
%As discussed in Section~\ref{s-recon}, the identification
%of {\it b}-jets is expected to improve the association of reconstructed
%jets to the partons from the decays of the top quarks. 

  To facilitate the measurement of the top quark mass, we apply
selection criteria which are expected to increase the fraction of $\ttbar$
events in the sample. We refer to these events as the Top Mass Candidate 
Sample, and they satisfy the following cuts:

\begin{enumerate}
 \item High-$\Et$ lepton trigger satisfied; The event should have
       an electron (muon) with $\et>20 \ \gev$ ($\Pt>20 \ \gevc$)
       and $|\eta |<1$. 
 \item \met, as calculated using the raw tower energies,
        is greater than 20 $\gev$. For events with a primary muon this
        \met includes a correction for the muon momentum.
 \item The candidate primary electron or muon track must be isolated and 
        of good quality (see Section~\ref{lepid}). Only one isolated lepton
       should be present.
 \item Candidate dilepton ($\tllx$) events, defined according to the selection
       criteria of Ref.~\cite{cdf-dilepton}, are rejected.
 \item Events with $Z$-boson candidates are removed. A $Z$-boson candidate
        is defined by two oppositely charged, same flavor
        high-$\Pt$ leptons ($\Pt>20\ \gevc$) that
        have an invariant mass between 75 and 105 $\gevcc$.
        Also, we remove the event if it includes a high-$\Pt$ 
        photon~\cite{phot} and 
        the $\ell \overline{\ell} \gamma$ invariant mass falls in the
        $Z$ mass window.
 \item The primary vertex of the event must be within $60 {\rm \ cm}$ 
       of $z=0.0$.  
 \item At least three jets with $\et>15 \ \gev $ and $|\eta|<2.0$.
 \item For events with exactly three jets satisfying criterion 7 above,
       we require at least one additional jet with 
       $\et>8 \ \gev$ and $|\eta|<2.4$.
       
 \item After the mass reconstruction is performed, events are required 
        to pass a goodness-of-fit cut,
       $\chi^2<10.0$, where the variable $\chi^2$ is defined in
       Section~\ref{s-algor}. 
%The cut is approximately 87\% efficient 
%       for $\ttbar$ events which pass criteria 1-8.
\end{enumerate}  

A sample of 324 events pass criteria 1--7, and are the same as those 
used in the CDF measurement of the $\ttbar$ production cross 
section~\cite{xsec}.  Criteria 1--9 are identical to those used in all our 
previous measurements of the
top quark mass~\cite{cdf-evidence,cdf-discovery}. After imposing
criteria 1--8, our sample consists of 163 events. The last
requirement removes 12 events, from which we obtain an inclusive 
sample of 151 $W$+multi-jet events. Thirty-four of the events
have SVX or SLT tagged jets. As discussed below, the Top 
Mass Candidate Sample is estimated to consist of approximately 74\% background.
Requiring the presence of $b$-tagged jets improves considerably the
signal-to-background ratio (see Section~\ref{s-optimize}).

\subsection{Mass subsamples}
\label{s-optimize}

To describe the mass subsamples which are used in this analysis,
it is helpful to decompose the Top Mass Candidate Sample into two
classes of events which are expected to have different 
signal-to-background ratios (S/B). Class I events have exactly three jets with 
$\Et>15\ \gev$ and $|\eta |<2$ and one or more additional jets 
with $\Et>8\ \gev$ and $|\eta |<2.4$.  Class II events 
have four or more jets with $\Et>15\ \gev$ and $|\eta |<2$. Because
of the larger amount of energy contained in the four leading 
(i.e., four highest $\Et$) jets, class II events have a larger
S/B than class I.

Previous measurements of the top quark mass at CDF used a combined
sample of $b$-tagged events~\cite{cdf-evidence,cdf-discovery} that
contained events from both class I and class II. Monte Carlo simulations 
show that the statistical uncertainty on the measured top quark
mass is reduced by 10\% by combining the results of separate fits on
three non-overlapping subsamples of events.
The first subsample consists of events that
have one and only one SVX tag. The second subsample consists of events in
which there are two SVX tags. The third one includes events that
have one or two SLT tags, but no SVX tags.  Further Monte Carlo
studies show that an additional 7\% improvement is obtained
by including the No Tag events from class II. The 75 No Tag events excluded 
from the Top Mass Candidate Sample are expected to have a 
background fraction of 93\%. Inclusion of
these events does not improve the statistical uncertainty on the top quark mass
measurement. To summarize, the four mass subsamples are~\cite{kirsten}:
%
%The statistical power of the data sample may be improved by 
%partitioning the data sample into several statistically
%independent subsamples, each characterized by
%the number and type (SVX, SLT) of observed $b$ tags. To 
%determine an optimal
%partitioning of the data sample, signal ({\small{HERWIG}}) and background
%({\small{VECBOS}}) Monte Carlo samples were used to generate a large
%number of simulated experiments. Each simulated experiment
%mixed a binomially-distributed number of background events 
%with top events according to the expected background fraction.
%The background fraction for a given sample was calculated as 
%discussed in Section~\ref{s-bgcalc}. In each of these experiments, the total 
%number of events in each subsample was constrained to the number 
%of observed data events. These studies showed that the smallest statistical
%uncertainty may be achieved by partitioning the data sample into 
%the following four non-overlapping subsamples~\cite{kirsten}: 
%
\begin{itemize}
 \item \underbar{SVX Double}: Events with two SVX tags;
 \item \underbar{SVX Single}: Events with one and only one SVX tag;
 \item \underbar{SLT}: Events with one or two SLT tags, but no SVX tags;
 \item \underbar{No Tags}: $\ge$ 4 jets with $\Et>15 \ \gev$ and $|\eta |<2$.
\end{itemize}
\noindent The numbers of data events in each of these subsamples are
shown in Table~\ref{t-sample}. In categorizing the events into the
subsamples, tags are only counted if they are on one of the four
highest $\Et$ jets. This choice is made because the four leading
jets are assumed to be the primary partons from the $\ttbar$ decay
(see Section~\ref{s-algor}).
%We refer to this sample selection as the {\it{Optimized}} analysis. 
Also shown in the table are the expected
S/B ratios, using the background estimates presented in Section~\ref{s-bgopt}.
The measurement of the top quark
mass in the lepton+jets channel is based on these four subsamples.
 
\begin{table}[ht]
\begin{center}
\begin{tabular}{ccc}
\hline\hline
Data Sample & Number of Events & Expected S/B \\ 
\hline
SVX Double  &  5  & 24 \\ 
SVX Single  & 15  & 5.3 \\
SLT         & 14  & 0.8 \\ 
No Tags     & 42  & 0.4 \\ 
\hline\hline
\end{tabular}
\caption{Subsamples used in the lepton+jets mass analysis
and the expected signal to background ratio (S/B) for each. See
Section~\ref{s-bgopt} for background estimates for these subsamples.}
\label{t-sample}
\end{center}
\end{table}

%% file: sim_and_back.tex
\chapter{SIMULATION AND BACKGROUND}
\label{sim_back}
This section describes the 
Monte Carlo methods used to simulate the signal and background events,
and the estimation of the background in the four mass subsamples.
For this purpose we use Monte Carlo programs that generate the 
signal and background processes contributing to the data sample, 
and a detector simulation which models the response of the detector 
to the final state particles. Unless otherwise noted, the Monte Carlo 
programs use the MRSD0$^{\prime}$~\cite{mrsd0} set of structure functions.
Detailed properties of $b$-hadron decay, based on observations from the CLEO 
experiment~\cite{cleo}, are included in all the Monte Carlo generators.
The response of the detector to the final state particles is parametrized
using distributions observed in data. See Section~\ref{s-recon} for details on
the calorimeter simulation.
%simulated using the CDF detector simulation and is tuned to reproduce observed 
%distributions in the data.
%~\cite{qflref} 
\section{Signal modeling}

   The simulation of $\ttbar$ events relies mainly on
the {\small HERWIG}~\cite{herwig} (Version 5.6) Monte Carlo program.
Additional checks are provided by both {\small PYTHIA}~\cite{pythia} 
(Version 5.7) and {\small ISAJET}~\cite{isajet} (Version 6.36).
{\small{HERWIG}} is based on the leading order QCD matrix elements for the 
hard process, followed by coherent parton shower evolution, cluster 
hadronization, and an underlying event model based on data.
{\small PYTHIA} is similar to {\small HERWIG} in that it is based on leading 
order QCD matrix elements; however, partons are fragmented using the Lund 
string model. {\small ISAJET} is a parton shower 
Monte Carlo program based on the leading-order QCD matrix elements for the
hard-scattering subprocess, incoherent gluon emission, and 
independent fragmentation of the outgoing partons.  

%  Detailed properties of $b$ hadron decay are included in all the
%Monte Carlo simulations based on observations from the CLEO 
%experiment ~\cite{cleo}. Unless otherwise noted, we use the 
%MRSD0$^{\prime}$~\cite{mrsd0} as the default set of
%structure functions.

\section{Background modeling}

  The Monte Carlo program used to study the kinematics of the 
background is {\small VECBOS}~\cite{vecbos}. This is a parton-level 
program based on tree-level matrix element calculations for
$W$+jets production. The simulated events produced by {\small VECBOS} 
contain a $W$ boson and up to four additional final state partons.
These partons are subsequently evolved and hadronized using a 
separate program~\cite{herfrag} derived from the parton 
shower model contained in the {\small HERWIG} Monte
Carlo generator. The CDF simulation program is then
used to simulate the detector response and produce the final sample 
of background events for further analysis. 

The {\small{VECBOS}}
events generated for this analysis use the $W+3$ parton matrix elements,
with the required additional jet being produced during parton
showering. The $Q^2$ scale of the hard scatter is set to the square of the 
average $\Pt$ ($\langle\Pt\rangle^{2}$) of the outgoing partons unless 
otherwise noted. 

The {\small VECBOS} Monte Carlo generator has been shown
to reproduce distributions of a wide range of kinematic variables
in a large sample of $W$+jets events~\cite{vecbos_w} in this
experiment. In addition, distributions of kinematic variables 
have been studied in $\ttbar$-depleted and $\ttbar$-enriched subsamples
of $W+\ge 3$ jet events in this experiment ~\cite{andy}. The
Monte Carlo simulations reproduce the distributions in both
subsamples when we use the expected fractions of {\small HERWIG}
(for $\ttbar$) and {\small VECBOS} (for background) events.
Further checks which demonstrate that {\small VECBOS} is 
appropriate for background modeling are given in Section~\ref{s-bkgd}.

\section{Background estimation}
\label{s-bgcalc}

  In the measurement of the top quark mass, we constrain the
fraction of background events in each of the mass subsamples to an 
expected value. The computation of the expected value for
each mass subsample is achieved by first computing the expected 
number of background events from relevant background processes 
for both class I and class II events (see Section~\ref{s-optimize}). 
Some of the background
processes are computed as absolute predictions while others are given as
a fraction of the number of background $W$-candidates in the data sample.
The expected $\ttbar$ and background fractions (which sum to unity) in 
the Top Mass Candidate Sample are
then estimated by using a maximum likelihood fit which
compares the observed rates of events with
SVX and SLT tags with predicted rates. The predicted rates, which
use estimates of the tagging probabilities for $\ttbar$ and background 
events, depend on these fractions. The $\ttbar$ fraction
is a free parameter in the fit, and is allowed to vary to optimize
the agreement between the observed and predicted numbers of tagged
events. The fitted
$\ttbar$ fraction in the Top Mass Candidate sample is then
combined with SVX and SLT tagging probabilities to evaluate the
expected $\ttbar$ and background contribution in each of the mass subsamples.
%A maximum likelihood
%technique is used to evaluate the background fraction in the Top Mass
%Candidate Sample, and subsequently in the mass subsamples.
The same principle has been used to measure the $\ttbar$ cross section 
using $W+\ge 3$ jets events ~\cite{xsec,ptohos}.

  The tagging probabilities we use, and the contributions of various
background channels, are similar to those in Ref.~\cite{ptohos}, but
are not identical because of differences in the event selections and
the exact tagging rules. The event selections used in this paper
require a fourth jet and impose a $\chi^2$ cut on the kinematic mass
fit (described in Section~\ref{s-algor}). The tagging rule used here,
requires that the SVX and SLT tags are counted only if
they are on one of the four leading jets in the event.
The resulting differences in tagging probabilities and backgrounds 
are determined using the {\small HERWIG} and {\small VECBOS} Monte Carlo
simulations.

%  The calculation of the expected background content in each subsample
%starts from the background estimate in the $W+\ge$3-jets sample given in 
%Ref.~\cite{xsec}. This sample includes events which pass criteria 1-7
%in Section~\ref{s-data} and consists of 324 events of which 34 are
%tagged by the SVX algorithm and 40 by the SLT
%algorithm. The expected background in these SVX and SLT
%tagged events is 9.2$\pm$1.5 and 22.6$\pm$2.8 events respectively.
%The extrapolation to the Top Mass Candidate Sample takes into account 
%the additional requirement of a fourth jet which reduces the sample 
%to 163 events. It also accounts for the $\chi^2<10$ cut on the kinematic
%fit of the event to the $\ttbar$ hypothesis, and the fact that 
%SVX and SLT tags are only counted if they are on one of 
%the four leading (highest $\Et$) jets. 

\subsection{Inputs into the background calculation}
\label{s-inputs}

The inputs into the calculation are the background processes, their 
expected rates, and the corresponding SVX and SLT tagging
probabilities. The rates and tagging probabilities are estimated
for both the class I and class II events of the Top Mass Candidate
Sample.
%and are defined as follows: class I events are those which have exactly 3 
%jets 
%with $\Et>15\ \gev$ and $|\eta |<2$ and 1 or more jets with $\Et>8\ \gev$ and 
%$|\eta |<2.4$; class II events have 4 or more jets with $\Et>15\ \gev$ and 
%$|\eta |<2$. 
Of the 151 events in the Top Mass Candidate Sample, 87 are in
class I and 64 in class II. 

   The background processes are classified into two categories: contributions
which are computed as an absolute number of events, and contributions
which are calculated as a fraction of the number of background candidate 
$W$+jets events ($N_W$) in the data sample. In the latter case, the 
contribution includes $Z$+jets events that pass the lepton+jets selection criteria.
The background processes considered are listed in Tables~\ref{bg_inputs1} 
and \ref{bg_inputs2} for the two classes (the processes are the same for both 
classes). The expected numbers of background events from the different 
processes are also given in the tables. 

  For the first six processes we have absolute predictions. 
%For processes 1-6 in class I 
%events, the expected background was estimated by scaling the estimated background from
%each source to $W+3$ jet events~\cite{xsec} by the fraction, $f, of $W$+3 jet events
%which have a fourth jet with $8 < \Et < 15 \gev$ and $|\eta |<2.4$. To estimate this
%fraction, we first compare the ratio of $W$+4 jet to $W$+3 jet events from {\small{VECBOS}}
%with that found from background processes 1-6. For processes 2-6, these ratios,
%are consistent, and therefore we use {\small{VECBOS}} to estimate the fraction $f$.
%For source 1, the ratio $W$+4 jet to $W$+3 jet events is nearly twice as large
%as that predicted by the {\small{VECBOS}} simulation, and therefore we do not use
%it by itself to evaluate $f$. To estimate $f$ for source 1, we consider source
%1 to be a convenient admixture of {\small{VECBOS}} (80\%) and {\small{HERWIG}} $\ttbar$
%(20\%) events in order to match the ratio of $W$+4 jet to $W$+3 jet events. We then
%calculate $f$ for {\small{VECBOS}} and {\small{HERWIG}} $\ttbar$ events and combine
%those fractions with an 80\% to 20\% weighting. For class II events, such a 
%procedure is not necessary, since background calculations in $W+4$ jet events 
%have been calculated. 
For the $W$+jets and $Z$+jets processes we have predictions for each process 
relative to their sum. The last two columns in Tables~\ref{bg_inputs1} 
and \ref{bg_inputs2} give the SVX and SLT tagging probabilities
per event for each background process. The probabilities in rows 1--13
are for cases where there is a real displaced vertex or a real soft lepton.
Each of the background processes can also  contribute fake SVX and SLT tags
(mistags), 
and these probabilities are given in row 14. In either case, the SVX and SLT 
tagging probabilities include the requirement that the tag is on one of the 
four leading jets and take into account the $\chi^2$ cut on the kinematic 
mass fit.

\begin{table}[htpb]
\begin{center}
\begin{tabular}{ccccc}
\hline\hline
Item & Background     &  Number of & $\epsilon_{SVX}$ & $\epsilon_{SLT}$ \\
\#   &  process        &   events   &     (\%)         &      (\%)        \\
\hline
\multicolumn{5}{c}{Absolute backgrounds} \\
\hline
1    & non-$W/Z$      & 5.7$\pm$0.8 & 4.3$\pm$2.2     &  2.5$\pm$1.8     \\
2    & $WW$           & 0.7$\pm$0.1 & 5.8$\pm$1.7     &  1.3$\pm$0.7     \\
3    & $WZ$           & 0.1$\pm$0.0 & 5.8$\pm$1.7     &  1.3$\pm$0.7     \\
4    & $ZZ$           & 0.0$\pm$0.0 & 5.8$\pm$1.7     &  1.3$\pm$0.7     \\
5    & $Z\to \tau\tau$& 0.9$\pm$0.1 & 3.5$\pm$2.5     &  4.6$\pm$4.6     \\
6    & Single Top     & 0.4$\pm$0.1 & 30.6$\pm$7.0    &  9.0$\pm$2.4     \\
\hline
\multicolumn{5}{c}{$W/Z$+jets backgrounds} \\
\hline
7    & $W\bbbar$  & $(0.028\pm 0.004)N_W^{I}$ & 22.7$\pm$3.1 & 7.0$\pm$1.9 \\ 
8    & $W\ccbar$  & $(0.056\pm 0.013)N_W^{I}$ &  5.7$\pm$1.0 & 5.5$\pm$1.2 \\ 
9    & $Wc$       & $(0.053\pm 0.016)N_W^{I}$ &  3.7$\pm$0.5 & 6.3$\pm$1.8 \\
10   & $Z\bbbar$  & $(0.005\pm 0.002)N_W^{I}$ & 22.7$\pm$2.0 & 7.0$\pm$1.9 \\
11   & $Z\ccbar$  & $(0.005\pm 0.002)N_W^{I}$ &  5.7$\pm$1.0 & 5.5$\pm$1.2 \\
12   & $Zc$       & $(0.001\pm 0.001)N_W^{I}$ &  3.7$\pm$0.5 & 6.3$\pm$1.8 \\
13   & $W/Z+u,d,s$& $0.85N_{W}^{I}$           &  0.0         & 0.0         \\
\hline
\multicolumn{5}{c}{Mistag probabilities }  \\
\hline
14   & 1-13      &                  &  0.4$\pm$0.1 & 3.2$\pm$0.4 \\
\hline\hline
\end{tabular}
\end{center}
\caption{Backgrounds which contribute to class I events in
the Top Mass Candidate Sample. Shown are
the contributing processes, their estimated contribution, and the
SVX and SLT tagging probabilities per event for each process. 
Backgrounds whose absolute rate is calculable (a total of
$N_{abs}^{I}$ events) 
are given by 1--6. Backgrounds that are given as fractions of the number of 
$W/Z$+jets events in the data sample are given by 7--13. $N_{W}^{I}$ is the 
total number of $W/Z$+jets background events in class I. 
All background processes contribute to SVX and SLT mistags, with the 
probabilities listed in row 14. There are 87 events in class I.}
\label{bg_inputs1}
\end{table}

\begin{table}[htpb]
\begin{center}
\begin{tabular}{ccccc}
\hline\hline
Item & Background     &  Number of & $\epsilon_{SVX}$ & $\epsilon_{SLT}$ \\
\#   &   process       &    events  &     (\%)         &     (\%)         \\
\hline
\multicolumn{5}{c}{Absolute background calculations} \\
\hline
1    & non-$W/Z$      & 5.5$\pm$1.7 &  4.3$\pm$2.2    &    2.5$\pm$1.8   \\
2    & $WW$           & 0.7$\pm$0.2 &  5.8$\pm$1.7    &    1.3$\pm$0.7   \\ 
3    & $WZ$           & 0.1$\pm$0.0 &  5.8$\pm$1.7    &    1.3$\pm$0.7   \\
4    & $ZZ$           & 0.1$\pm$0.0 &  5.8$\pm$1.7    &    1.3$\pm$0.7   \\
5    & $Z\to \tau\tau$& 0.7$\pm$0.3 &  3.5$\pm$2.5    &    4.6$\pm$4.6   \\
6    & Single Top     & 0.3$\pm$0.1 & 30.6$\pm$7.0    &    9.0$\pm$2.4   \\
\hline
\multicolumn{5}{c}{$W/Z$+jets backgrounds} \\
\hline
7    & $W\bbbar$  & $(0.054\pm 0.012)N_W^{II}$ & 27.4$\pm$2.7 & 7.5$\pm$2.6 \\ 
8    & $W\ccbar$  & $(0.087\pm 0.025)N_W^{II}$ &  6.0$\pm$1.0 & 5.6$\pm$1.2 \\ 
9    & $Wc$       & $(0.073\pm 0.022)N_W^{II}$ &  3.8$\pm$0.5 & 6.3$\pm$1.8 \\
10   & $Z\bbbar$  & $(0.003\pm 0.003)N_W^{II}$ & 27.4$\pm$2.7 & 7.5$\pm$2.6 \\
11   & $Z\ccbar$  & $(0.003\pm 0.003)N_W^{II}$ &  6.0$\pm$1.0 & 5.6$\pm$1.2 \\
12   & $Zc$       & $(0.001\pm 0.001)N_W^{II}$ &  3.8$\pm$0.5 & 6.3$\pm$1.8 \\
13   & $W/Z+u,d,s$& $0.78 N_{W}^{II}$          &  0.0         & 0.0         \\
\hline
\multicolumn{5}{c}{Mistag probabilities }       \\
\hline
14   & 1-13      &                        &  0.4$\pm$0.1 & 4.2$\pm$0.5 \\
\hline\hline
\end{tabular}
\end{center}
\caption{Backgrounds which contribute to class II events in
the Top Mass Candidate Sample. Shown are
the contributing processes, their estimated contribution, and the
SVX and SLT tagging probabilities per event for each process. 
Backgrounds whose absolute rate is calculable (a total of
$N_{abs}^{II}$ events) 
are given by 1--6. Backgrounds that are given as fractions of the number of 
$W/Z$+jets events in the data sample are given by 7--13. $N_{W}^{II}$ is the 
total number of $W/Z$+jets background events in class II. 
All background processes contribute to SVX and SLT mistags with the 
probabilities shown in row 14.  There are 64 events in class II.}
\label{bg_inputs2}
\end{table}

  The expected backgrounds and tagging probabilities are calculated as follows.
The non-$W/Z$ background is calculated directly from the data ~\cite{xsec}. 
The $WW$, $WZ$, and $ZZ$ background rates are evaluated by multiplying the 
acceptances for these processes as determined from the {\small PYTHIA} Monte 
Carlo simulation by their production cross sections ~\cite{wzxs}.
The $Z\to \tau\tau$ background is estimated using the {\small PYTHIA} 
Monte Carlo simulation. The normalization is obtained by scaling the
number of reconstructed $Z\to\ell\ell +\ge 1$-jet events in the simulation 
to the number observed in the Run 1 data sample. For single top quark 
production, we use the {\small PYTHIA} and {\small HERWIG} Monte Carlo programs
to evaluate the acceptances for the $W^*\to tb$ and $W$-gluon fusion 
processes respectively. The production cross sections are normalized to 
the published theoretical values ~\cite{single_top}.

  The expected fractions of $W\bbbar$ and $W\ccbar$ events
in the data sample are evaluated using the {\small HERWIG} and {\small VECBOS}
Monte Carlo programs. For each jet multiplicity bin, the expected 
background is given by the product of the corresponding background fraction,
tagging probability and the number of $W$-candidate events.
%
%These Monte Carlo simulations are used to 
%evaluate the fraction of events  
%in each jet multiplicity bin which pass the analysis cuts described in
%Section~\ref{s-data}. The background is then given by the
%product of these fractions with the number of $W$-boson candidates in
%the data sample in
%the respective jet multiplicity bin. 
The $Wc$ background is estimated
from {\small{HERWIG}} in an analogous way to what is done for
the $W\bbbar$ and $W\ccbar$ backgrounds. The  $Z\bbbar$, $Z\ccbar$
and $Zc$ backgrounds are calculated using a combination of {\small{HERWIG}},
{\small{PYTHIA}} and {\small{VECBOS}}. The simulations show that 
in both the $Z$+1 jet and $Z$+2 jet multiplicity bins $Z\bbbar$ events are 
approximately twice as likely to
pass our kinematic cuts as $W\bbbar$. The corresponding ratio for $Z\ccbar$ to $W\ccbar$
is approximately 1, and $Zc/Wc$ is about 0.3. We assume that these scalings 
also hold in the higher jet multiplicity bins. 
The $Z\bbbar$, $Z\ccbar$ and $Zc$ background rates are thus obtained
by scaling the $W\bbbar$, $W\ccbar$, and $Wc$ rates by $2.0\pm 0.5$, $1.0\pm0.3$, 
and $0.3\pm 0.15$ respectively. The overall $Z/W$ normalization is determined
from the data sample, and is 0.092$\pm$0.020 for events in class I
and 0.030$\pm$0.030 for events in class II. 

  The SVX and  SLT tagging probabilities in lines
1-13 in Tables~\ref{bg_inputs1} and \ref{bg_inputs2} give the probability
per event, that one or more jets will be tagged due to the decay of a 
long-lived particle (i.e., a $b$ hadron, a $c$ hadron, or a $\tau$). 
For backgrounds which are computed using Monte Carlo programs,
the tagging probabilities are evaluated by simulation of the detector's
response to the final state particles of each of the background processes.
For SVX tags, the probabilities are calculated using only jets which have 
an uncorrected $\Et>15~\gev$ and $|\eta|<2$. For SLT tags the probabilities
include all jets which have an uncorrected $\Et>8~\gev$ and $|\eta|<2.4$.
The tagging probabilities for $W+u,d,s$ are set to zero since these
events have a negligible contribution from long-lived particles.

  The SVX and SLT mistag probabilities (line 14 in Tables~\ref{bg_inputs1} 
and \ref{bg_inputs2}) are estimated 
by applying  ``mistag-matrices'' to the jets in each event of the
Top Mass Candidate Sample. The mistag matrices ~\cite{cdf-evidence} 
for SVX and 
SLT tags are measured from inclusive jet data and
describe the probability for a jet that does not contain heavy
flavor to be tagged by the SVX and SLT algorithms
respectively. Monte Carlo simulations show a lower mistag rate in
background events than in $\ttbar$ events, with a ratio of 0.70$\pm$0.05
for both SVX and SLT tags. This ratio is included in the
mistag probabilities shown in Tables~\ref{bg_inputs1}-\ref{ttbar-eff}.
The effect of using equal mistag probabilities for $\ttbar$ and background has 
been investigated, and the resulting background numbers change by a 
negligible amount.

  Tagging probabilities for $\ttbar$ events were determined using the 
{\small{HERWIG}} Monte Carlo program. Additional checks of these probabilities 
were provided by both the {\small{PYTHIA}} and {\small{ISAJET}} simulations. 
The probabilities for tagging at least one $b$-quark jet in a $\ttbar$ event
are shown in Table~\ref{ttbar-eff}. Also shown are the probabilities for
tagging a jet which does not contain heavy flavor (mistags).
As before, the SVX and SLT tagging probabilities 
include the requirement that the tag is on one of the four leading
jets and require the $\chi^2$ cut on the kinematic mass fit.

%  The final set of inputs are the probabilities for the events
%to pass the $\chi^2$ cut on the kinematic mass fit (see Section~\ref{s-algor}). 
%The probabilities for  $\ttbar$ and background events are shown in 
%Table~\ref{chi-eff}. They are determined by applying the kinematic mass fit
%to simulated events generated with the
%{\small{HERWIG}} and {\small{VECBOS}} Monte Carlo simulations. 
%The background numbers presented in Tables~\ref{bg-inputs1} and \ref{bg-inputs2}
%do not include these $\chi^2$ cut probabilities.

\begin{table}[htpb]
\begin{center}
\begin{tabular}{cccccc}
\hline\hline
         & \multicolumn{5}{c}{Tagging probabilities per $\ttbar$ event} \\
         & \multicolumn{2}{c}{$\epsilon_{SVX}(\%)$} && 
           \multicolumn{2}{c}{$\epsilon_{SLT}(\%)$} \\
\cline{2-3}\cline{5-6}
         & Class I      & Class II       && Class I       & Class II      \\
\hline
Real tags & 44.8$\pm$4.5 &  49.9$\pm$5.0  &&  14.9$\pm$1.5 &  14.8$\pm$1.5 \\
Mistags & 0.6$\pm$0.1  &  0.7$\pm$0.1   &&   4.8$\pm$0.5 &   6.4$\pm$0.7 \\
\hline\hline
\end{tabular}
\end{center}
\caption{SVX and SLT tagging probabilities in $\ttbar$ events for class I and 
class II events. Shown are the probabilities for tagging one or more jets 
which contain $b$ or $c$ quarks (real tags) and the probabilities for tagging 
one or more jets which do not contain $b$ or $c$ quarks (mistags).}
\label{ttbar-eff}
\end{table}

%\begin{table}[ht]
%\begin{center}
%\begin{tabular}{|c|c|c|}
%\hline
%        & \multicolumn{3}{c|}{$\chi^2$ Cut Probability (\%)} \\
%\hline
%Tag Type           &    $\ttbar$      &  Background  \\
%\hline
%SVX Only &    87$\pm$1      &  78$\pm$2    \\
%SLT Only &    94$\pm$2      &  85$\pm$2    \\
%{\small{SVX+SLT}}  &    77$\pm$3      &  77$\pm$3    \\
%No Tag           &    96$\pm$1      &  95$\pm$1    \\
%SVX Double &    ??$\pm$?      &  ??$\pm$?    \\
%\hline
%\end{tabular}
%\end{center}
%\caption{Probabilities for various subsamples of events to pass the
%$\chi^2$ cut on the kinematic mass fit (see Section~\ref{s-algor}
%for details on the kinematic mass fitter).}
%\label{chi-eff}
%\end{table}
  
\subsection{$\ttbar$ and background fractions in each event class}
\label{est_top}

We first estimate the fractions of background and $\ttbar$ events in
each of the
two event classes defined in the preceding section. 
%
% The procedure is
%  the same for each class. Then we use these background fractions, plus
%  relevant tagging probabilities, to estimate background fractions
%  in each of the four mass subsamples defined in Section~\ref{s-optimize}.
%
For each event class, we compare the expected rates of tags with
the observed rates in each of four subsamples. The subsamples are 
events with (i) only SVX tags, (ii) only SLT tags, (iii) both SVX and SLT 
tags, and (iv) No Tags. The division into these subsamples was chosen to 
optimize, according to Monte Carlo studies, the background fraction estimate, 
and is not identical to the mass subsample division.  Note that for subsample 
(iii) the tags can be on the same jet or on different jets.

The expected numbers of events in each of these subsamples (indexed by $j$) 
can be calculated as a function of the numbers of $\ttbar$ events 
($N_{\ttbar}$) and non-top $W$+jets events in the event class, using an 
expression of the form:
\begin{equation}
N^{exp}_{j} = a_j\times  N_{\ttbar} + \sum_k c_j^k\times N_{abs,j}^k + 
\sum_i d_j^i b_j^i\times N_W  
\label{eq:bg-eq} 
\end{equation}
Here the first term gives the expected contribution from $\ttbar$ events,
and the last two terms give the expected number of events from background
processes.
The indices $k$ and $i$ refer to the background 
processes 1-6 and 7-13, respectively, in Tables~\ref{bg_inputs1} and 
\ref{bg_inputs2}. The parameter $a_j$ is the 
(SVX or SLT) tagging probability for $\ttbar$ events 
in the $j^{th}$ subsample, while $c_j^k$ and $b_j^i$ are the 
tagging probabilities, including those for mistags, for background 
processes $k$ and $i$.
The quantities represented by $d_j^i$ 
%is the acceptance for the $i^{th}$ $W$+jets background source to the $j^{th}$ 
%subsample and 
are the coefficients of $N_W^I$ and $N_W^{II}$ in
Tables~\ref{bg_inputs1} and ~\ref{bg_inputs2}.
%The parameters $b_j^i$ and $c_j^k$ are the 
%tagging probabilities for background processes 1-6 and 7-14, and are given 
%in columns 3 and 4 of the tables. 
The parameter $N_{abs,j}^k$ is the expected number of background events 
from the $k^{th}$ process. Equation~\ref{eq:bg-eq}
applies separately to both class I and class II events. 

The tagging probabilities in the expression above are derived from the values
in Tables~\ref{bg_inputs1}--\ref{ttbar-eff}, apart from some correlation terms.
Correlation terms between real and mistag probabilities and between SVX and SLT
tag probabilities are included in the calculations, but these terms are 
relatively small and their effect on the final result is negligible.

  To determine the background and $\ttbar$ contributions to class I and 
class II events, we constrain
the total number of $\ttbar$ and background events (i.e., summed over the
subsamples) to be equal to the observed number of events
in each class. Then we have just one parameter for each class, the fraction,
$f_{\ttbar}$, of $\ttbar$ events (or, equivalently, the fraction of background
events). A given value of $f_{\ttbar}$ determines values of $N_{\ttbar}$ and
$N_W$ to be used in Eq.~\ref{eq:bg-eq}. A maximum likelihood method is used to
determine a best estimate of $f_{\ttbar}$. The likelihood has the form:

%    For any given fraction, $f_{\ttbar}$, of $\ttbar$ events we calculate 
%  the expected number of events in subsamples (i)-(iv), with the sum
%  constrained to the observed sum. The calculation uses the rates and
%  probabilities given in Table~\ref{bg_inputs1} or ~\ref{bg_inputs2} and 
%  ~\ref{ttbar-eff}. Correlation terms
%  between real and fake probabilities, and between SVX and SLT probabilities
%  are included, but these terms are relatively small and their effect on the
%  result is negligible. A maximum likelihood method is used to determine
%  a best estimate of $f_{\ttbar}$.  The likelihood has the form:

\begin{equation}
 L = \prod_i {F_j^i(f_{\ttbar})}
\label{eq:likef} 
\end{equation}

\noindent where the $i^{th}$ event falls into subsample $j$ and the expected 
fraction of events in subsample $j$ is $F_j^i(f_{\ttbar})$.

The results of the maximum likelihood fit are 
$f_{\ttbar} = 0.13^{+.07}_{-.06}(stat.)\pm .01(syst.)$ for class I, and 
$f_{\ttbar} = 0.45^{+.12}_{-.11}(stat.)\pm .05(syst.)$ for class II. 
The statistical uncertainties correspond to changes in $\ln L$ from the maxima
by 0.5 units. The systematic uncertainties result from adding in quadrature 
the many contributions due to changing all the relevant input rates and 
probabilities one at a time by their stated uncertainties. 
These $f_{\ttbar}$ values imply that $\ttbar$ events comprise 
$11.5^{+6.4}_{-5.2}$ of the 87 class I events and $28.5^{+8.2}_{-7.6}$ of the 
64 class II events.  The numbers of $\ttbar$ and background events are 
summarized in Table~\ref{tmcs-makeup}.

To check that the model we are using is reasonable, we compare the 
expected numbers of events in each subsample with the observed numbers.  
The comparison is presented in Table~\ref{nev_cat},
%  where the expected sums are constrained to the observed sums, 
and shows reasonable agreement between expected and observed numbers.

\begin{table}[htpb]
\begin{center}
\begin{tabular}{ccccc}
\hline\hline
  Process    &    Class I          &       Class II     \\
\hline
\rule{0mm}{5mm} $\ttbar$    & $11.5^{+6.4}_{-5.2}$ &
  $28.5^{+8.2}_{-7.6}$ \\
\rule{0mm}{5mm} Absolute Backgrounds & $7.9\pm 0.9$         & $7.4\pm 1.8$    \\
\rule{0mm}{5mm} $W/Z$+jets  & $67.6^{+5.2}_{-6.4}$ & $28.1^{+7.6}_{-8.2}$ \\
\hline\hline
\end{tabular}
\end{center}
\caption{Estimated composition of the Top Mass Candidate Sample for
class I and class II events using the background likelihood fit
described in the text. Shown are the expected contributions from $\ttbar$
events, absolute backgrounds (as listed in lines 1-6 in
Tables~\ref{bg_inputs1} and \ref{bg_inputs2}) and $W/Z$+jets events.
The sum of each column is constrained to the number of observed 
events in the Top Mass Candidate Sample.}
\label{tmcs-makeup}
\end{table}

\begin{table}[htpb]
\begin{center}
\begin{tabular}{ccccc}
\hline\hline
 & \multicolumn{2}{c}{Observed}    & \multicolumn{2}{c}{Expected} \\
Subsample             & Class I    & Class II & Class I &  Class II \\
\hline
Only SVX tags         &      3     &   10     &     5.6 &   12.4    \\
Only SLT tags         &      6     &    8     &     4.2 &    4.8    \\
Both SVX and SLT tags &      3     &    4     &     1.1 &    3.0    \\
No tags               &     75     &   42     &    76.0 &   43.8    \\
\hline
Total                 &     87     &   64     &     87  &   64      \\
\hline\hline
\end{tabular}
\end{center}
\caption{The number of observed and expected events in the four subsamples.
The expectation values are based on the background likelihood fit described
in the text. The events are separated into class I and class II events.}
\label{nev_cat}
\end{table}

\subsection{$\ttbar$ and background events in the $\ell$+jets mass subsamples}
\label{s-bgopt}

Having found the numbers of  $\ttbar$  and background events for the samples
in class I and class II, we can go to the next step, i.e., compute the 
expected numbers of top and backgrounds events in the mass subsamples.
To arrive at estimated $\ttbar$ fractions in the mass subsamples, we need 
probabilities for two SVX tags in an event.  We must also combine the
$\ttbar$ fractions for class I and class II events in each tagged subsample.
The untagged mass subsample only contains class II events.

For most of the background channels the probabilities for two real SVX
tags (i.e., tags due to $b$-hadron, $c$-hadron, or $\tau$ decays) are very
small or zero. The non-negligible probabilities are given in 
Table~\ref{two-tag-prob}. Our calculations for the SVX Double subsample do
allow appropriately for real and fake tags in all channels. The 
probabilities for events to enter into one of the four mass subsamples
use the probabilities in Tables~\ref{bg_inputs1}, ~\ref{bg_inputs2}, 
and ~\ref{two-tag-prob}, and are computed as follows:
\begin{eqnarray}
 P({\rm SVX\; Single})     & = & P({\rm SVX}) - P({\rm SVX\; Double}) \\
 P({\rm SLT\; (no\; SVX)}) & = & P({\rm SLT}) - P({\rm SVX}\otimes{\rm SLT}) \\
 P({\rm No\; Tag})         & = & 1 - P({\rm SVX}) - P({\rm SLT\; (no\; SVX)})
\end{eqnarray}
The symbol $\otimes$ in the second line is used to
signify the probability of obtaining both an SVX and SLT tag
in the same event.

\begin{table}[htpb]
\begin{center}
\begin{tabular}{cccc}
\hline\hline
      & \multicolumn{3}{c}{Double SVX Tag Probability per Event (\%)}\\
  Process           &\rule{10mm}{0mm}&    Class I     & Class II      \\
\hline
 $W(Z)\bbbar$       &             &   1.9$\pm$0.5  &  3.7$\pm$1.0  \\
%%%% $W(Z)\ccbar$   &             &   0.1$\pm$0.1  &  0.1$\pm$0.1  \\   
 $\ttbar$           &             &   12.0$\pm$2.4 &  16.4$\pm$3.2 \\
\hline\hline
\end{tabular}
\end{center}
\caption{The probability per event to have two 
SVX-tagged jets for $W(Z)\bbbar$ 
%and $W(Z)\ccbar$ 
background processes and for $\ttbar$ events. Double SVX-tag 
probabilities for all other background processes are negligible and are set to zero.
The probabilities are evaluated for class I and class II events.}
\label{two-tag-prob}
\end{table}

The computation of the expected $\ttbar$ fraction in each of the mass 
subsamples proceeds as follows. First, for each mass subsample, we calculate 
the expected $\ttbar$ fraction in each event class. Then, the $\ttbar$ 
fractions for class I and II events are combined into a single $\ttbar$
fraction.  For each class, the expected $\ttbar$ fraction, $g^m$, in mass 
subsample $m$ is given by the following expression:

\begin{equation}
 g^m = {N_{\ttbar, exp}^m \over N_{tot, exp}^m} \\
 \label{eq:fr-m}
\end{equation}

\noindent An expression of this form applies to both class I and class II
events in each mass subsample. The numerator, $N_{\ttbar}^m$ is the expected 
number of $\ttbar$ events in mass subsample $m$, and the denominator
is the expected total number ($\ttbar$ $+$ background) of events.
The expected total number of events in 
subsample $m$ is calculated using an expression of the form shown in 
Eq.~\ref{eq:bg-eq} (replace $j$ with $m$, and use the tagging probabilities
appropriate for the mass subsamples). The $\ttbar$ fractions for
each event class in mass subsample $m$ are then combined into a single 
$\ttbar$ fraction, $f^m_{\ttbar}$, using the following expression:

\begin{equation}
f^m_{\ttbar} = {N^m_I g^m_I + N^m_{II} g^m_{II}\over N^m_I + N^m_{II} }.
\label{eq:scale-eq}
\end{equation}

  \noindent Here, $N^m_I$ and $N^m_{II}$ are the observed numbers of events, and 
  $g^m_I$ and $g^m_{II}$ are the predicted fractions of $\ttbar$ events in the 
  two event classes in subsample $m$. The expected number of $\ttbar$
  events is given simply by the numerator of Eq.~\ref{eq:scale-eq}.
%These fractions are computed by
%  dividing the expected number of $\ttbar$ events in subsample $m$
%  by the total expected number 
%  of events in the same subsample. These expected numbers are calculated from
%  the estimated number of $\ttbar$ and background events (Table~\ref{tmcs-makeup}), 
%  and the appropriate tag probabilities and event rates 
%  (see Sections~\ref{s-inputs}-\ref{est_top}). 
  For the No Tag mass subsample we have  $f^m_{\ttbar} = g^m_{II}$,
  because only class II events contribute.
    Table~\ref{total-ev} shows the observed number of events,
  the expected total number of events ($\ttbar$ + background) and the
  expected contribution from $\ttbar$ alone. Note that the total number of 
 $\ttbar$ events in each class is the same as that of Table~\ref{tmcs-makeup}
  as expected.

\begin{table}[htpb]
\begin{center}
\begin{tabular}{cccccccc}
\hline\hline
             & \multicolumn{3}{c}{Class I} && \multicolumn{3}{c}{Class II} \\ 
\cline{2-4}\cline{6-8}
             &  Total  & Total  & $\ttbar$ && Total  & Total  & $\ttbar$   \\
Subsample    &   Obs.  &  Exp.  &   Exp.   &&  Obs.  &  Exp.  &   Exp.     \\
\hline
SVX Double   &     3   &  1.5   &  1.4     &&   2    &  4.8   &  4.7       \\
SVX Single   &     3   &  5.3   &  3.7     &&  12    & 10.7   &  9.6       \\
SLT (no SVX) &     6   &  4.2   &  1.2     &&   8    &  4.8   &  2.8       \\
No Tags      &    75   & 76.0   &  5.2     &&  42    & 43.8   & 11.4       \\
\hline  
Total        &    87   &  87    & 11.5     &&  64    & 64     & 28.5       \\
\hline\hline
\end{tabular}
\end{center}
\caption{The number of observed events, $N_{obs}$, in the mass subsamples,
the total expected number of events, and the expected number of $\ttbar$ 
events.  Events in class I with No Tags are not used in the top quark mass 
analysis.}
\label{total-ev}
\end{table}

\begin{figure}[ht]
%\leavevmode
%\begin{center}
\hspace{0.5in}
\epsfysize=6.5in
%\epsffile{backg.ps}
%\epsffile{logl_v_nb_new.ps}
\epsffile{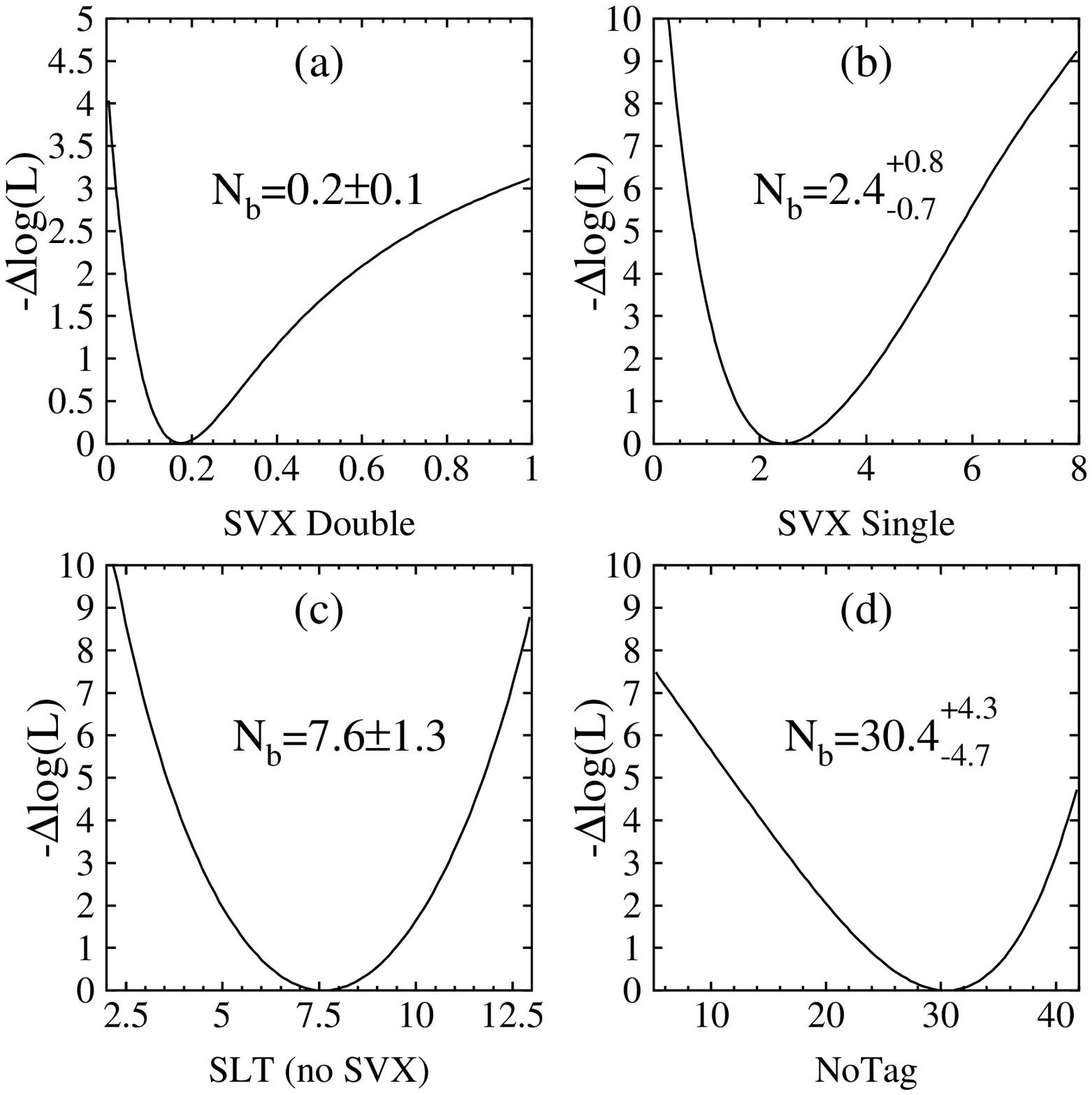}
%\end{center}
\caption{The negative log-likelihood function for obtaining a
 given number of background events in each mass subsample: 
%The subsamples correspond to the four mass subsamples 
(a) SVX Double tags, (b) SVX Single tags, 
(c) SLT (no SVX) tags, and (d) No Tag events.} 
\label{backg}
\end{figure}

The $f^m_{\ttbar}$ have both statistical and systematic
uncertainties. The statistical uncertainties are asymmetric, and are
convoluted with the systematic uncertainties separately for classes I and II,
and the results are in turn convoluted.  The systematic uncertainties are
assumed to be Gaussian. The end result is a likelihood function for
each $f^m_{\ttbar}$, which is used in the mass likelihood fit described 
in Section~\ref{s-like}. These negative-log-likelihood distributions 
as a function of the expected number of background events are shown in 
Fig.~\ref{backg}.

Finally, the estimated composition of each mass subsample can be calculated 
from the $f^m_{\ttbar}$ values and the various tagging probabilities
and event rates. The result is shown in Table~\ref{bg-table}.
The contributions from mistags are included in the sums for each process. 
From the Table we see that 80\% of the background is from $W$+jets and 
$Z$+jets, and another 15\% is from non-$W/Z$ events, i.e., from multijets 
(including $\bbbar$ events). The remaining 5\% is from diboson events, 
$Z\to\tau\tau$, and single-top production. The background fraction per 
subsample varies from 4\% for SVX Double tagged events to 73\% for No Tag 
events.

\begin{table}[htpb]
\begin{center}
\begin{tabular}{ccccccc}
\hline\hline
Item &                 &  SVX   & SVX     & SLT      &           &         \\
\#   & Process          & Single & Double  & (no SVX) &  No Tags  &  Total  \\
\hline
1  & non-$W/Z$         &  0.5   &   0.0   &   1.0    &   4.6     &  6.1    \\
2  & $WW$              &  0.1   &   0.0   &   0.1    &   0.6     &  0.8    \\
3  & $WZ$              &  0.0   &   0.0   &   0.0    &   0.1     &  0.1    \\
4  & $ZZ$              &  0.0   &   0.0   &   0.0    &   0.1     &  0.1    \\
5  & $Z\to \tau\tau$   &  0.1   &   0.0   &   0.2    &   0.5     &  0.8    \\
6  & Single Top        &  0.2   &   0.0   &   0.1    &   0.2     &  0.4    \\
7  & $Wc+Zc$           &  0.2   &   0.0   &   0.8    &   1.7     &  2.7    \\
8  & $W\bbbar+Z\bbbar$ &  0.8   &   0.2   &   0.4    &   1.1     &  2.5    \\
9  & $W\ccbar+Z\ccbar$ &  0.4   &   0.0   &   0.8    &   2.0     &  3.2    \\
10 & $W/Z+u,d,s$       &  0.2   &   0.0   &   4.1    &  19.6     & 23.9    \\
\hline
\multicolumn{2}{r}{\rule{0mm}{5mm}Background sum} 
 & $ 2.4^{+0.8}_{-0.7}$  &  $0.2\pm 0.1$  &  $7.6\pm 1.3$  &  
   $30.4^{+4.3}_{-4.7}$  &  40.7   \\
\hline
11 & $\ttbar$          & 12.6   &   4.8   &   6.4    &  11.6     & 35.3    \\
\hline
\multicolumn{2}{r}{Observed events} &  15 &  5  &  14 & 42   &   76   \\
\hline\hline
\end{tabular}
\end{center}
\caption{Expected composition (in events) for the four mass subsamples
from various processes. The $W$+jets and $Z$+jets processes have been summed
together. The No Tag subsample only includes contributions from class II,
as only these events are used in the top quark mass analysis.}
\label{bg-table}
\end{table}

%\subsection{Summary of Background Estimation}
%
%    The background in the four mass subsamples is estimated by comparing the
%observed number of tags with expectations based on expected background rates
%and tagging probabilities for $\ttbar$ and background events. The expected
%background rates and the tagging probabilities are estimated either directly from
%the data or from Monte Carlo programs and a detector simulation. For some
%of the background processes, absolute rates are estimated, while others are
%expressed as a fraction of $N_W$, the number of $W$+jets events in the data sample.
%Based on the background rates and tagging probabilities, we estimate the number
%of $W$+jets events in the data sample by minimizing a likelihood function
%with respect to $N_W$. The estimate
%of $N_W$ is then combined with appropriate tagging probabilities 
%to compute the expected background fraction in each mass subsample. 
%The likelihood function is then used to generate probability distributions as
%a function of the expected background fraction. These distributions are
%used in the fit for the top quark mass to constrain the background to its
%expected value (see Section~\ref{s-like}).

%% file: chapter_5_mod3.tex
\chapter{CORRECTIONS TO RAW CALORIMETER ENERGIES}
%\section{Corrections to Raw Calorimeter Energies}
\label{s-recon}

Calorimeter information is used to estimate the jet momenta and the
 net transverse momentum of the particles recoiling against 
the $\ttbar$ system. 
 This section details how those estimates are made.
The signal from each calorimeter tower is converted into a raw~\cite{defraw} 
energy estimate. Tower energies are then used to evaluate the total
energy in the event and other quantities used in the top
mass analysis. The raw measurements are corrected for non-instrumented 
regions, 
non-linear response of the calorimeter, multiple interactions at high 
luminosity, and other effects, before a constrained fit is applied to the 
$\ttbar$ candidate events. Also in this section checks of the jet
 energy scale are discussed, this being the source of the 
largest systematic uncertainty in the measurement of the top quark mass.

\section{Jet corrections and their uncertainties}
\label{s-jetcor}

%The jet clustering algorithm estimates a raw jet energy 
%by adding the energy in
%the calorimeter towers included in the jet cluster. 
The raw momentum of a jet is calculated by adding vectorially the momenta from 
all the towers belonging to the jet cluster (see Section~\ref{s-jetrecon}).  
Tower momenta are calculated 
from tower energies with the assumption that they are energies of particles 
with zero mass~\cite{scale} that originate from the reconstructed primary 
vertex and are located at the center of the tower.
%
%******** This is not clear at all *************
% The jets we measure are 
%used to determine the top mass from its decay
%products, i.e., $t \into W b$ with one $W \into q \overline{q}'$. Since
%the top quark decays before it hadronizes we need to determine 
%the energies of the
%quarks ('partons') which are the decay products of the $\ttbar$ pair. 
%The jets observed in the detector which come from the $\ttbar$ decay
%The daughter partons from the $\ttbar$ decay fragment into the observed jets 
%For
%this purpose corrections are applied
%to the raw jet momenta, to obtain correct parton momenta which are the 
%inputs to the mass fitting
%programs. 
%
  To measure the top quark mass from candidate $\ttbar$ events, 
corrections are applied to the raw jet momenta in order to obtain
estimates of the momenta of the daughter partons in the $\ttbar$ decay.
The corrections occur in two stages.
\begin{itemize}
  \item A set of ``flavor-independent'' corrections \cite{three_jet} is applied
        to all jets with raw $\Et>8\ \gev$.
  \item A second set of corrections, specific to $\ttbar$ events, is applied 
        to the leading four jets which are assumed to be the daughter jets 
        from the $\ttbar$ decay. These corrections are applied after the 
        flavor-independent corrections, and map the measured jet momenta to 
        the momenta of the partons in the $\ttbar$ decay.
\end{itemize}
A description of the corrections to the raw jet momenta
is the focus of this section.

\subsection{Flavor-independent jet corrections}
\label{s-fijetcor}

  To account for detector and reconstruction effects, raw jet 
transverse momenta
%\cite{defraw} 
are corrected using a set of ``flavor-independent'' 
jet corrections~\cite{three_jet}. 
The following expression includes all the corrections applied:
\begin{equation}
 P_T(R) = (P_T^{raw}(R) \times f_{rel} -UEM(R)) \times f_{abs}(R) - UE(R) + OC(R).
\label{eq:jet-eq}
\end{equation}
\noindent
The parameter R=$\sqrt{(\Delta \eta)^2 + (\Delta \phi)^2}$ is the cone radius 
chosen for the jet measurement; R=0.4 for this
analysis. The corrections are described below:
% and their effects are shown in Figure~\ref{f-jetcor}.

\begin{itemize}
 \item $f_{rel}$, the relative energy scale, corrects for
       non-uniformities in calorimeter response as a function of $\eta$.   
 \item $UEM(R)$ takes into account energy due to multiple interactions 
       in the event.
 \item $f_{abs}(R)$, the absolute energy scale, maps the raw jet energy
   observed in a cone of radius R into the average true jet energy. 
   This average is determined in the central calorimeter assuming a
   flat $\Pt$ spectrum.
 \item $UE(R)$ takes into account the energy due to the underlying event,
       i.e., the energy from the primary $\ppbar$ interaction due to 
       fragmentation of partons not associated with the hard scattering, 
 \item $OC(R)$, corrects for the energy expected to be
       outside the cone radius of 0.4.
\end{itemize}

\noindent The $f_{abs}(R)$ and the $OC(R)$  corrections
are functions of the transverse momentum of the jet. The relative correction
is primarily dependent on the pseudorapidity of the jet, with only a weak
dependence on the jet momentum.

The reconstruction of jets starts with the raw clustered energy, 
$\Pt^{raw}(0.4)$. 
An uncertainty of $\pm$1\% is assigned to the stability of the calorimeter
over the course of the data taking period. This systematic uncertainty
was evaluated by comparing the response of the calorimeter to single
charged tracks between data from Run 1 and data from the 1988--1989 run,
which was used for the energy calibration discussed later.
No systematic difference was observed. Also the 
raw inclusive jet cross section~\cite{incjet} obtained with 
the 1988-1989 data run was compared with that of the Run 1a data 
(after correcting for multiple interactions)
and it was found that the ratio was consistent with unity at the 5\% level. 
Because of the rapidly falling $\Et$ spectrum, this corresponds to an upper 
limit on a difference in the energy scale of 1\%.

The relative correction is derived from di-jet balancing data and
corrects for the relative response of the different calorimeter sections
to that of the calorimeter in the central region ($0.2<|\eta |<0.7$)
~\cite{scale}. The plug ($1.1<|\eta |<2.4$)
and forward ($2.4<|\eta |<4.2$) regions are thus calibrated.
The precision to which this calibration is known is
limited mostly by the number of di-jet events available. The 
effects of different resolutions of the central and plug 
calorimeters on the energy measurements were studied using Monte 
Carlo simulation and are properly included.
The uncertainty is larger near the cracks between  the different 
detectors due to smaller statistics and worse energy resolution.
Table~\ref{t-relsys} gives the uncertainty (in \%) on the relative 
corrections for various detector $\eta$ ranges. 

\begin{table}
\begin{center}
\begin{tabular}{cc}
\hline\hline
                  & \multicolumn{1}{l}{Uncertainty on} \\
$|{\eta}|$ range  & \multicolumn{1}{l}{relative correction}  \\
\hline
   $0.0 - 0.1$      &   2.0 \% \\
   $0.1 - 1.0$      &   0.2 \% \\
   $1.0 - 1.4$      &   4.0 \% \\
   $1.4 - 2.2$      &   0.2 \% \\
   $2.2 - 2.6$      &   4.0 \% \\
   $2.6 - 3.4$      &   0.2 \% \\
\hline\hline
\end{tabular}
\end{center}
\caption{The percentage uncertainty on the relative jet energy correction for
various detector $\eta$ ranges. The cracks in $\eta$ between different 
detectors are located near $\eta$=0, 1.2, and 2.4, and have larger 
uncertainties than the regions away from the cracks.}
\label{t-relsys}
\end{table}

The corrections for multiple interactions ($UEM$) in the same event and the 
underlying event ($UE$) in the primary interaction are derived
from minimum bias data. 
The average number of interactions  in Run 1a (N$_v$=0.6) is different
from that of Run 1b (N$_v$=1.8), hence a different procedure is used for the 
two samples. For the Run 1a sample, 0.72 $\gevc$ is subtracted
from the jet $\Pt$ after the absolute correction and accounts for 
both effects on average. For Run 1b,
the effects of the underlying event and additional interactions are
separated. To account for multiple interactions, prior to the
absolute correction, 0.297 $\gevc$ is subtracted from 
the jet $\Pt$ for each additional reconstructed vertex in the event. 
This correction is obtained by studying the amount of energy 
in the event as a function of the number of vertices over the course of the
run. For the underlying event ($UE$), we subtract 0.65 $\gevc$ from each 
jet after the absolute correction. 

\begin{figure}[htbp]
%%\leavevmode
\begin{center}
\hspace{0.5in}
\epsfysize=6.5in
\epsffile[0 72 612 720]{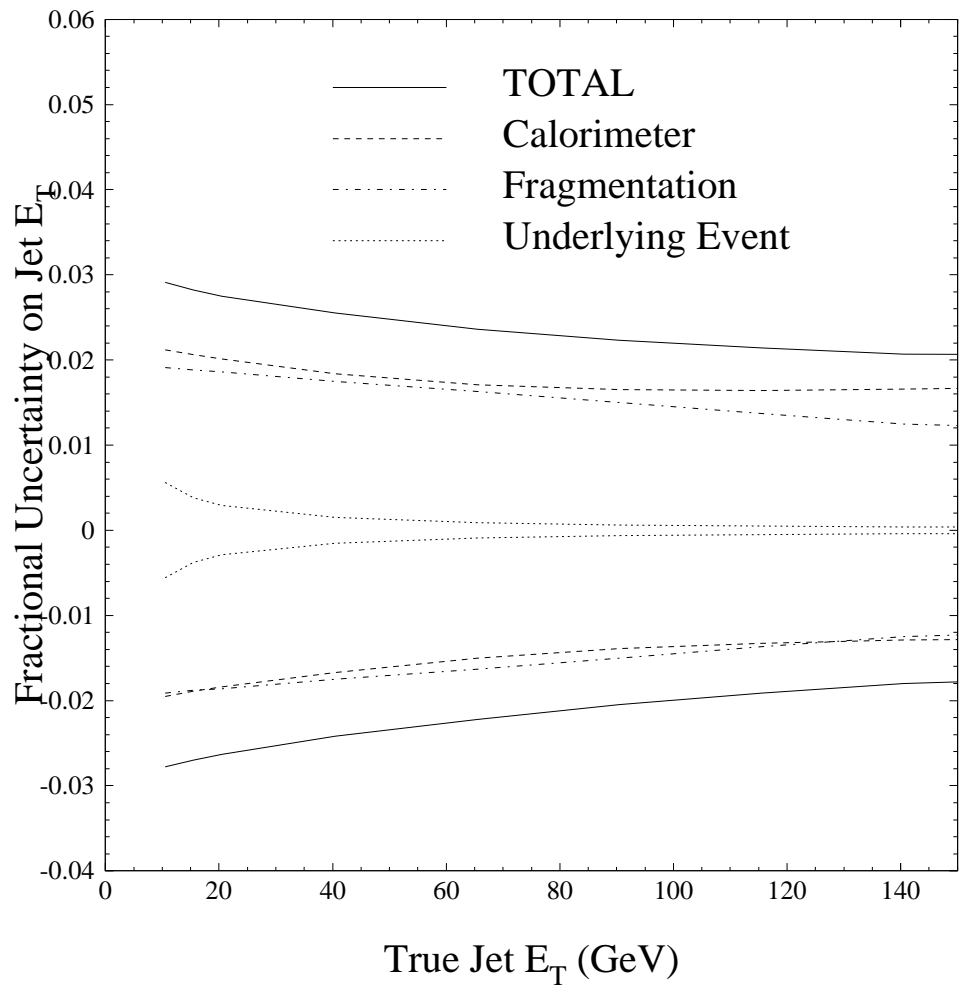}
%\epsffile[0 72 612 720]{$PRD_FIG/behrends.ps}
%\epsffile[0 72 612 720]{$SG5_TEX/behrends_tt.ps}
\end{center}
\caption{Uncertainty in jet $\Et$ scale as measured
with a jet clustering cone of size 0.4. 
The vertical axis shows the extent to which the measured
jet $\Et$ response varies due to different systematic effects.}
\label{f-behr}
\end{figure}

  The uncertainty on the $UEM$ correction is estimated to be 
100 MeV/c for each vertex in the event. The uncertainty in the $UE$
correction is evaluated by looking at variations in the energy density
at $\pm$90$^{\rm o}$ with respect to the two jets in di-jet events when
varying the maximal $\Et$ threshold on the third jet from 5 to 15 
$\gev$. Based on these studies, we assign a $\pm$30\% relative uncertainty to the 
underlying event correction~\cite{scale,three_jet}. For jets with 
$\Pt>$ 20  $\gevc$
the uncertainty is typically less than 0.5\% of the jet's 
$\Pt$, as shown in Fig.~\ref{f-behr}. 

\begin{figure} [htbp]
%\leavevmode
\epsfysize=6.5in
\hspace{0.5in}
%\epsffile[0 72 612 720]{f_jetcor}
%\epsffile[0 72 612 720]{tprd_jtc_cor.psn}
\epsffile[0 72 612 720]{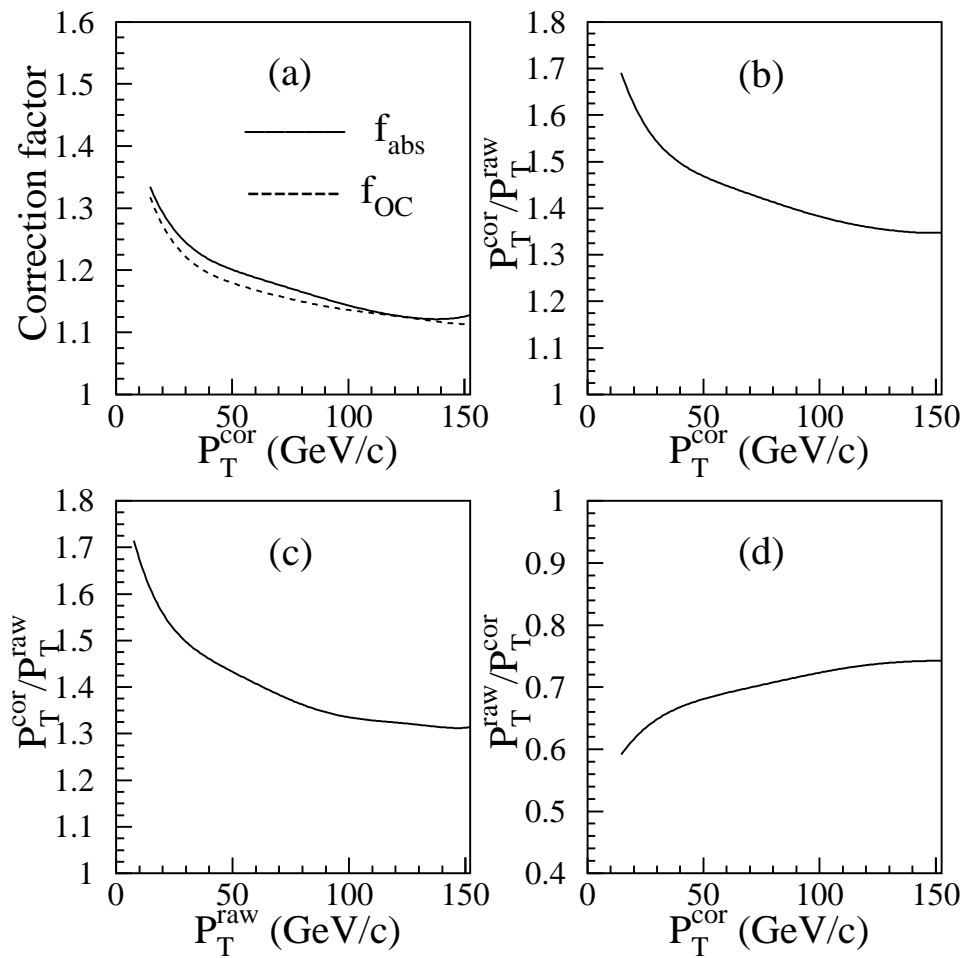}
%\epsffile[0 72 612 720]{$PRD_FIG/tprd_jtc_cor.ps}
\caption{``Flavor-independent'' jet corrections, for a jet clustering
 cone of R=0.4. (a) Absolute correction, $f_{abs}$, and 
   out-of-cone correction factor, 1+ OC/$\Pt$, 
    versus corrected jet $\Pt$, $\Pt^{cor}$ .
(b) Total correction,  $\Pt^{cor}$/$\Pt^{raw}$, as a function of $\Pt^{cor}$.
(c) total correction, $\Pt^{cor}/\Pt^{raw}$, as a function of $\Pt^{raw}$.
(d) fraction of measured momentum, $\Pt^{raw}/\Pt^{cor}$ versus  
    $\Pt^{cor}$. }
\label{f-jetcor}
\end{figure}

The absolute
correction is derived from data and Monte Carlo plus detector simulation.
The simulation includes many features of the CDF calorimeters, the
main ones being: non-linearity, cracks and less 
sensitive regions, single tower thresholds.
The response of the calorimeter to 
incident pions and electrons is studied using testbeam data, minimum
bias runs, special runs which triggered on events containing single
isolated tracks, as well as standard data runs.
The detector simulation has been tuned to agree with these data. 
The step from individual particle response to jets is achieved
by tuning the Monte Carlo ({\small{ISAJET}}) fragmentation 
parameters to reproduce a number
of distributions observed in di-jet data: number of charged particles,
spectra and invariant mass of charged particles, and the
ratio of charged to neutral energy~\cite{three_jet}. 
The derived correction then accounts for non-linearity of the calorimeter,
energy losses near the boundaries of different
calorimeter wedges, response variation
as a function of the position along the wedge and all the other
effects included in the simulation. 
The absolute
correction, $f_{abs}(0.4)$, as a function of corrected jet P$_T$, P$_T^{cor}$,
 is shown in Fig.~\ref{f-jetcor}(a). 

The systematic uncertainty in the absolute correction
is attributed to (a) calorimeter response, and (b) fragmentation related 
effects~\cite{scale,three_jet} (see Fig.~\ref{f-behr}). 
%The parameters which describe the calorimeter's response are tuned in 
%the detector simulation to reproduce the detected energy in testbeam 
%and collider data. 
The parameters that describe the calorimeter's
response to incident electrons, photons and pions 
have uncertainties due to finite statistics and assumptions which
are made. For example, at low momentum ($|p|<5\ \gevc$),
the largest source of uncertainty in the charged pion response comes
from the estimation of the amount of energy in the shower from
$\pi^0$'s. Additional uncertainty comes from the uncertainty in
the relative response across the face of a calorimeter cell and
the energy deposition in cracks between calorimeter cells. 
The uncertainty in the calorimeter's response to photons is 
assigned to be the same as for electrons. Uncertainty
in the fragmentation parameters comes from the modeling of the 
tracking efficiency in jets, and the level of agreement between
the simulation and data.

   The contributions to the jet $\Et$ uncertainty from these
sources are evaluated by shifting the input values of these
parameters by $+1$ and $-1$ standard deviation ($+1\sigma$ and $-1\sigma$), 
and calculating the resulting shift in the reconstructed jet energies.
For (a) we separately vary the pion, electron, and photon 
responses by $+1\sigma$ and $-1\sigma$, and add the resulting
shifts in the jet energies in quadrature.
%The resulting deviations from the three sources are added 
%in quadrature to obtain the total uncertainty from calorimeter response.
For (b), we vary the charged tracking efficiency by its uncertainty
and reevaluate a new set of fragmentation parameters. These new
fragmentation parameters are in turn varied one at a time, and the
resulting deviations in the jet energies are added in quadrature.

%The uncertainty in the jet $\Et$ scale due to uncertainties in the
%fragmentation parameters are evaluated by
%The fragmentation parameters, which describe the jet shapes observed in data,
%are also a source of systematic uncertainty in the jet $\Et$. This
%contribution is evaluated by 
%first varying the charged tracking efficiency
%by $+1\sigma$ and $-1\sigma$, and reevaluating a new set of 
%fragmentation parameters. The jet energy scale uncertainty from this source 
%is taken as the quadrature 
%sum of the deviations in the energy scale resulting from varying each 
%parameter individually. 

   The systematic uncertainties in the jet $\Et$ scale from the sources (a)
and (b), as well as from the UE correction 
are shown in Fig.~\ref{f-behr}. The total systematic uncertainty from these
three sources is obtained by adding in quadrature the 
three curves, and is shown as the solid curve in Fig.~\ref{f-behr}. 

  The out-of-cone correction was derived from a Monte Carlo simulation
and accounts for the energy falling outside the jet cone~\cite{scale}.
This study was done with light quarks; the $\ttbar$ specific
corrections take into account differences with heavy flavor jets.
The amount of energy outside the cone of R=0.4 is related to 
emission of low energy gluons from the initial partons, and is referred 
to as ``soft gluon'' radiation. The correction factor, $f_{OC}$ =
1+$OC$(0.4)/$\Pt(0.4)$, is a function of the jet $\Pt$  
corrected for all other effects  and is
given by the equation,

\begin{eqnarray}
 f_{OC}\; =\; 1.0\, +\, {23.0\,(1.0-0.915\, e^{-0.0074\,\Pt})\over \Pt}.
\label{eq:foc}
\end{eqnarray}

\noindent The correction factor is shown in Fig.~\ref{f-jetcor}(a). 

\begin{figure}[htbp]
%\leavevmode
%\begin{center}
\epsfysize=6.5in
\hspace{0.5in}
%%%\epsffile[0 72 612 720]{mss_ann_wj.eps}
%\epsffile[0 72 612 720]{$SG5_TEX/w1jet_oc_8.ps}
\epsffile[0 72 612 720]{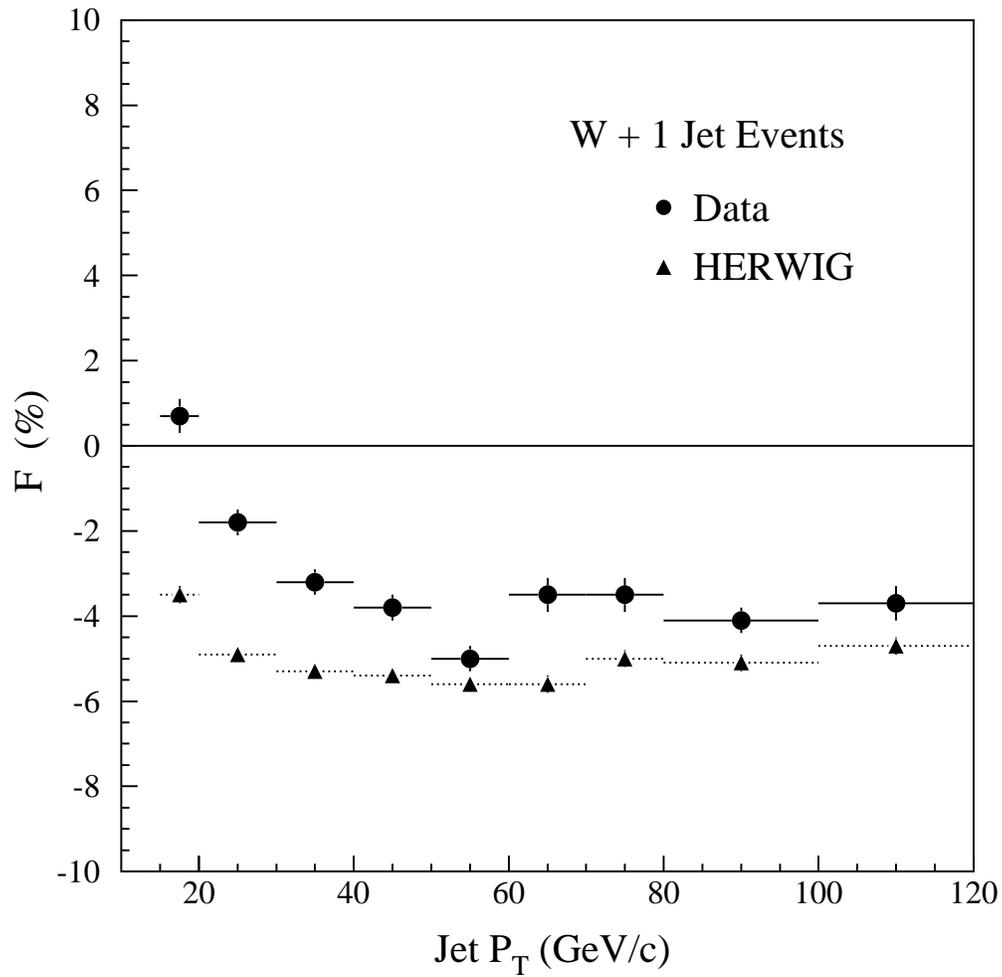}
%\epsffile[0 72 612 720]{$PRD_FIG/w1jet_oc_8.ps}
%\end{center}
\caption{Fractional difference in corrected jet $\Pt$ obtained using 
cone radii of 0.4 and 1.0 as
a function of the corrected jet $\Pt$ from $W$+1 jet events.
The circles are the results from the data sample and the
triangles are from a sample of {\small{HERWIG}} 
Monte Carlo events which have been processed through the CDF detector
simulation.}
\label{f-w1j}
\end{figure}

\begin{figure}[htbp]
\epsfysize=6.5in
\hspace{0.5in}
%\epsffile[0 72 612 720]{mss_data_annul_pt.eps2}
%\epsffile[0 72 612 720]{$SG5_TEX/w1jet_oc_df.ps}
\epsffile[0 72 612 720]{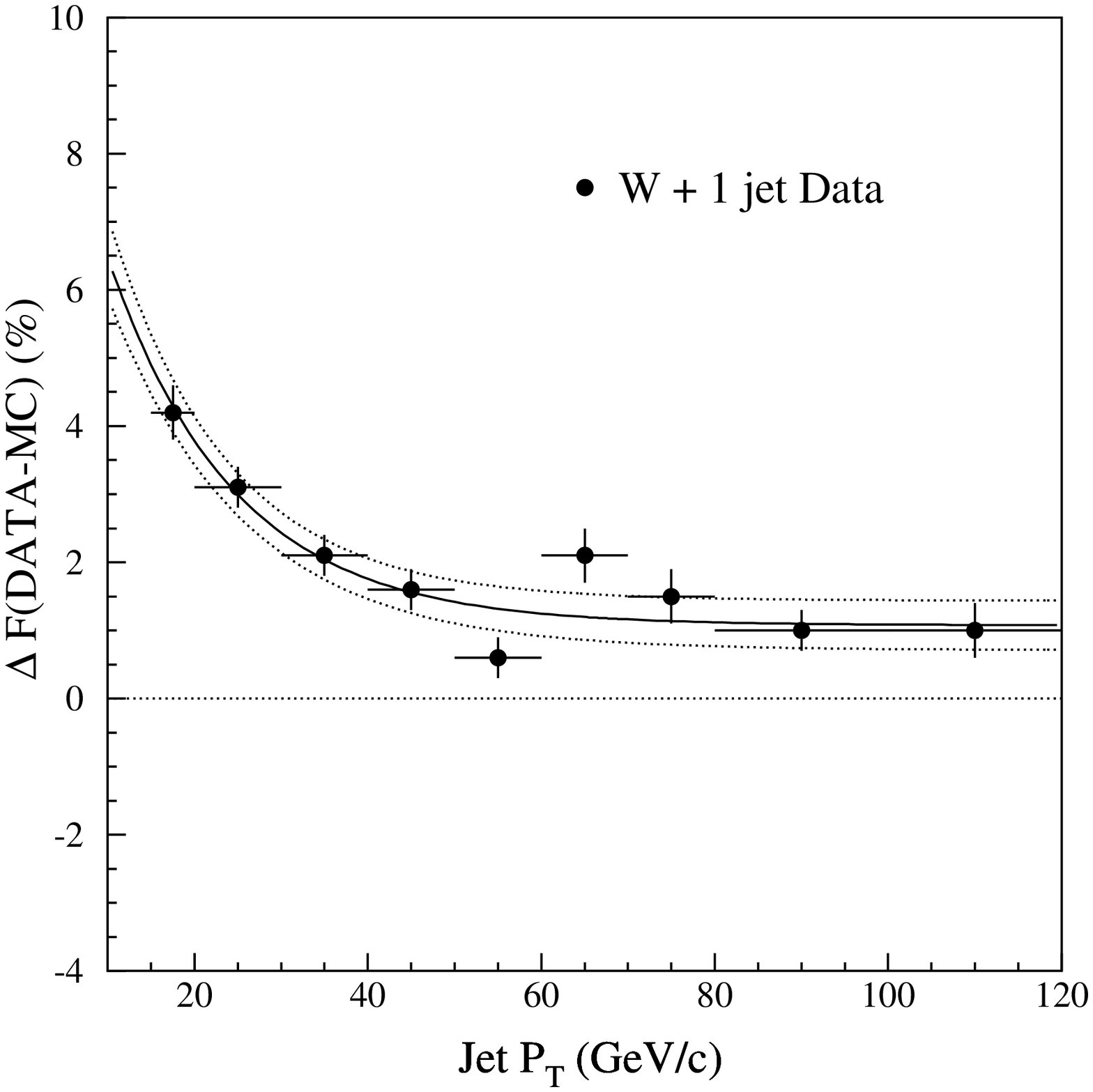}
%\epsffile[0 72 612 720]{$PRD_FIG/w1jet_oc_df.ps}
\caption{ Systematic uncertainty on the out-of-cone correction as obtained in
          $W$+1 jet events. The abscissa is the fully corrected jet
          $\Pt$ using a clustering cone of 0.4. The vertical axis is the
          difference between data and Monte Carlo of the variable $F$
          described in the text. The full curve represents a fit
          through the data points, the dotted curves were obtained
          using the one standard deviation values of the fit parameters. }
\label{f-anndata}
\end{figure}

  The systematic uncertainty on the jet momentum from the OC correction
 originates
from the uncertainty in modeling the radiation of low energy gluons in
parton showers.  To estimate this uncertainty, we use 
$W$+1 jet data and a {\small{HERWIG}} Monte Carlo
simulation of $W$+1 jet events to compare 
the energy contained in an annulus 
with radii of 0.4 and 1.0 around the jet direction. 
%Samples of $\ttbar$ events generated using the {\small{PYTHIA}}
%Monte Carlo simulation show that turning off the final state
%radiation has a large impact on the width of a jet.
We define a variable $F$, 
\begin{equation}
F \; =\; \frac{\Pt(1.0)-\Pt(0.4)}{\Pt(0.4)}
\end{equation}
\noindent where $\Pt(0.4)$ and $\Pt(1.0)$ are the jet momenta
corrected using the corrections described above (note that Eq. 5.2 is
used for R=0.4; for R=1.0 the correction is much smaller). The quantity 
F is the fractional difference of the momentum in an annulus with
radii between 0.4 and 1.0, calculated for each event using the
calorimeter towers in that annulus or using the average
OC correction. A comparison of data and Monte Carlo
tests the agreement between the Monte Carlo soft gluon radiation
modeling and what is observed in the data in that annulus.
Figure~\ref{f-w1j} shows the mean value of $F$ as 
a function of the corrected $\Pt$ (corrected using a cone size of 0.4)
for data and Monte Carlo. There is a clear difference between the two 
distributions. 
%The direction of this difference implies that the
%{\small{HERWIG}} $\Pt(0.4)$ jets are receiving a larger correction
%than the jets in the data sample. This implies that there is a
%smaller fraction 
%of the jet $\Pt$ in the cone $R<0.4$ for the {\small{HERWIG}} simulation 
%than the data. As a result, one concludes that the {\small{HERWIG}} jets
%are broader than the ones in the data sample, or, in other words there is 
%more soft gluon radiation in the {\small{HERWIG}} simulation than in the data
%sample. 
This implies that the jet shapes in data and Monte Carlo disagree at
the few \% level. The difference between {\small{HERWIG}} and data is shown in 
Fig.~\ref{f-anndata}. We take this difference as the uncertainty 
on the out-of-cone
correction. Its effect on the top quark mass measurement is referred 
to as the systematic uncertainty from soft gluon radiation.

  Similar distributions have been obtained 
for other sets of data, namely $Z+1$ jet data and jet data with
two $b$-tagged jets. Since the statistics for the latter sets of data are
low, only the $W$+1 jet data are used. 
A fit to the points of  Fig.~\ref{f-anndata} gives a maximum (upper
dotted curve)
uncertainty of $\delta \Pt/\Pt = \exp{(2.467-0.074\Pt)}+1.438$ (in \%).
It can be seen that for jets typical of those produced in
$\ttbar$ events ($\approx 30-90 \ \gev $ for \mtop$=175 \ \gevcc $), the
difference between {\small{HERWIG}} and data is $< 2\%$.  For 
softer jets, the difference is closer to 4\%. 
%We have
%compared the distribution observed in $W$+1 jet data with
%$\gamma$+1-jet and $Z$+1-jet data samples and have found
%reasonable agreement with $W$+1-jet data (see Fig.~\ref{f-anndata}). 
%Because of larger QCD background in the $\gamma$+1-jet
%data sample and lower statistics in the $Z$+1-jet data sample, we 

  The systematic uncertainty assigned to the soft gluon radiation
accounts for differences in the energy contained in the annulus 
$0.4<R<1.0$ between data and the Monte Carlo simulation. For the additional
energy which falls outside a cone of 1.0, we assign an uncertainty
of $\pm$ 1 $\gev$. We refer to this energy as ``splash-out''. 
%and, we believe this estimate of the uncertainty to be conservative.

\begin{figure}[htbp]
\epsfysize=6.5in
\hspace{0.5in}
\epsffile[0 72 612 720]{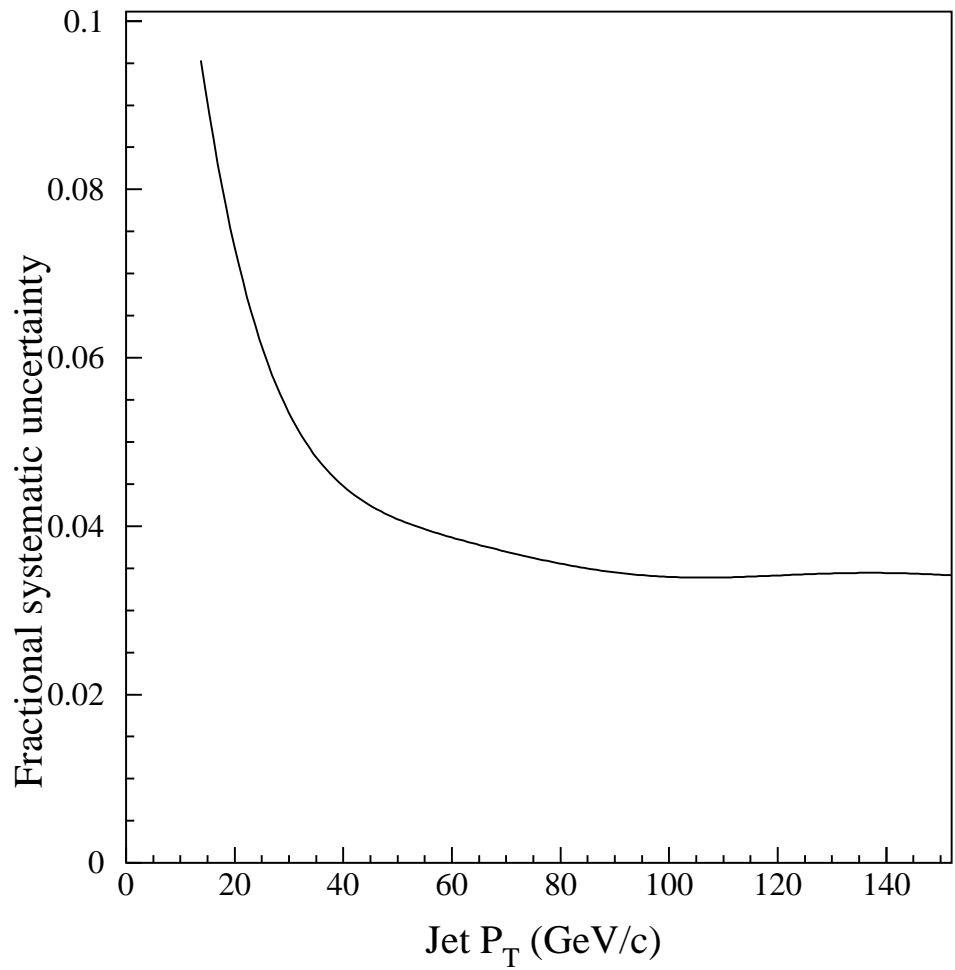}
%\epsffile[0 72 612 720]{$PRD_FIG/tprd_jtc_sys.ps}
%\epsffile[0 72 612 720]{tprd_jtc_sys.ps}
\caption{Total systematic uncertainty on jets as a function of
          the corrected jet $\Pt$.}
\label{f-jetsys}
\end{figure}

In summary, Figure~\ref{f-jetcor} shows some of the flavor-independent
jet corrections and their $\Pt$ dependence.  
%The relative correction
%is nearly flat with jet $\Et$, and so it is presented as a function of
%pseudorapidity. 
Figure~\ref{f-jetcor}(a) shows the absolute and out-of-cone correction
factors as a function of the corrected jet $\Pt$.
They vary from $\approx$1.3 at $\Pt=15$ $\gevc$ to $\approx$1.12 
for $\Pt>100$ $\gevc$. Figure~\ref{f-jetcor}(b) shows the 
ratio of the fully corrected jet $\Pt$ ($\Pt^{\rm{cor}}$) to the raw 
jet $\Pt$ ($\Pt^{\rm{raw}}$) as
a function of the fully corrected jet $\Pt$. Jets from $\ttbar$ events 
typically have a $\Pt$ of $\approx30-90 \ \gevc$, for
which the average jet correction factor is $\approx 1.45$.
Figure~\ref{f-jetcor}(c) shows the correction factor as a function of
$\Pt^{\rm{raw}}$. Finally, Fig.~\ref{f-jetcor}(d) shows the fraction 
of momentum
measured in the detector before the jet corrections as a function of the
corrected jet $\Pt$. Figure~\ref{f-jetsys} shows the overall  
systematic uncertainty as a function of the corrected $\Pt$ of the jets. 
In the 30-90 $\gevc$ range, the systematic uncertainty on jet
energies is about 4\%.

\subsection{Checks on the jet $\et$ scale and its uncertainty}
\label{etchecks}
  
The procedures used to obtain the jet corrections and their systematics 
have been checked by applying them to $Z+1$ jet events, where the $Z$-boson 
decays into either $e^+e^-$ or $\mu^+\mu^-$.
The energy scale for electrons and the momentum scale for muons are known 
to a precision of 0.14\% and 0.065\% respectively~\cite{wmass}. 
In the absence of initial state 
radiation, such events are expected to have zero net transverse 
momentum. The jet in each event is corrected according to the previous 
prescription, and the quantity
\begin{equation}
F_b \; =\; \frac{\Pt (Z)-\Pt ({\rm jet})}{ \Pt (Z)} 
\label{eq:f_b}
\end{equation}
\noindent is calculated, where $\Pt (Z)$ is in the range 30-150 $\gevc$. 
The lower limit was chosen to avoid biases due to the sample selection. 
The jet recoiling against the $Z$ boson is required to have an uncorrected 
$\Et \ge 8~ \gev$ and $|\eta| <$ 2.4.
To test the jet energy scale we need a clean environment, i.e., events in which
there is only one jet recoiling against the $Z$ boson. We therefore require
that any additional calorimeter cluster have an uncorrected energy 
$\Et< 6.0~ \gev$ (at any $\eta$).

\begin{figure}[htbp]
\epsfysize=6.5in
\hspace{0.5in}
%%%\epsffile[0 72 612 720]{zj1_ptbal_data_mc.ps}
\epsffile[0 72 612 720]{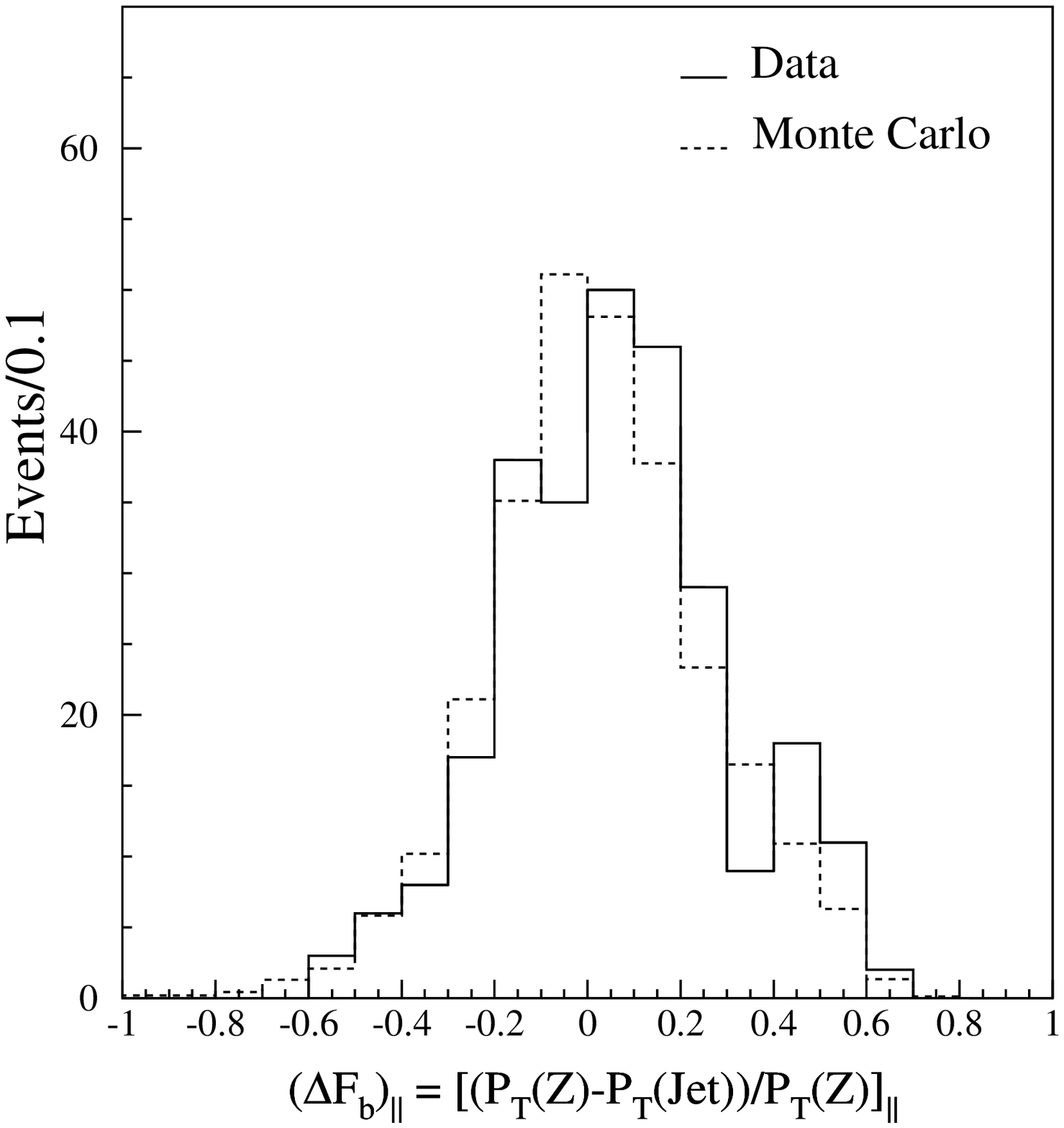}
%%\epsffile[0 72 612 720]{$PRD_FIG/dmc_zbal_6.ps}
\caption{ Parallel component of transverse momentum imbalance between 
the $Z$ and the jet
in reconstructed $Z+1$ jet candidate events. Both data (solid) and
Monte Carlo (dashed) are shown (see text). }
\label{f-zjbal}
\end{figure}

To separate detector effects from those due to gluon radiation in the 
initial state, we use the component analysis first suggested in 
Ref~\cite{bagnaia}. We compute the direction of the bisector between the 
$Z$ and the jet directions 
in the transverse plane. The ``parallel component'' of $F_b$ is then 
defined to be the component perpendicular to the bisector. Balancing the
jet against the $Z$ along this component will give information about
the jet energy scale. 
Figure~\ref{f-zjbal} shows the distribution 
of this component of $F_b$
in $Z+1$ jet events for data and Monte Carlo.
The difference in the medians of the two distributions is:
\begin{equation}
(\Delta F_b)_{\parallel} = (3.2  \pm 1.5(stat) \pm 4.1(syst)) \%. 
\label{eq:df_b}
\end{equation}
\noindent The 
%deviation of $\Delta F_b$ from zero is consistent with the 
4.1\% systematic uncertainty was calculated using the jet 
energy uncertainties discussed in the previous section. 
We conclude that any possible energy scale shift detected by this check is
compatible with zero within the evaluated uncertainties.
%The positive offset may be attributed to unmeasured initial state 
%radiation (also known as ``$k_{T}$-kick''). Studies have
%been performed which show that the Monte Carlo simulation
%models the $k_{T}$-kick quite well. This implies that 
%the 3.2$\pm$1.5\% difference shown in Fig.~\ref{f-zjbal} can be 
%attributed to a possible difference in the energy scale between 
%the Monte Carlo simulation and data.

%  The presence of $W^{\pm}$ bosons decaying to light
%quarks in $\ttbar$ decays~\cite{wboson} can also be used to 
%check the jet energy scale. 
%One uses the same $\chi^2$ described in Section~\ref{s-algor},
%with the exception that the constraint on the di-jet invariant 
%mass is removed. A fit to the di-jet invariant mass distribution
%yields a W mass of 77.2$\pm$3.5(stat.)$\pm$2.9(syst) $\gevcc$.
%The systematic uncertainty is dominated by the jet energy scale 
%uncertainty which amounts to 2.8 $\gevcc$. This result may
%be compared to the world average of 80.41$\pm$0.10 $\gevcc$ \cite{pdg}.
%While the data is statistically consistent with the world average
%value, the jet energy scale systematic uncertainty can account
%for nearly the entire difference.
%By comparing the di-jet invariant mass in data 
%with templates 
%
%in which the jet energy scale is shifted by varying 
%amounts, we measure an overall jet energy scale of $1.04\pm 6.2\%$.
%This value is consistent with the previously mentioned checks.

\subsection{Jet momentum corrections for $\ttbar$ events}
%\section{Corrections to Input Quantities }
\label{s-correct}

%The jet corrections are designed to
%provide an estimate of parton kinematics under the assumption
%that each of the four leading jets correspond to the parton level
%decay products of the $\ttbar$ system.  Jets associated with
%{\it b}-quarks in top decays are significantly different
%than those from light quarks and gluons. Therefore 
%the jet corrections depend on the assumed associations made 
%between jets and partons in reconstructing an event.

%Corrections to jets proceed in two steps. First, 
%an average correction is made (see Section~\ref{s-jetrecon}), 
%and then jets are corrected for aspects particular to the 
%hypothesis of a $\ttbar$ event. 
%
%Since the average correction
%used di-jet Monte Carlo samples with a flat $\Et$ distribution
%to tune the absolute energy scale, these corrections do not 
%fully account for the complex multi-jet final states of $\ttbar$ 
%events. We therefore invoke a set of top-specific corrections
%which address the differences between $\ttbar$ events and the
%di-jet events which are used in determining the ``flavor-independent'' 
%corrections.
%

\begin{figure}[htbp]
%\leavevmode
%\begin{center}
\epsfysize=6.5in
\hspace{0.5in}
\epsffile[0 72 612 720]{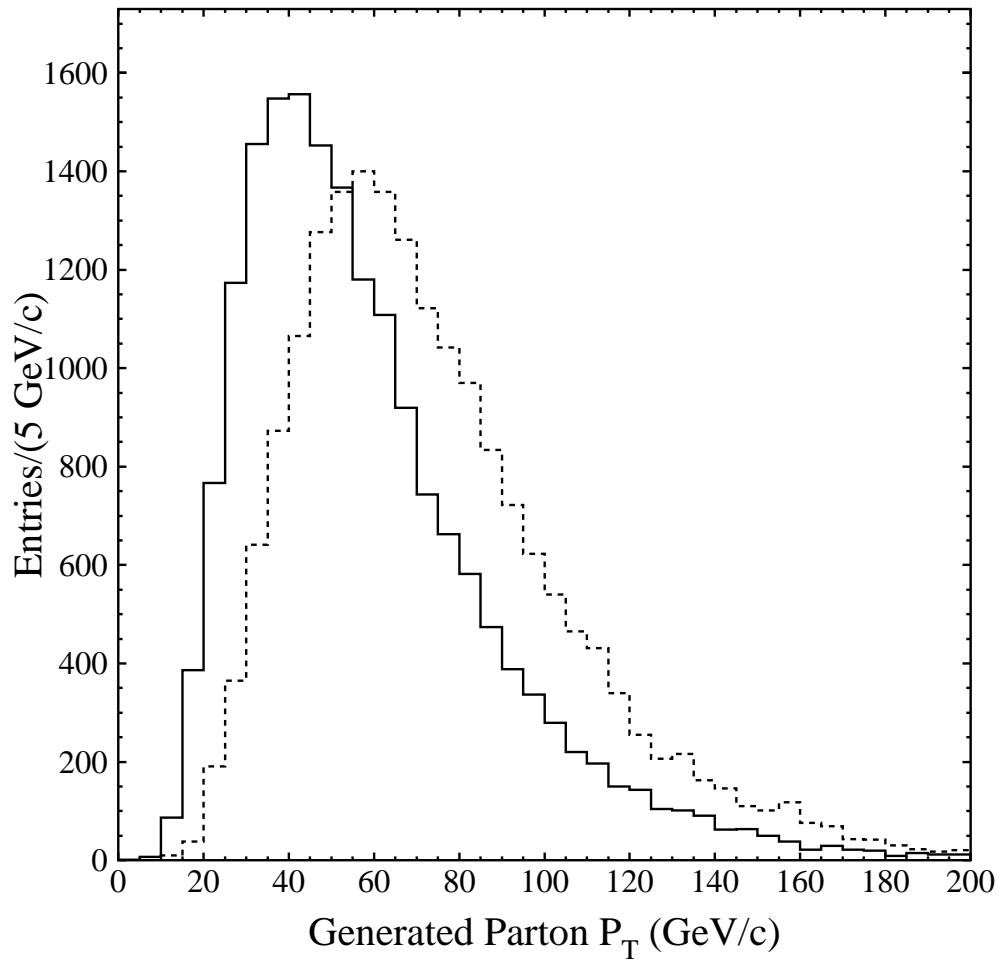}
%\epsffile[0 72 612 720]{$PRD_FIG/myptparton.ps}
%\epsffile{ptparton.eps}
%\end{center}
\caption{$\Pt$ distributions for partons from top quark decays
obtained from the {\small{HERWIG}} Monte Carlo program after simulation
of detector response and including the effects of the
Top Mass Candidate Sample data selection.  The solid line
indicates the distribution for light quarks from the
$W\rightarrow q\bar{q}^{\prime}$ decay and the dashed line
is the distribution for $b$ quarks.}
\label{f-top_pt}
\end{figure}

  The $\ttbar$ specific jet momentum corrections are designed to make
an average correction to the jet momenta to obtain an estimate of
the original parton momenta~\cite{cdf-evidence}. 
The $\Pt$ spectra of partons from {\small{HERWIG}} generated $\ttbar$ events
which pass our experimental selection cuts
%which match to one of the four highest $\Pt$ jets
are shown in Fig.~\ref{f-top_pt}. The $\ttbar$-specific corrections account 
for (a) the difference in the $\Pt$ spectrum
between top induced jets and the flat spectrum 
used to derive the flavor-independent corrections, (b) the energy lost
through semileptonic $b$ and $c$-hadron decays, and (c) the multi-jet
final state of $\ttbar$ events as compared to di-jet final state used
to derive the flavor-independent corrections.
The correction for these three effects are derived using
the {\small{HERWIG}} Monte Carlo generator with an input top quark
mass of 170 $\gevcc$. The generated events are processed using the CDF
simulation and reconstructed in the same way as the data sample.
An average correction factor is determined by first matching (in $\eta-\phi$
space) the reconstructed
jets with the generated partons, and then comparing the reconstructed
jet $P_T$ (after the flavor-independent corrections) with the original
parton $P_T$. The correction is given by the median of the distribution
of $\Delta = (\Pt (parton)-\Pt (jet))/\Pt (jet)$.
This is done as a function of the reconstructed jet $\Pt$.

\begin{figure}[htbp]
\epsfysize=6.5in
\hspace{0.5in}
%\epsffile[0 72 612 720]{aa_corrections.eps2}
\epsffile[0 72 612 720]{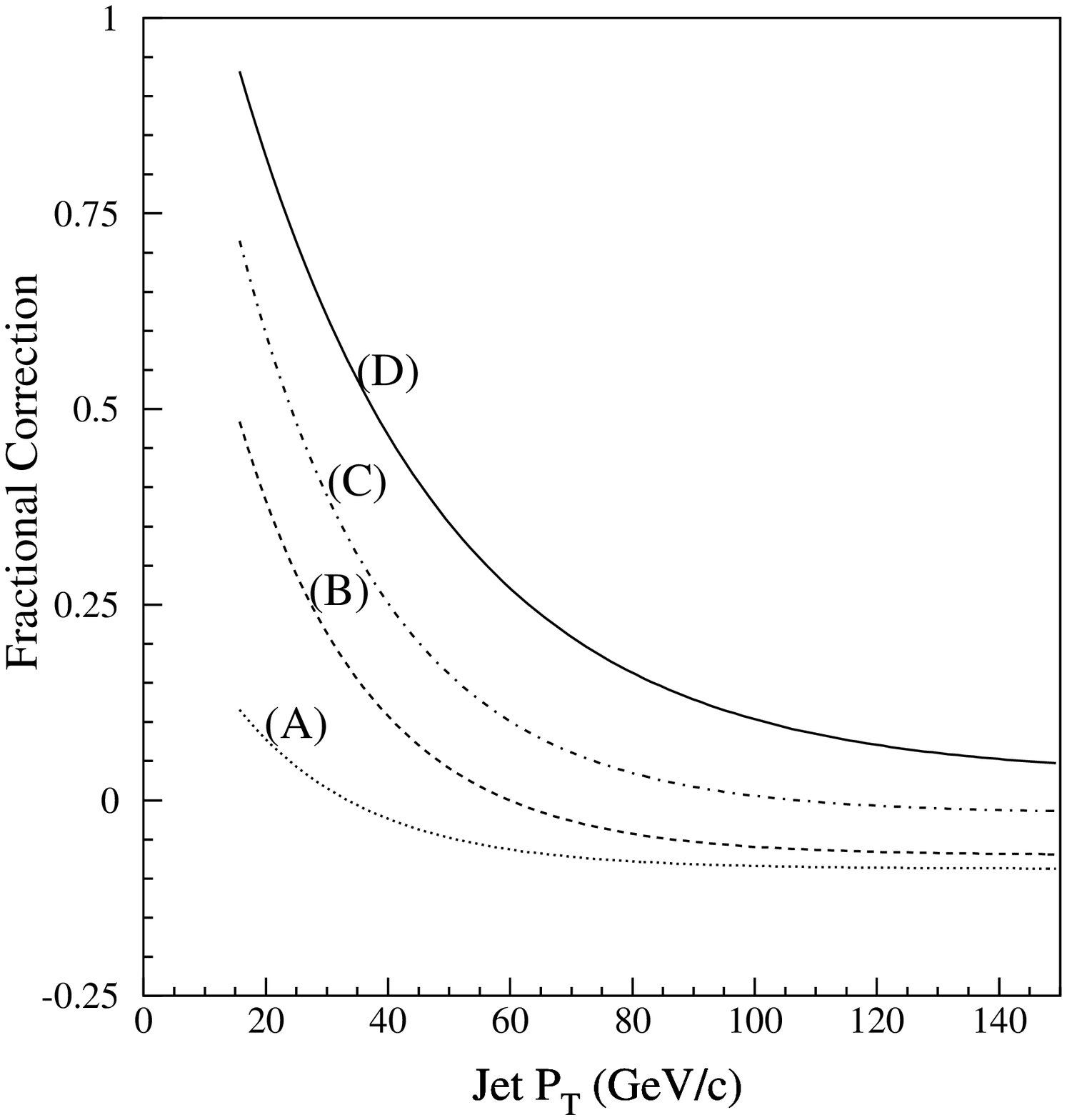}
%\epsffile[0 72 612 720]{$PRD_FIG/jet_aa_corr.ps}
\caption{The $\ttbar$-specific corrections applied to
jets according to available $b$-jet information.
The curves show the fractional change to
the corrected jet $\Pt$ after all ``flavor-independent'' jet
corrections have been applied. The curves
are for: (A) jets from the decay of $W$ bosons,
(B) jets from all $b$ quarks (no selection on decay mode), 
(C) jets from $b$ quarks containing an electron, and (D) jets from $b$ quarks
containing a muon.}
\label{f-aa1}
\end{figure}

Figure~\ref{f-aa1} shows the size of the $\ttbar$-specific correction factors 
for four types of jets: (A) jets from hadronic 
$W$ decays, (B) average $b$ jets ( no selection on decay mode) (C) 
$b$ jets 
containing an electron, and (D) $b$ jets containing a muon. 
The general shape of each curve is primarily a result of the difference
between using a flat jet $\Pt$ spectrum and the spectrum appropriate
for top decays. In particular, this difference is responsible for 
the rising values of the curves at low $\Pt$, and the asymptotic
values at large $\Pt$. The larger corrections applied to the $b$
jets with a soft lepton are a consequence of the amount of energy
carried off by undetected neutrinos, 
%Jets from $W$-decays have only
%a small contribution from undetected muons and neutrinos, whereas jets
%from $b$-quarks contain these leptons in the final state.
and, for jets containing a $b \rightarrow \mu \nu X$
decay, of the fact that muons deposit only $\approx$2 $\gev$, on
average, in the calorimeter.
%This occurs because the jet energies are not corrected for muons
%(using the momentum measured by the CTC) which typically deposit 
%only $\approx$2 GeV of energy in the calorimeters.

% The purpose of the jet energy corrections is to bring the mean
%reconstructed jet energy into agreement with the initial parton energy.
%Because of the finite number of sampling layers in the calorimeters,
%detector cracks and boundaries, and other detector related effects,
%there is a limited energy resolution which can be achieved. 
%The resulting fits give the uncertainty in the jet energy as a function
%of the jet $\Et$. 

\begin{figure}[htbp]
%\leavevmode
%\begin{center}
\epsfysize=6.5in
\hspace{0.5in}
%\epsffile[0 72 612 720]{aa_sigma.ps}
\epsffile[0 72 612 720]{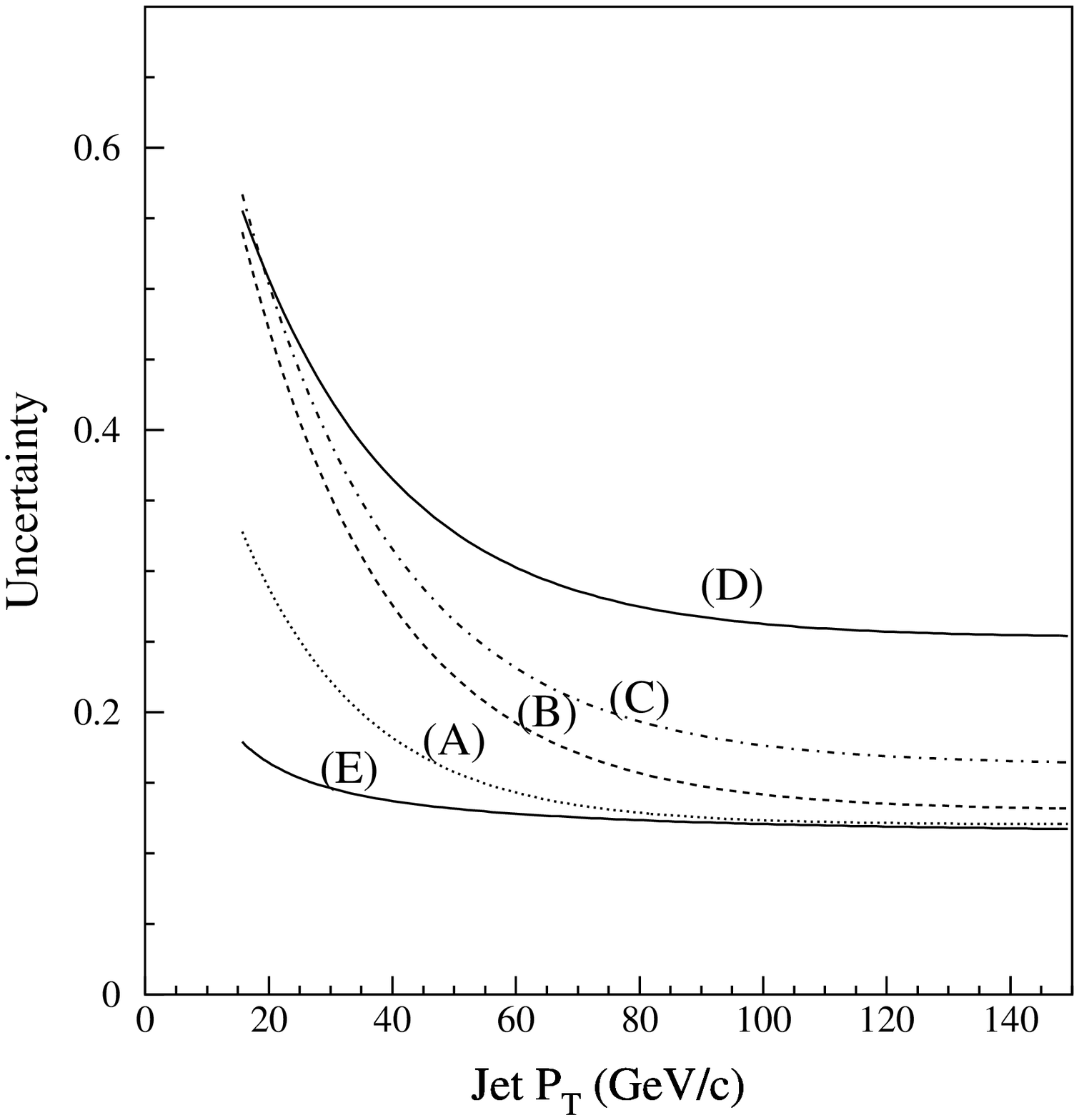}
%\epsffile[0 72 612 720]{$PRD_FIG/jet_aa_res.ps}
\caption{Fractional uncertainty in the estimated parton $\Pt$
as a function of the jet $\Pt$ after the flavor-independent 
jet corrections are applied. The uncertainty shown
on the vertical axis is given as a fraction of the jet $\Pt$.
Curves (A) through (D) have the
same meaning as for the previous figure. The curve labeled (E) is 
used for the jets beyond the four highest-$\Pt$ jets and is applied
only to the $\Pt$ within the cone of radius R=0.4.}
\label{f-aa2}
\end{figure}

The flavor-independent and $\ttbar$-specific corrections bring
the median reconstructed jet $\Pt$ into agreement with the initial parton 
$\Pt$ in $\ttbar$ events. The uncertainty on the jet $\Pt$ after these 
corrections is given by the $\sigma$ 
of the $\Delta$ distribution, defined as
one half of the separation of the 16th and 84th percentiles of the distribution.
For each bin of reconstructed
jet $\Pt$, we obtain the $\sigma$  of the $\Delta$ distribution, which is
then parametrized as a function of the reconstructed jet $\Pt$.
These uncertainties are shown in Fig.~\ref{f-aa2} for 
jets from $W$ decay and $b$ jets. As above, we display curves for
generic $b$ jets (no selection on decay mode), for jets 
containing an electron, and for jets containing a  muon. 
%The jet $\Pt$ uncertainty is larger for $b$-quark jets and increases as the 
%size of the corresponding correction increases (see Fig.~\ref{f-aa1}). 
These jet $\Pt$ uncertainties are input into the 
kinematic mass fitter (see Section~\ref{s-algor}) and dictate how much 
the jet energies can be altered to accommodate the applied constraints.

  The jet corrections described above are applied only to the four 
highest $\Pt$ jets in the event, which are assumed to be daughters 
of the $t$ and $\overline{t}$ decays. 
Any additional jets beyond the leading four jets are corrected only
with the ``flavor-independent'' corrections (excluding the out-of-cone
corrections, see Section~\ref{s-uclus}) and are assigned an
uncertainty of $0.1\Pt \oplus 1 ~\gevc$. This curve is also shown in 
Fig.~\ref{f-aa2}.

%\subsection{Uncertainty on the jet energy scale}
%\label{s-det_eff}
\subsection{Summary of systematic uncertainties on jet energy measurements} 

  A number of corrections are performed to estimate the original
parton momenta from the observed jets.  
%These corrections were
%outlined in Section~\ref{s-recon} and Section~\ref{s-correct}.
%Uncertainties in these corrections generally arise from either 
%the assumptions used to derive the correction or from finite 
%statistics. 
%We categorize these uncertainties into two main sources: 
%{\it jet energy scale} and {\it soft gluons effects}.
The jet energy scale uncertainty is evaluated 
from the uncertainties in the corrections for calorimeter stability, multiple 
interactions, calorimeter response, jet fragmentation,
underlying event, out-of-cone correction, and splash out. 
Figure~\ref{f-jetsys} shows the dependence of the overall jet energy scale
uncertainty on the corrected jet $\Pt$. The total systematic uncertainty
varies between 7\% for jets with corrected $\Pt$ of 20 $\gevc$ and
3.5\% for jets with $\Pt$=150 $\gevc$.

We do not assign a separate systematic uncertainty to the top specific 
corrections.  Such uncertainties may arise from modeling of initial and final
state gluon radiation, and modeling of the primary parton collision. We
discuss these uncertainties in Section~\ref{s-sys}.
%The systematic uncertainty from the modeling
%of the primary parton collision
%is evaluated by comparing the {\small{HERWIG}} Monte Carlo 
%to the {\small{PYTHIA}} Monte Carlo (see Section~\ref{s-sys}).

%Previously published work~\cite{jetshape} relying
%on charged tracks inside of jets indicated that jet shapes as
%modeled by the parton shower model in the {\small{HERWIG}}
%simulation appear to
%be slightly broader than those observed in data. The jets in
%reference~\cite{jetshape} were, however, of higher jet $\et$ than those
%typically found in this analysis ($~\sim 30-70 \ \gev$). We have
%therefore reevaluated the jet $\Et$ uncertainty from ``soft gluon''
%effects by studying the energy flow around jets
%in samples of events containing $Z^{0}$, $W^{\pm}$, and
%$\gamma$ bosons produced in association with single jets.

% Differences in the fragmentation 
%properties between gluons and quarks and between $b$ quarks and 
%light quarks have also been considered. Figure~\ref{f-anndata} shows
%the fractional $\et$ distribution obtained from the 3 boson+jet
%samples ($\gamma , Z^{0}, {\rm and \ } W^{\pm}$), each of which has
%varying compositions of jets induced from gluons and quarks.  No
%evidence of significant differences is seen.  Figure~\ref{f-bjann}
%shows the fractional $\et$ distribution for $\bbbar$ events which
%also indicates qualitative agreement with the results obtained from
%the $W + {\rm 1 \ jet} \ $ sample.
%Varying the form of
%the parameterization or using $Z+{\rm 1 \ jet} \ $ samples yield
%similar results.

\section{Measurement of other calorimeter variables }
\label{s-uclus}

To measure the top mass we apply energy-momentum conservation to the process
$\ppbar \to \ttbar + X$, with subsequent decay of the $t$ ($\overline{t}$) 
into $W+b$ ($\overline{b}$) (see Section~\ref{s-algor}).  Here, $X$ is the 
unspecified particles which recoil against the $\ttbar$ system.  The 
calorimeter provides the measurement of $X_T$, the transverse momentum of $X$.
The quantity $X_T$ is computed from the energy left over after the lepton and 
the four jets from the $\ttbar$ system are removed from the total measured 
energy. This leaves two terms:
\begin{equation}
\vec{X}_T =  \vec{U}_T + \sum_{i=5}^{N_{jets}}\vec{E}_{T}(jet)
\label{eq:x_t}
\end{equation}

\begin{figure}
\epsfysize=6.5in
\hspace{0.5in}
\epsffile[0 72 612 720]{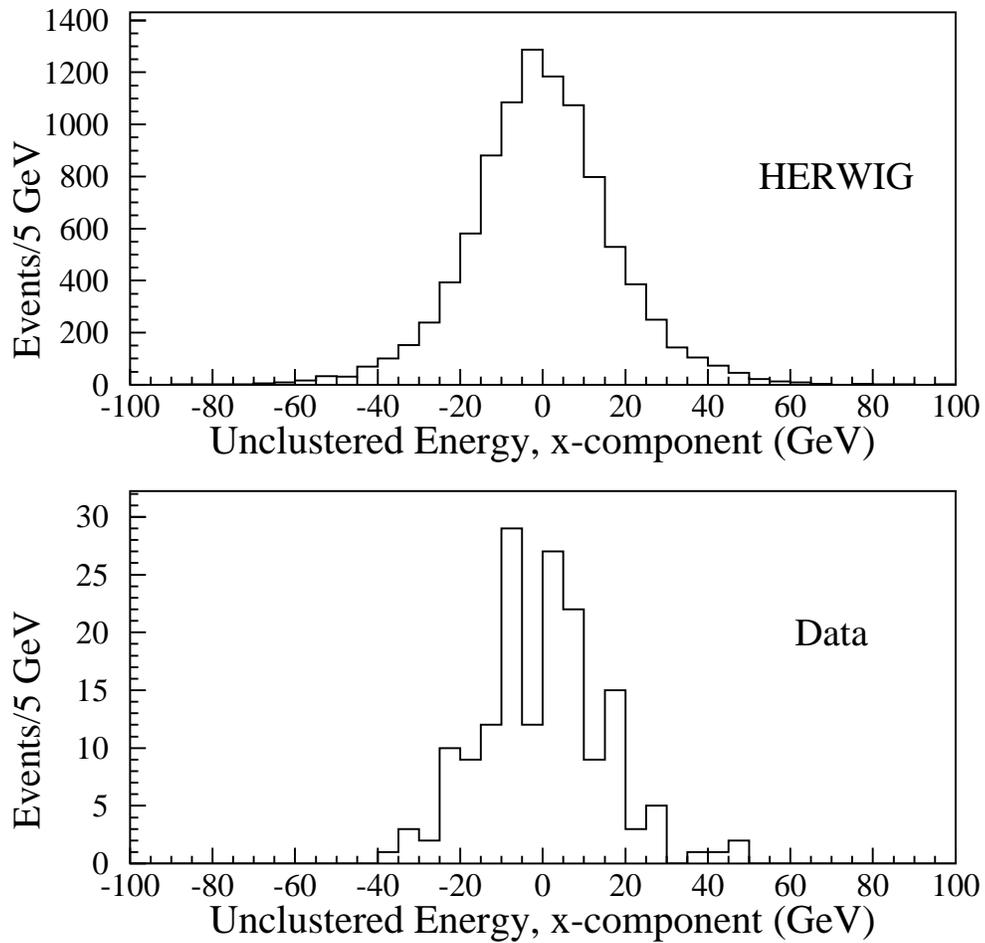}
%\epsffile[0 72 612 720]{$PRD_FIG/uncexy_data_mc175.ps}
\caption{Unclustered energy for the mass sample used here and for 
$\ttbar$ Monte Carlo (\mtop=175 $\gevcc$). Only the $x$ component is shown.   }
\label{u_nc}
\end{figure}

\noindent
%$X_T$ has an x and a y component. 
Each component of the unclustered energy, $\vec{U}_T$, is defined as the 
vector sum of the energies 
in the calorimeter towers after excluding the primary lepton and
all the jets with raw $\Et>8~\gev$ and $|\eta|<$ 3.4 in the event. 
Using a $\ttbar$ Monte Carlo (\mtop=175 $\gevcc$) we find
a distribution in $U_{x}$ with $\langle U_{x}\rangle\sim 0$ and 
$\sigma = 15.8~\gev$ for events which enter into the mass subsamples.
The same distribution for the data has a mean consistent
with 0 and a $\sigma = 14.9~\gev$. The Monte Carlo and data
distributions in $U_{x}$ are shown in Fig.~\ref{u_nc}. Similar results are 
obtained for the $y$ component.

Each component of the unclustered energy is corrected with a single 
factor $f_{u.e.}$ = 1.6,
based on studies of the recoil energy that the calorimeter measures in
$Z$-boson events with no extra jets, where the $Z$ boson is well measured by 
the two leptons it decays into~\cite{et_nc}. 
The final mass value is not sensitive to the value of $f_{u.e.}$.
For example, using  $f_{u.e.}$ = 2.0 makes a negligible
change in the reconstructed top quark mass (0.2 $\gevcc$), hence it is not
included in the Table of systematics in Section~\ref{sumsyst}.
The uncertainty with which each component of $\vec{U}_T$ is measured
is taken to be $100\%$ (added in quadrature to 1 GeV).
The jets beyond the
four with the highest $\Et$ are corrected only within the cone of 0.4,
so as to avoid counting the out of cone energy twice (it is already included 
in the unclustered energy). The uncertainties on these jet energies 
were discussed in Section~\ref{s-correct} and shown in Figure~\ref{f-aa2}.

Another quantity that can be estimated from the calorimeter
measurement  is the $\met$. It is calculated using the following
expression:
\begin{equation}
% -\vec{\met} = \vec{E}_T^{tot} = \vec{E}_T(lepton)+\sum_{i=1}^{4} 
-\vec{\rm \raisebox{.3ex}{$\not$}E}_{T} =
\vec{E}_T(lepton)+\sum_{i=1}^{4} {\vec{E}_T(jet)} + \vec{X}_T
\label{eq:met_eq}
\end{equation}
The above expression shows that the $\met$ measurement is highly 
correlated with the jet energy measurements, and therefore it is not 
considered as an independent measurement in the mass fitting.  
As discussed in Section~\ref{s-algor}, it is only used as a
starting value for the neutrino's transverse momentum when the overall
mass fit is performed.

%% file: event_fitting.tex
\chapter {MASS FITTING}
\label{s-algor}

  The kinematics of events in the decay channel
$\ppbar\to\tljx$ are over-constrained by the number of measured
quantities and the number of applicable energy-momentum conservation 
equations of production and decay. This allows for complete 
reconstruction of the four-momenta of the particles in the
decay chain and hence an event-by-event top mass determination. 

In this section we discuss the methods used for event reconstruction
and then study the validity of the 
algorithms using $\ttbar$ Monte Carlo events.
Effects due to combinatorics, wrong parton assignments 
and shapes of backgrounds on the top mass measurement are also discussed.

\section{Event reconstruction}
 \label{ev-algor}

  The first step in the reconstruction is the estimation of the 
four-momenta of the decay products of the $\ttbar$ pair: the
lepton and the four jets. Electron and muon measurements, resolutions 
and identification are discussed in section~\ref{lepid}.
The four leading jets, as defined in Section~\ref{s-data}, are assumed to be 
the $q$, $\qbar$, $b$, and $\bbar$ quarks from the $\ttbar$ decay chain. 
According to a {\small{HERWIG}}
Monte Carlo plus detector simulation, this assumption is correct 
55--72\% of the time, depending on the number and type (SVX or SLT)
of tags (see Table~\ref{mc_ev_frac}). The momenta of the reconstructed
jets are corrected as described in Section~\ref{s-recon}.
The direction (i.e., $\eta$ and $\phi$) of each parton is assumed
to be the same as the direction of the associated jet. The masses of the
partons are assumed to be 0.5 $\gevcc$, except for $b$ and $\bbar$ quarks
which are assigned a mass of 5.0 $\gevcc$. The resolutions on the
jet energy measurements are discussed in Section~\ref{s-recon}.

%We begin discussion of the mass reconstruction method with
%a general description of the event fitting algorithm. 

%The four-momenta 
%and uncertainties of the jets and the isolated lepton (from
%the assumed W-decay) are the input into the algorithm. From
%these quantities, the momentum of the neutrino is evaluated
%as discussed in the previous section.
%Together with the kinematic constraints imposed by the
%$\ttbar$ hypothesis, the algorithm leads to the
%construction of a $\chi^2$ whose minimization yields 
%fitted values for the momenta of all particles in
%the decay chain. The performance of the algorithm is evaluated 
%using Monte Carlo samples of signal and background events.

%\section{Chisquare Definition}

  The mass fitting algorithm applies the constraints implied
by the production and decay of a $\ttbar$ pair to evaluate an
event-by-event mass. The hypothesis of standard model $\ttbar$ 
implies the production process

\begin{equation}
p\bar{p} \to t + \bar{t} + X, 
\label{eq:ttbar_decay}
\end{equation}

\noindent followed by the decays

\begin{eqnarray}
 t       & \to & W^+ + b, \\
 \bar{t} & \to & W^- + \bar{b}, \\
 W^{\pm} & \to & \ell^{\pm} + \nu, \\
 W^{\mp} & \to & q + \overline{q}^{\prime}.
\label{eq:decay_prod}
\end{eqnarray}

      The quantity $X$, in $p\bar{p} \to t + \bar{t} + X$, represents the unspecified 
  particles recoiling against the $\ttbar$ system. Only two components of
$X$ are measured, as discussed in Section~\ref{s-uclus}.

   An estimate of the top quark mass is obtained on an event-by-event
basis after minimizing a $\chi^2$. In general, the $\chi^2$ definition
is not unique, in that any formulation which expresses the constraints 
implied by the measurements and four-momentum conservation is equally
valid. We have chosen a particular formulation of the $\chi^2$, which
is minimized using the program {\small{MINUIT}}~\cite{minuit}. An alternate 
method, the {\small{SQUAW}} kinematic fit~\cite{dahl},
has also been used and the results are essentially identical.
We describe here both of these fitters.

\subsection{Mass fitting using {\small{MINUIT}} }
The $\chi^2$ expression which uses the {\small{MINUIT}} minimization
routines applies energy and momentum constraints to the above production and
decay chain to obtain six effective constraints: (1,2) the two transverse 
momentum components of the $\ttbar +X$ system must be zero, (3) the invariant 
mass of the $\ell\nu$ system must equal the $W$-boson mass, ${\rm M_W}$, (4)
the invariant mass of the $q\overline{q}^{\prime}$ system must equal 
${\rm M_W}$, and (5,6) the two three-body invariant masses must each 
equal the top quark mass, ${\rm M_t}$.
The relevant unmeasured quantities are then the three momentum
components of the neutrino and the top quark mass. The system may
therefore be solved by minimizing a two-constraint chisquare.
The chisquare expression used to obtain the present results is:
%\noindent Here, $X$ denotes the unspecified debris from the fragmentation of 
%the spectator partons. The (corrected) recoil momentum $X$ is given by
%
%$$\vec{X} = 1.6\times\vec{U} + \sum_{i=5}^{Njets} {\vec{\Pt}_i^{cor}} $$
%
%The first term includes a factor of 1.6 which is the approximate size of 
%the generic jet corrections for raw $\Et\approx 8 \ \gev$. The second
%term in $\vec{X}$ adds the transverse momentum from any additional jets 
%beyond the leading four, and are corrected only with the relative and
%generic jet corrections.
%  In fitting each event to the hypothesized decay, the four largest
%$\Et$ jets are presumed to have come from the $\ttbar$ decay. While
%the jet energies are directly measured (see Section~\ref{s-jetrecon}),
%the neutrino's three-momentum is not measured (we assume the neutrino
%is massless). However, five constraints may be obtained from the
%hypothesized decay chain: the transverse momentum components of the
%$\ttbar + X$ system must be zero; the invariant mass of the $\ell^{\pm}\nu$
%and $j_1 j_2$ must equal the W-boson mass $M_W$, and the mass of the top
%quark must be equal to that of the anti-top quark. The system is therefore
%twice overconstrained, and is solved by minimizing a $\chi^2$.
%We have studied various formulations of this $\chi^2$ and different 
%methods of minimization~\cite{cdf-evidence},
%none of which give results that differ significantly from those 
%described here. We minimize the following $\chi^2$:
\begin{eqnarray}
%\chi^{2} & = & \sum_i^{N_{\ell , jets}}
\chi^{2} & = & \sum_{\ell , jets}
        \frac{\left( \hat{P}_T - P_T \right)^2}{\sigma^{2}_{P_T}} 
 +\sum_{i=x,y}\frac{\left(\hat{U}_i^{\prime}-U_i^{\prime}\right)^2}
 {\sigma^{2}_{U_i^{\prime}}}
     +\frac{\left( M_{\ell\nu}-M_W\right)^2}{\sigma^{2}_{M_W}} \nonumber\\
    & & \mbox{} + \frac{\left( M_{jj} - M_W \right)^2}{\sigma^{2}_{M_W}} + 
        \frac{\left( M_{\ell\nu j} - M_t \right)^2}{\sigma^{2}_{M_t}} +
        \frac{\left( M_{jjj} - M_t \right)^2}{\sigma^{2}_{M_t}}.
\label{eq:chi2_eq}
\end{eqnarray}
The notation is as follows: $\ell$ signifies the primary lepton in 
the event, $\nu$ refers to the inferred neutrino, and $j$ refers to one of
the four leading jets in the event.
The first sum is over the primary lepton and all
%${\rm b}$, $\bbar$, ${\rm q}$,   ${\rm{\overline{q}^{\prime}}$ partons,
jets with raw $\Et>8\ \gev$ and $|\eta |<2.4$.
%The quantity $\hat{P}_T$ represents the transverse
%momentum of the objects as altered by the fit procedure, whereas $\Pt$ 
%represents the input value. 
The second sum is over the transverse components of the unclustered 
energy~\cite{unclus}, discussed in Section~\ref{s-uclus}, plus those of the
energies of jets with  $2.4<|\eta|<3.4$. The hatted symbols 
in the sums represent quantities altered by the fit procedure, whereas 
unhatted symbols represent the input values.
The uncertainties on the
energy of the primary lepton, the jets, and the unclustered energy are 
discussed in Sections~\ref{lepid}  and ~\ref{s-recon}.
 The W-boson mass, ${\rm M_W}$, is taken to be 
80.4 $\gevcc$~\cite{wmass}, $\sigma_{\rm M_W}$ is set to 
2.1 $\gevcc$~\cite{wwid}, and $\sigma_{M_t}$ is set to 2.5 $\gevcc$. 
The results are 
insensitive to the values used for  $\sigma_{\rm M_W}$ and
$\sigma_{M_t}$. The 
quantity $M_{\ell\nu}$ is the invariant mass of the primary lepton and the
neutrino and $M_{\ell\nu j}$ is the invariant mass of the primary lepton,
neutrino, and one of the four leading jets. Of the remaining three
jets, we assign two to the decay products of
the $W$ boson in order to calculate $M_{jj}$. The third jet
is then combined with the other two jets to form the three
body mass $M_{jjj}$. The issue of combinatorics is discussed in
Section~\ref{s-combinatorics}.

The first two constraints are that the total
transverse momentum components of the $\ttbar +X$ system 
are zero. These constraints are imposed 
by setting the neutrino transverse momenta to exactly balance
the sum of the current $\hat{P}_T$ and $\hat{U_T^{\prime}}$ values.
The other four constraints appear as explicit terms in the $\chi^2$.
This $\chi^2$ yields two minima which correspond to the two 
solutions for the neutrino longitudinal momentum in the $W$ decay.
This is referred to as the $P^{\nu}_{z}$ ambiguity.
After minimizing this $\chi^2$ with respect to the 
collective set of transverse momenta, $\hat{P}_{T}$, for the
jets and the charged lepton, the unclustered energy, $\hat{U}_T$, 
the $z$-component of the neutrino momentum, and the top mass,
$M_t$, for the event, we obtain an event-by-event determination
of the top quark mass.

%    In general, this $\chi^2$ yields two minima, i.e, two solutions, for
%  any given set of input values. The two minima correspond to the two
%  solutions for the neutrino longitudinal momentum in the $W\to l\nu$ decay
%  when given the lepton four-momentum and the neutrino transverse momenta.
%  Therefore the two solutions are referred to as the $\Pz^{\nu}$ ambiguity.

%\noindent where the sums are over the primary lepton,
%all identified jets with $\et> 8 \ \gev$, $|\eta|<2.4$,
%and the transverse components of the unclustered energy ({\it U}).
%$\hat{E}_T$ and $\hat{U}$ represent the transverse energies of the objects
%as altered by the fit procedure, whereas $\Et$ and $U$ represent
%measured values, corrected for detector and physics effects.
%The W-boson mass, $M_W$, is taken to be 80.41 $\gevcc$, and its width,
%$\sigma_{M_W}$ is set to 2.12 $\gevcc$. $M_t$ is 
%the fitted top quark mass for the event, and
%$\sigma_{M_t}=2.5 \ \gevcc$ is the width of the top quark
%(we find the results to be insensitive to this choice as long as 
%$\sigma_{M_t}$ is larger than $\sim 1~\gevcc$).
%The $M_{xx(x)}$ are invariant masses of the designated objects.

\subsection{The SQUAW fitter }
 \label{s-squaw}
The {\small{SQUAW}} fitter is a general kinematic fitting program that can
be used for any production and decay processes, provided that there
are enough constraints~\cite{cdf-evidence,dahl}. It has been used to measure 
the top quark mass in the lepton+jets channel and for the all-hadronic
decay channel~\cite{all-hadron}.

In brief, it applies energy-momentum conservation to the five 
processes~(\ref{eq:ttbar_decay})--(\ref{eq:decay_prod}),
thus providing 20 equations, i.e., 20 constraints, for the measured quantities
and their uncertainties. It uses the measured $W$ mass, 
M$_W$ = 80.4~$\gevcc$. In the fit an uncertainty is assigned to the W mass
in order to take into account the expected W width of $2.1~\gevcc$. Additional
ingredients of the kinematic fit:
\begin{itemize}
\item The measured quantities are: the lepton, the four leading jet momenta, 
      and $X_T$.
\item For each event there are 18 unknowns. These are: energy and $P_{z}$ of 
      $X$ (2), 3-momenta of $t$ and $\overline{t}$ plus the top mass (7), 
      3-momenta of the $W$ bosons (6), and the 3-momenta of the $\nu$ (3).
\end{itemize}
This is then a 5-vertices, 2-constraints fit, 5V-2C in {\small{SQUAW}}'s 
language. Notice that the $\nu$ momentum is considered an unknown quantity. 
This is because the $\met$ is highly correlated with the jet momentum 
measurements.  The calculated value of $\met$ is used as
a starting point to help with the convergence of the fit.
Lagrange multiplier techniques are used to solve the 20 equations.
The final $\chi^{2}$ has contributions from all 20 equations.

One of the differences with the {\small{MINUIT}} algorithm is that 
SQUAW works with the 4-vectors, hence it allows the angles of the
lepton and jets to vary within their uncertainties.  The momentum
magnitude and angles are assumed to be uncorrelated.

 The results of the two methods for a given event are very close. In 
the 76 event data sample the masses obtained with the two methods 
(using the mass from the lowest $\chi^2$ solution in each case)
differ on the 
average by 0.1 $\gevcc$, and in 70\% of the events the absolute value
of the mass differerence is less than 0.5 $\gevcc$. 

%\subsection{Alternative formulation of the chisquare}
%
%    An alternative formulation of the $\chi^2$ expression to be minimized
%  has been tried. That alternative is the formulation in the kinematic
%  fitting program {\small{SQUAW}} ~\cite{cdf-evidence,dahl}
%    Some of the ways in which the {\small{SQUAW}} program
%  differs from that described
%  above are as follows. It allows the angles of the four partons
%  and the primary lepton to vary within their assigned errors. 
%  It sets the two three-body masses to be equal (i.e., does not use $\sigma_{M_t}$).
%  The program inputs into the $\chi^2$ the recoiling system, $X$, as a single entity, 
%  rather than its 
%  component jets and unclustered energy. {\small{SQUAW}} adds the constraints of energy and
%  longitudinal momentum conservation at production, at the cost of
%  adding as unmeasured quantities the energy and longitudinal momentum
%  of the entity $X$. {\small{SQUAW}} makes the approximation that the two 
%  transverse
%  components of $X$ are uncorrelated, as opposed to the approximation that
%  just the unclustered energy transverse components are uncorrelated.
%    Despite the above differences, the results from the two formulations
%  are almost identical. 

\section{Combinatorics}
\label{s-combinatorics}
There is always some ambiguity in how to assign the four leading
jets to the four relevant partons. If none of the jets is tagged
as a $b$ candidate, by either the SVX or SLT algorithm, then there
are 12 different ways of assigning jets to the $b$ and $\bbar$ partons.
Combined with the $P^{\nu}_{z}$ ambiguity, there are then 24 combinations,
or configurations, per event. If one jet is tagged as a $b$ candidate,
we require that it is assigned to a $b$ or $\bbar$ parton, and this
reduces the number of allowed combinations to 12.
If two jets are $b$ tagged, there are four combinations.
Of the above combinations the solution with the lowest 
$\chi^2$ is chosen, and
that solution is required to have $\chi^2<10$. The latter requirement
defines criterion 9 of the Top Mass Candidate Sample described
in Section~\ref{s-data}. We have not found a satisfactory method
for improving the top mass resolution by including any solutions 
with $\chi^2$ values larger than the lowest one, and therefore we
take the lowest $\chi^2$ solution as the best estimate of the top
mass for each event.

%It is not known if a reconstructed jet in the calorimeter was
%initiated by the fragmentation of either a light quark, a 
%heavy quark or a gluon. As a result, the mass fitting algorithm must
%consider all unique ways for assigning the 4 highest $E_T$ jets
%to the 4 daughter partons from the $\ttbar$ decay. This leads to
%a total of 12 combinations. Including a factor
%of 2 for the two neutrino $P_Z$ solutions (see Section~\ref{s-correct}),
%a total of 24 combinations
%are tried. If $b$-tagging information (either SVX or SLT tags)
%is available, and we impose the requirement that $b$-tagged jets must be
%considered as $b$ quarks in the fit, the number of allowed combinations
%is reduced to 12 (for 1 $b$-tagged jet) or 4 (for 2 $b$-tagged jets).
%The solution with the lowest $\chi^2$ among the allowable configurations
%is chosen as our best estimate of the top mass for a given event.
%This solution is required to have $\chi^2<10$.
%Using $b$-tagging information not only improves the S/B (as
%discussed in Section~\ref{s-optimize}), but also improves the
%fraction of correct jet-parton matches, particularly for SVX
%tagged jets. The performance of the
%mass fitter, as it relates to correct jet-parton matches and 
%available tagging information is discussed in Section~\ref{s-fperform}.

\section{Impact of gluon radiation}
\label{s-gluonrad}

  A substantial fraction of $\ttbar$ events are expected to contain
extra jets resulting from gluon radiation. From a {\small{HERWIG}} 
Monte Carlo plus detector simulation, we find that $\approx$40\%
of events have one or more jets which do not correspond to the
partons from the $\ttbar$ decay. These extra jets may be
produced during the production of the $\ttbar$ 
pair (initial state
radiation) or in the decay stage (final state radiation) ~\cite{lynorr1}.
From a theoretical perspective, whether or not the extra jet(s) are 
to be included in the fit depends on whether the gluon was radiated 
during production of the $\ttbar$ pair or during its decay. If the 
radiation comes from the production stage, then it should not be 
included in the mass fit. If the radiation is produced from a quark in 
the decay stage, then it should be included as one of the decay 
products~\cite{lynorr1}. 

   From an experimental perspective, the radiation results in jets which 
may or may not have been produced in the $\ttbar$ decay process.
On an event-by-event basis, production and decay stage radiation cannot 
be differentiated from each other or, for that matter, from the partons 
from the $\ttbar$ decay (unless the jet is $b$ tagged).
Gluon jets which come from decay stage radiation are more correlated
with the partons emerging from the hard scatter, and therefore one
can consider merging jets which are close in $\eta -\phi$ space.
It is also possible to try all unique permutations of four jets among all 
the reconstructed jets. However, taking a fifth jet into consideration
increases the number of combinations by a factor of 3, 4, and 5 for
the 2, 1, and 0 $b$-tag cases respectively. This increase in the number
of solutions reduces the probability
for choosing the correct jet assignment. The mass reconstruction 
presented here does not implement either of these possibilities. 
Our approach is to assume the model of initial and final
state radiation in the Monte Carlo simulation is correct, and to associate
a systematic uncertainty with this assumption. 

\section{Results of the kinematic fit on simulated $\ttbar$ events}
\label{s-fperform}

\begin{figure}[ht]
\epsfysize=6.5in
\hspace{0.5in}
%\epsffile[0 72 612 720]{cumul.ps}
\epsffile[0 72 612 720]{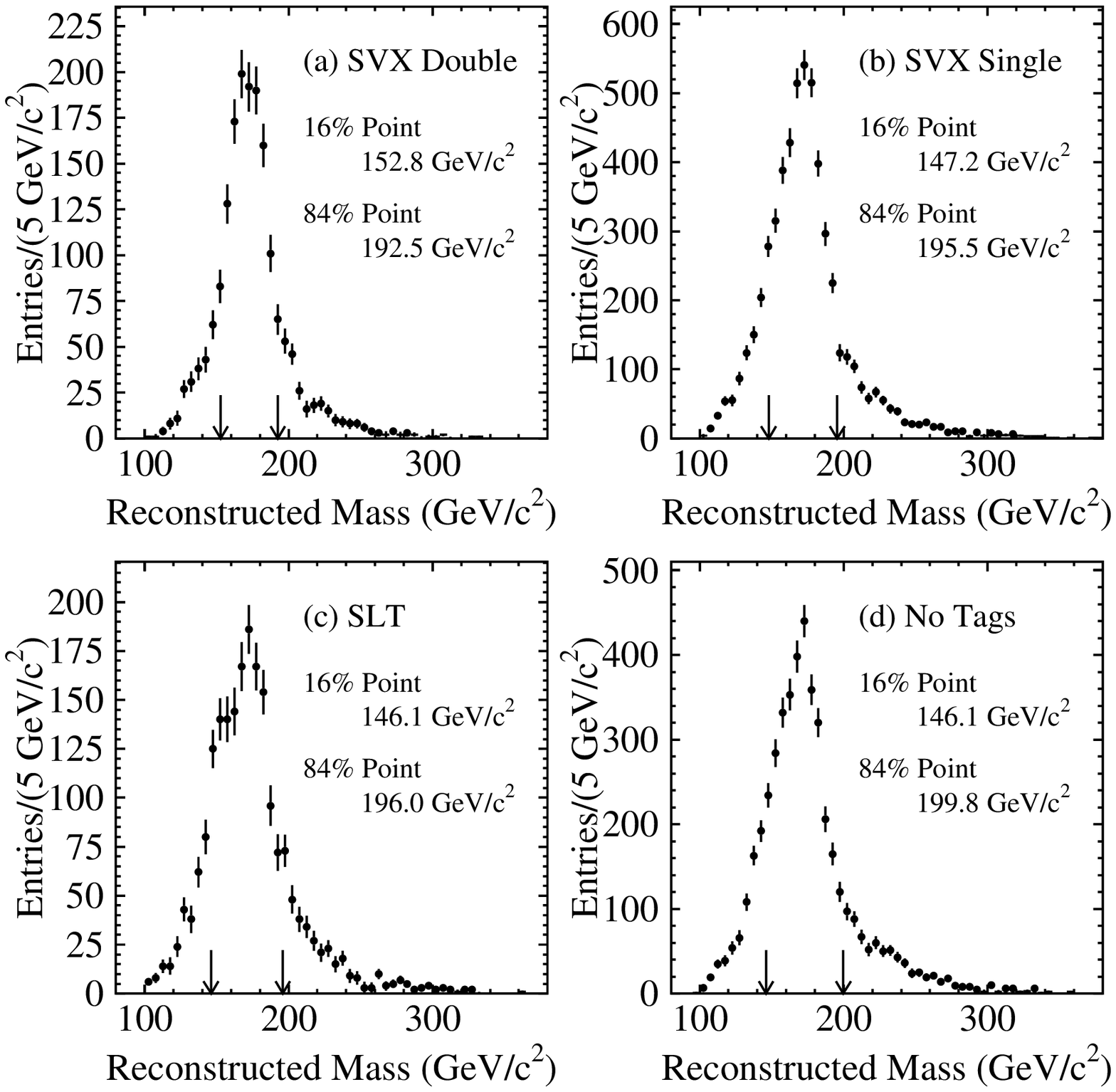}
%\epsffile[0 72 612 720]{$PRD_FIG/mass_subsamp.ps}
\caption{Reconstructed-mass distributions for {\small{HERWIG}} $\ttbar$
events (\mtop=175 $\gevcc$) for the mass subsamples: (a) SVX Double,
(b) SVX Single, (c) SLT (no SVX), and (d) No Tags.  The $16^{\rm th}$ and 
$84^{\rm th}$ percentiles for each distribution are indicated by the 
arrows on the figures along with their values.}
\label{cumul}
\end{figure}

The reconstructed-mass distribution obtained by fitting simulated $\ttbar$
events depends on the intrinsic resolution of the detector,
and, more importantly, the ability to correctly associate
the daughter partons from a $\ttbar$ decay with the observed jets.
Both combinatorics and gluon radiation play a role in degrading
the resolution of the top quark mass measurement. In this
section, we discuss the performance of the mass fitter 
by dividing events (which enter into one of the four mass subsamples) into 
three categories:
\begin{enumerate}
  \item {\bf Correctly Assigned Events:}
          Each of the four leading jets are within $\Delta R <0.4$ of 
          a parton from the $\ttbar$ decay and are correctly
          associated with the appropriate quark by the lowest
          $\chi^2$ solution satisfying any imposed tagging
          requirements. The jet-parton match is required to be
          unique.
  \item {\bf Incorrectly Assigned Events:}
          Each of the four leading jets are within $\Delta R<0.4$ of 
          a parton from the $\ttbar$ decay and each jet-parton match
          is unique, but the configuration with the lowest $\chi^2$
          is not the correct one.
  \item {\bf Ill-Defined Events:}
          The four leading partons from the $\ttbar$ decay
          cannot be uniquely matched ($\Delta R<0.4$) to the four leading 
          jets in the event. Such events often 
          have extra jets produced from either initial state
          or final state radiation.
\end{enumerate}
The fractions of events falling into each of these categories
are estimated using a {\small{HERWIG}} $\ttbar$ Monte Carlo plus
detector simulation. 
%The fractions
%in each category depend on the detector acceptance, resolution, 
%and the amount of gluon radiation. 
These fractions depend on the $b$-tagging information in the event. 
For example, having two $b$-tagged jets in an event reduces the
probability that one (or more) of the leading four jets is a gluon 
jet. The fractions of events falling into categories 1--3 above, and the 
width of the reconstructed-mass distribution for each of the four mass 
subsamples are shown in Table~\ref{mc_ev_frac}. The widths are calculated
as half the difference between the $16^{\rm th}$ and $84^{\rm th}$ percentiles
of the reconstructed-mass distributions.
The reconstructed-mass distributions for the four mass subsamples
are shown in Fig.~\ref{cumul}.
%Events are required to pass all the kinematic cuts
%and that the $\chi^2$ for the best combination 
%be less than 10. The results are shown in Table~\ref{mc_ev_frac}. 

\begin{table}[ht]
\begin{center}
\begin{tabular}{ccccc}
\hline\hline
        & \multicolumn{3}{c}{Event fractions (\%)} & Width  \\ 
\cline{2-4}
Data sample &     1     &    2      &     3     & ($\gevcc$) \\
\hline
SVX Double  & $49\pm 2$ & $23\pm 2$ & $28\pm 2$ & 19.9    \\           
SVX Single  & $30\pm 1$ & $26\pm 1$ & $44\pm 1$ & 24.2    \\      
SLT         & $26\pm 2$ & $31\pm 2$ & $43\pm 2$ & 25.0    \\ 
No Tags     & $23\pm 1$ & $32\pm 1$ & $45\pm 1$ & 26.9    \\
\hline\hline
\end{tabular}
\end{center}
\caption{Fractions of $t\bar{t}$ events falling into categories 1--3 
described in the text. The last column shows the width of 
the distribution of reconstructed masses for each subsample.
The width is taken to be half the difference between the
$16^{\rm th}$ and $84^{\rm th}$ percentiles of the relevant mass distribution.}
\label{mc_ev_frac}
\end{table}

\begin{figure}[ht]
%\leavevmode
%\begin{center}
\epsfysize=6.5in
\hspace{0.5in}
\epsffile[0 72 612 720]{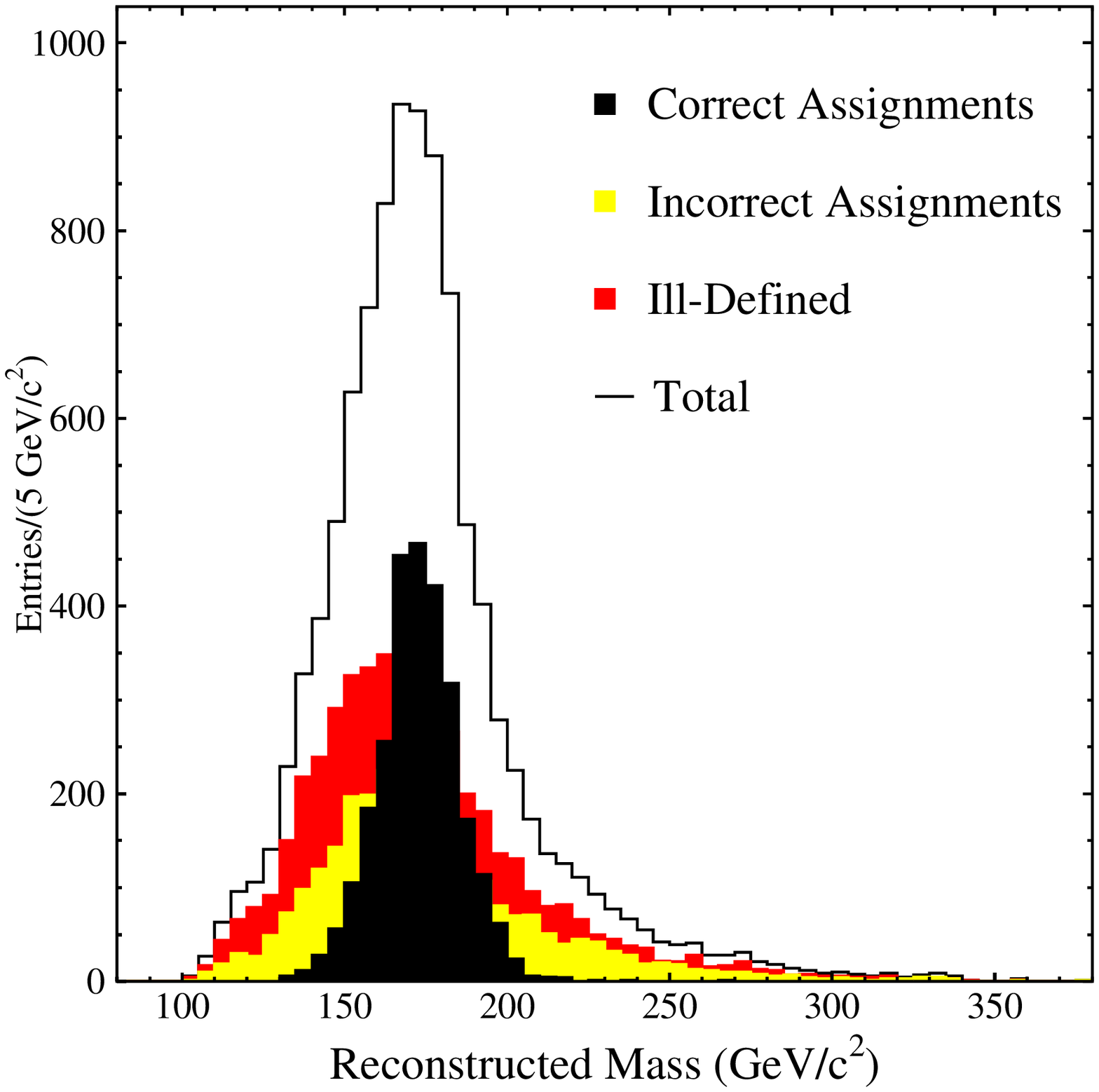}
%\epsffile[0 72 612 720]{$PRD_FIG/tag_breakdown.ps}
%\epsffile{tag_breakdown_f3.eps}
%\end{center}
\caption{Reconstructed mass for $\Mt=175 \ \gevcc$ $\ttbar$ events which
enter into any of the four subsamples.
The black filled histogram shows the distribution
for those events for which the selected jet-parton configuration was also
the correct one (category 1).  The lightly shaded histogram
shows the distributions for which a correct assignment could be defined,
but was not selected (category 2). The darker shaded histogram shows the 
distribution for events where a correct assignment was ill-defined 
(category 3). The solid line shows the three distributions combined.}
\label{f-frecon}
\end{figure}

As Table~\ref{mc_ev_frac} shows, the fraction of correctly assigned
jets increases as the number and purity of $b$ tags increase. 
Figure~\ref{f-frecon} shows the reconstructed-mass distributions for events
in each of these three categories. 
%The events are required to have
%at least one $b$-tagged jet among the leading 
%four jets. 
When the correct jet-parton assignments
are made (category 1), the resolution is $\approx 13\ \gevcc$, while 
for categories 2 and 3 it is $\approx$36 and 34 $\gevcc$ respectively.
As Fig.~\ref{f-frecon} demonstrates, the mass resolution is
dominated by incorrect assignment of jets to partons from the 
$\ttbar$ decay. 
%However, we observe that the 
%reconstructed mass distribution for misassigned events peaks at
%nearly the same mass as correctly assigned events. 
%The width
%of the mass distributions for each subsample depends on the fraction 
%of events in each category and the corresponding resolution. 
For Double SVX tagged events, where nearly half of the events
have the four leading jets correctly assigned to the $\ttbar$ decay products,
we obtain the best resolution on the reconstructed mass. 
%The
%reconstructed mass distributions for the four mass subsamples are
%shown in Fig.~\ref{mass_subsamp}. The arrows indicate the points at
%which 16\% and 84\% of the entries are below. The widths given in
%Table~\ref{mc_ev_frac} are one half the difference between the 84\%
%and 16\% points ($\pm$34\% about the median).

%
%%%%%%%%%%%%%%%%%%%%%%%%%
%

\begin{figure}[ht]
\epsfysize=6.5in
\hspace{0.5in}
\epsffile{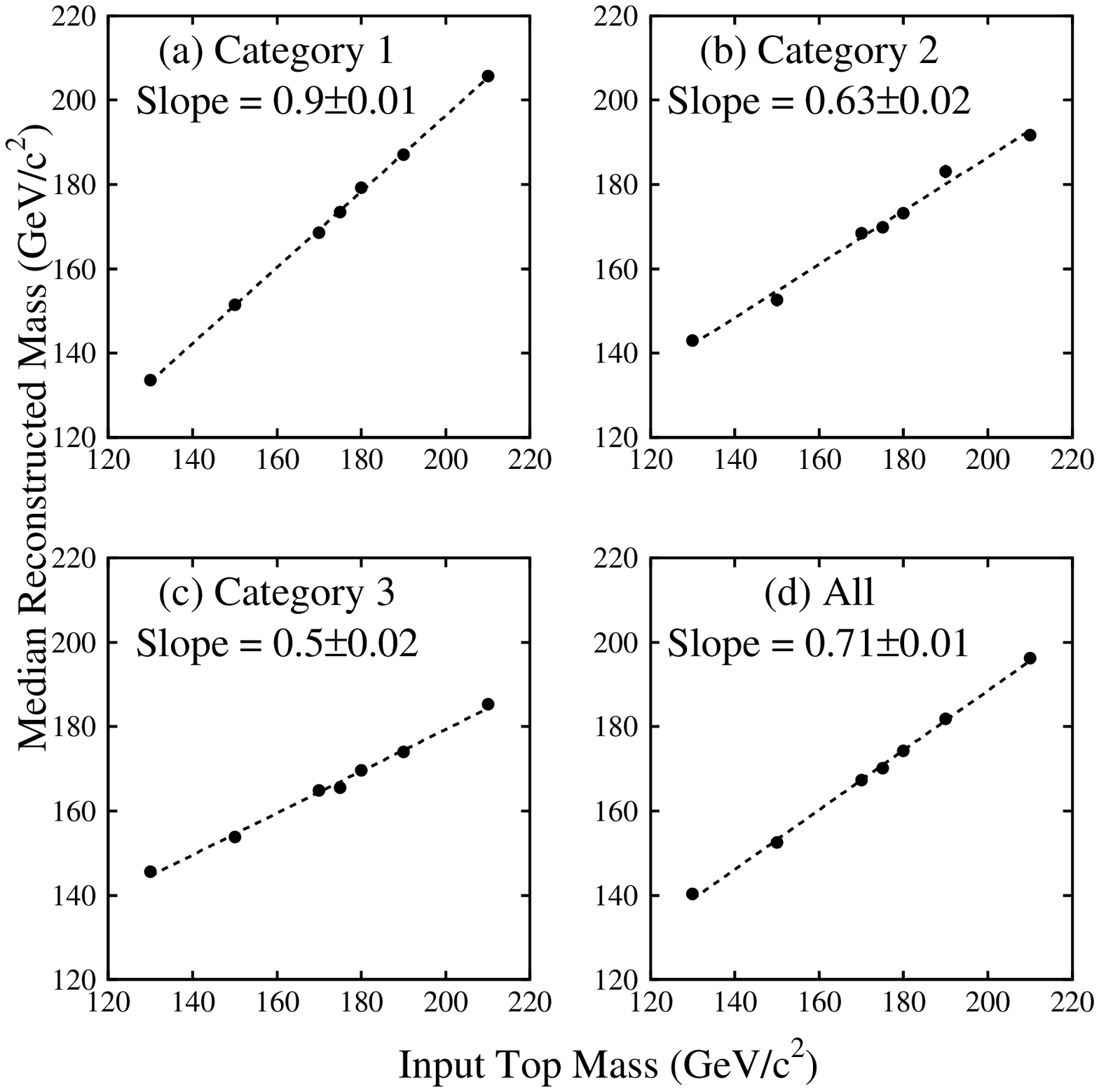}
%\epsffile{$PRD_FIG/med_vs_input_top.ps}
\caption{Median of the reconstructed-mass distribution as a function
of the input top quark mass used in the simulation. The medians are
evaluated from mass distributions which include events from all 
four mass subsamples. The figure 
demonstrates the relative sensitivity of the reconstructed-mass
distribution to the input top quark mass. We show the distributions
for events in (a) category 1, (b) category 2, (c) category 3, and
(d) the three categories combined. The slopes indicated were evaluated using
a linear fit (dashed line) to the data points.}
\label{med_vs_input_top}
\end{figure}

%\begin{figure}[ht]
%\leavevmode
%\begin{center}
%\epsfysize=6.0in
%\epsffile[0 72 612 720]{mass_subsamp.ps}
%\epsffile{tag_breakdown_f3.eps}
%\end{center}
%\caption{Reconstructed-mass distributions for $\Mt=175 \ \gevcc$ $\ttbar$ events 
%which are in the four mass subsamples. The arrows indicate the 16\% and 84\%
%points in the cumulative distribution of the respective reconstructed
%mass distribution. The widths quoted in the figures are one-half of
%the difference between the 84\% and 16\% points.}
%\label{mass_subsamp}
%\end{figure}
%
%
   A priori, it is not obvious whether events which have the jets misassigned
to the $\ttbar$ daughter partons contain information on the top quark
mass. This is quantified by studying the sensitivity of the 
distribution of reconstructed masses to changes in the input value of the 
top quark mass. We examine the events in categories 1--3 separately
in order to determine if the misassigned events contribute information
to the top quark mass measurement. For each category of events, we 
evaluate the rate of change of the median of the reconstructed-mass
distribution as we vary the input value of the top quark mass. 
Larger changes in the median imply greater sensitivity to the top
quark mass. Figure~\ref{med_vs_input_top} shows the median 
reconstructed-mass as a function of the input top quark mass. 
Events from all four mass subsamples 
%of the Optimized analysis 
are included in the
distributions. The four distributions
correspond to events in (a) category 1, (b) category 2, (c) category 3,
and (d) the three categories combined. We find that the events in which the
jets are correctly assigned to the partons have the largest slope
(0.90), while incorrectly assigned events have a slope of 0.62 and
ill-defined events have a slope of 0.48.
Correctly assigned events (category 1) do not have a slope
of 1.0 because the top-specific corrections (see Section~\ref{s-correct})
are derived using a specific input top quark mass of 170 $\gevcc$. 
%This 10\% reduction in the slope results in a slight increase in the expected
%statistical uncertainty on the measured top quark mass.
%Based on the widths presented in Table~\ref{mc_ev_frac}, we make
%two conclusions. First, the
%events with correct jet-to-parton assignments result in significantly
%better resolution than events which have incorrect jet-to-parton
%assignments. Secondly, we find that 
We conclude that the events with incorrect jet-to-parton 
assignments do in fact contain information on the top quark mass,
since the slope is not zero. However, because of the smaller slope
and larger width of the reconstructed-mass distribution, incorrect
combinations degrade the resolution of the top quark mass measurement. 
%It is now clear that events with both correct and incorrect jet-to-parton
%assignments contribute to the precision with which we measure the top
%quark mass. 
The slopes for each of the four subsamples in each 
category are shown in Table~\ref{slope_tab}.  The slopes vary from 
a maximum of 0.81 for SVX Double tags to a minimum of 0.62 for SLT 
tagged events. 
%A larger slope implies better discrimination
%between different top quark mass values. 
Since SVX double-tagged events have the largest slope, narrowest
width and lowest background, they generally yield the
best precision on the top quark mass measurement (for equal size
subsamples).

\begin{table}[ht]
\begin{center}
\begin{tabular}{ccccc}
\hline\hline
           & \multicolumn{4}{c}{Slopes} \\ 
\cline{2-5}
Data sample&        1       &         2      &        3       & Combined \\
\hline
SVX Double & $0.89\pm 0.02$ & $0.48\pm 0.08$ & $0.57\pm 0.05$ &$0.81\pm 0.03$\\
SVX Single & $0.90\pm 0.01$ & $0.60\pm 0.04$ & $0.52\pm 0.02$ &$0.72\pm 0.02$\\
SLT        & $0.93\pm 0.02$ & $0.68\pm 0.05$ & $0.38\pm 0.04$ &$0.62\pm 0.03$\\
No Tags    & $0.90\pm 0.01$ & $0.62\pm 0.03$ & $0.47\pm 0.03$ &$0.68\pm 0.02$\\
\hline\hline
\end{tabular}
\end{center}
\caption{Rate of change (``slope'') of the median reconstructed mass with 
the input value of the top quark mass, for the four mass subsamples.
For each subsample, we show the slope for the three categories of 
events defined in the text, both separately and combined.}
\label{slope_tab}
\end{table}

\subsection{Mass reconstruction in other $\ttbar$ decay channels}
\label{s-nonljt}

  Although the fitting procedure  assumes that the candidate $\ttbar$ events
have decayed through the 
${\ttbar \rightarrow (e~{\rm or }~\mu) \nu q \bar{q}^{\prime} b \bar{b}}$
channel, there is a non-negligible
contribution from top events decaying through other channels.
The additional acceptance from other decay channels 
comes mostly from events where either an electron or a $\tau$ from 
the $t$ or $\tbar$ decay is misconstrued as a jet or from events with
a leptonically decaying $\tau$. In either case, two $b$ jets are still 
present. The fourth jet can be produced through gluon radiation. 
Table~\ref{t-nljt} gives the expected contributions of various decay 
channels to the candidate $\ttbar$ sample and to the subsample with 
at least one SVX or SLT tag. It shows an 11\% contribution from $\tau$
events and 4\% contribution from $ee$, $e\mu$ and $\mu\mu$ events. 

\begin{figure}[b]
%\leavevmode
%\begin{center}
\epsfysize=6.5in
\hspace{0.5in}
%\epsffile{nonljt.eps}
\epsffile[0 72 612 720]{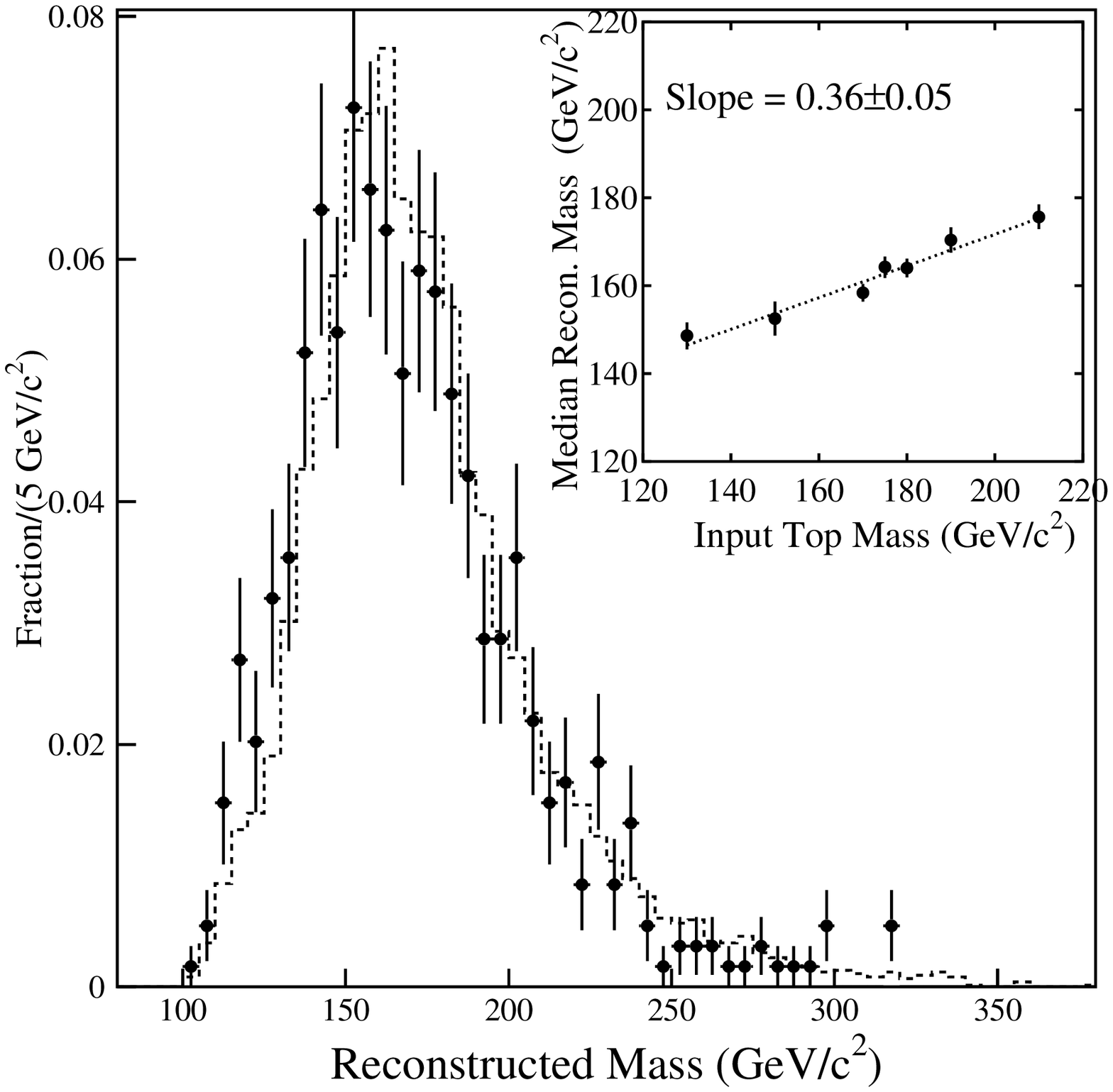}
%\epsffile[0 72 612 720]{$PRD_FIG/nonljt.ps}
%\end{center}
\caption{Reconstructed masses for $\ttbar$ events (\mtop=175 $\gevcc$)
which decay into decay channels other than $e$+jets or $\mu$+jets,
but fit the hypothesized 
${\ttbar \rightarrow (e~{\rm or }~\mu) \nu q \bar{q}^{\prime} b \bar{b}}$
decay chain (points). Most of these events
are due to $W$ decays into $\tau$ leptons (see Table~6.3). 
 Events from all four mass subsamples
%of the Optimized analysis 
are included.  For comparison, the
dashed distribution corresponds to events decaying through the 
${\ttbar \rightarrow (e~{\rm or }~\mu) \nu q \bar{q}^{\prime} b \bar{b}}$
channel, but the lowest $\chi^2$ solution is incorrect (i.e., events in
categories 2 and 3). The inset shows how the
median of the reconstructed-mass distribution changes with the 
input value of the top quark mass used in the simulation. }
\label{f-nljtmss}
\end{figure}

Figure~\ref{f-nljtmss}
shows the reconstructed-mass distribution for events from these
decay channels. 
%other than $e$+jets or $\mu$+jets, i.e., the channels
%shown in Table~\ref{t-nljet}.
%
%showing 
%that kinematically they appear similar to events decaying in the
%$\tljx$ mode, but not meeting the hypothesized decay chain. 
%
The inset shows 
how the median of the reconstructed-mass distribution changes with the input
value of the top quark mass used in the simulation. The 
relatively low, but non-zero value of the slope indicates that these
events also provide information about the top quark mass. The signal
templates, to be discussed in Section~\ref{s-like}, include
contributions from these channels, hence we do not expect any bias
on the fitted top mass from these events.
%Clearly
%these events contain less information about the top quark mass
%than events which decay through the channel $\tljx$.

\begin{table}[ht]
\begin{center}
\begin{tabular}{lcc}
\hline\hline
Channel                & Top Mass Candidate Sample &    Tagged Events    \\
\hline
e+jets                 &    $0.423 \pm 0.008$      &  $0.424 \pm 0.010$  \\
$\mu$+jets             &    $0.426 \pm 0.008$      &  $0.430 \pm 0.010$  \\
$e-\tau(had)$          &    $0.017 \pm 0.002$      &  $0.017 \pm 0.003$  \\
$e-\tau(lep)$          &    $0.007 \pm 0.001$      &  $0.007 \pm 0.002$  \\
$ee$                   &    $0.012 \pm 0.002$      &  $0.011 \pm 0.002$  \\ 
$e-\mu$                &    $0.023 \pm 0.002$      &  $0.024 \pm 0.003$  \\ 
$\mu-\tau(had)$        &    $0.017 \pm 0.002$      &  $0.017 \pm 0.003$  \\
$\mu-\tau(lep)$        &    $0.005 \pm 0.001$      &  $0.006 \pm 0.002$  \\
$\mu-\mu$              &    $0.004 \pm 0.001$      &  $0.004 \pm 0.001$  \\
$\tau(had)+\tau(lep)$  &    $0.002 \pm 0.001$      &    -                \\
$\tau(lep)$+jets       &    $0.063 \pm 0.004$      &  $0.058 \pm 0.005$  \\
%$\tau(lep)+\tau(lep)$ &    $0.000 \pm 0.000$      &  $0.001 \pm 0.001$  \\ 
\hline\hline
\end{tabular}
\end{center}
\caption{Fractional contribution (according to Monte Carlo simulation)
of lepton+jets events to $\ttbar$ events in the Top Mass Candidate 
Sample and the Tagged subsample.  The Tagged subsample includes events with 
at least one $b$-tagged jet.  Similar numbers are found for other subsamples 
({\it i.e.} SVX Double, SVX Single and SLT). 
%but are not included for the sake of brevity.
{\it{lep}} and {\it{had}} denote leptonic and hadronic decays,
respectively, for the $\tau$ lepton. A dash indicates that no events  
were found in the category.}
%No events with in either the fully hadronic channel or where one $W$
%decayed hadronically and the other through
% $W\rightarrow \tau\nu,\tau\rightarrow {\rm hadrons}$.}
\label{t-nljt}
\end{table}

\section{Mass reconstruction in non-$t\bar{t}$ events}
\label{s-bkgd}
\begin{figure}[ht]
\leavevmode
\epsfysize=6.5in
\hspace{0.5in}
\epsffile[0 72 612 720]{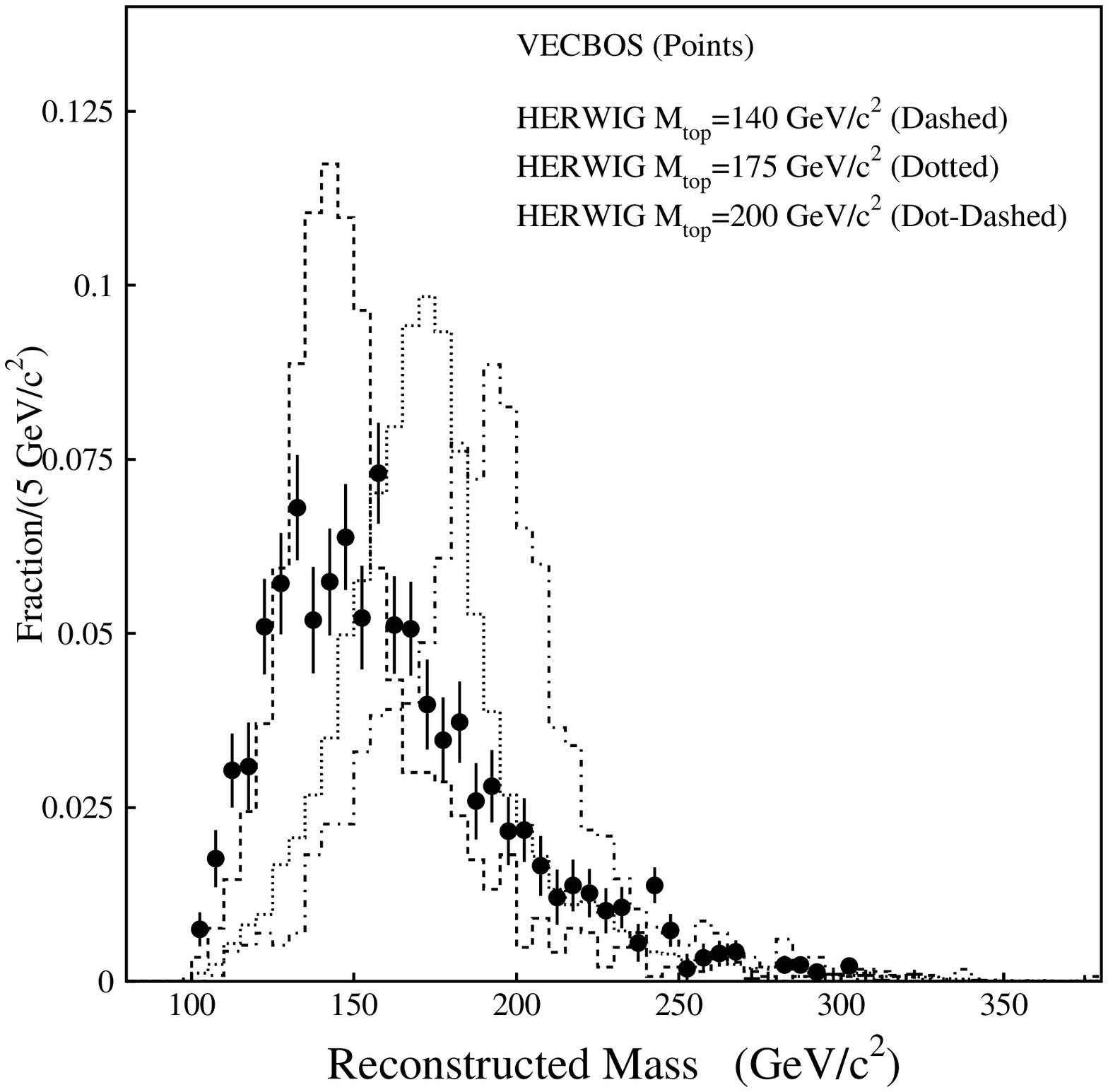}
%\epsffile[0 72 612 720]{$PRD_FIG/vecstd.ps}
\caption{Reconstructed-mass distribution for 
background events from the {\small{VECBOS}} Monte Carlo simulation (points).
Also shown are the reconstructed-mass distributions from the
{\small{HERWIG}} Monte Carlo simulation using input top quark
masses of 140, 175, and 200 $\gevcc$. In all cases, events are required to 
have at least one SVX or SLT tagged jet. Each distribution is normalized
to have unit area.}
\label{vecstd}
\end{figure}

  Non-$\ttbar$ events are also present in the data samples.  For
all the samples considered, the dominant background is expected
to be from production of $W$ bosons in association with
extra jets.  The background shape is modeled with the
{\small{VECBOS}} Monte Carlo simulation. As with $\ttbar$ events,
we fit the background events using the $\chi^2$ defined 
in Section~\ref{s-algor}. Since the sample
of events does not contain $\ttbar$, one does not expect any
resonant peaks in the reconstructed-mass spectra. 
%However, a 
%peak at low reconstructed mass occurs because of the event 
%selection and the kinematic constraints in the $\chi^2$.
%Assuming that the top decay width~\cite{twidth} is smaller than
%the resolution on the reconstructed mass,
%we expect $\ttbar$ events to have a more pronounced peak
%in the reconstructed mass spectrum than do background events.
The reconstructed-mass spectrum for {\small{VECBOS}} events 
which have at least one $b$-tagged jet is shown in Fig.~\ref{vecstd}.
This distribution is compared to the distributions for $\ttbar$
events with input top quark masses of 140, 175, and 200 $\gevcc$.
It is observed that for a top quark mass of 140 $\gevcc$, the
signal and background peak at nearly the same value of reconstructed
mass. However, the $\ttbar$ events are more sharply peaked than
background, and therefore there is still shape discrimination between
the two. As the top quark mass increases, the reconstructed-mass 
distribution for $\ttbar$ events is clearly separated from the
background. Since we include a background constraint in the
top quark mass likelihood fit (see Section~\ref{s-like}), differences in
shape between signal and background events are not required.
However, the shape differences do improve the
resolution on the top quark mass measurement.

\subsection{Comparisons of {\small{VECBOS}} with data}
\label{vecbos-comp}
\begin{figure}[ht]
%\leavevmode
%\begin{center}
\epsfysize=6.5in
\hspace{0.5in}
%\epsffile{noniso.eps}
\epsffile[0 72 612 720]{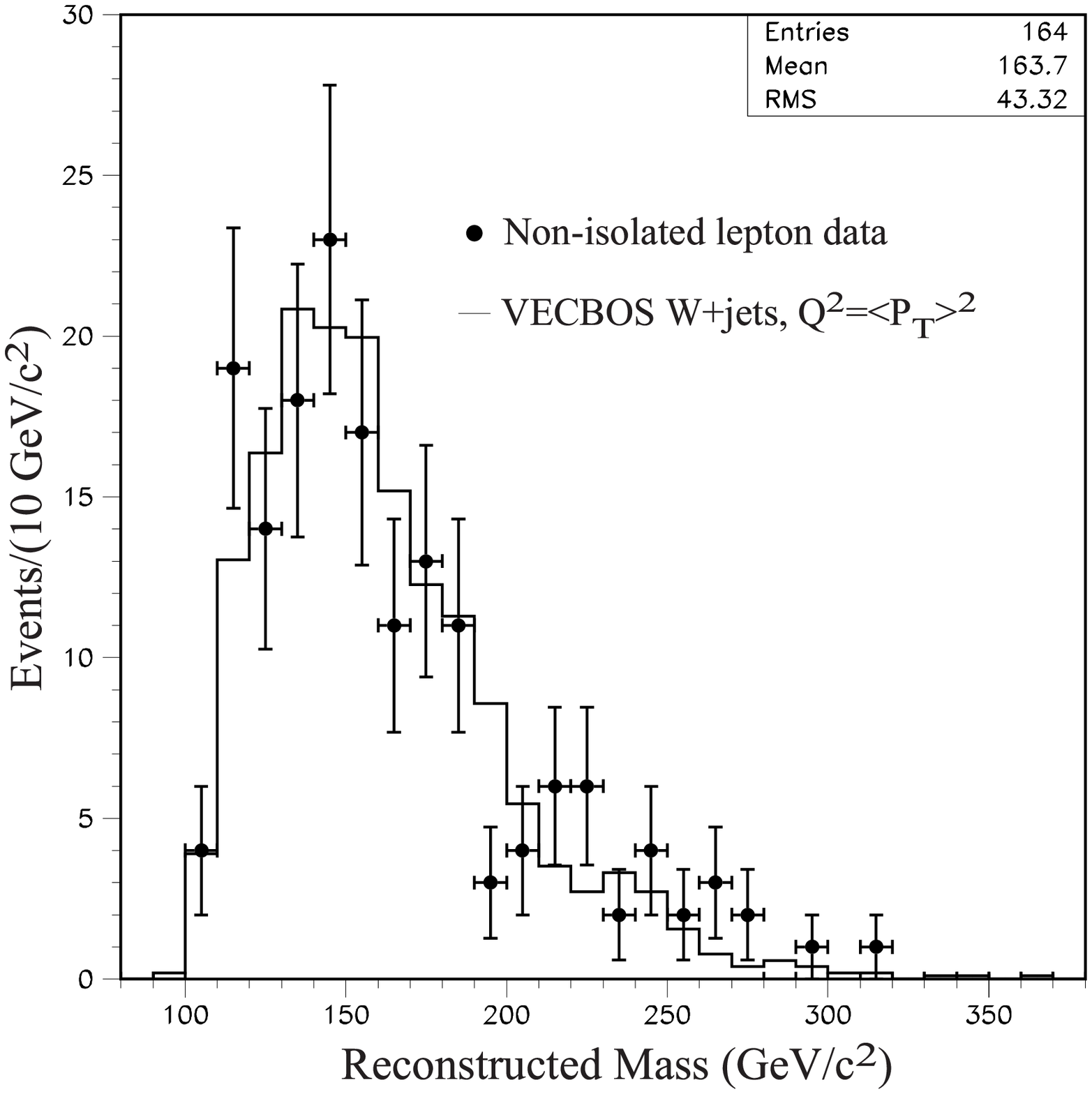}
%\epsffile[0 72 612 720]{$PRD_FIG/noni.eps}
%\end{center}
\caption{Reconstructed-mass distribution for events containing at
least four jets, \met, and a non-isolated lepton. The expected
fraction of $\ttbar$ in this sample is $\approx$9\%. The points
are data and the histogram is the {\small{VECBOS}} distribution.}
\label{noniso}
\end{figure}

\begin{figure}[ht]
%\leavevmode
%\begin{center}
\epsfysize=6.5in
\hspace{0.5in}
%\epsffile{loosewj.eps}
\epsffile[0 72 612 720]{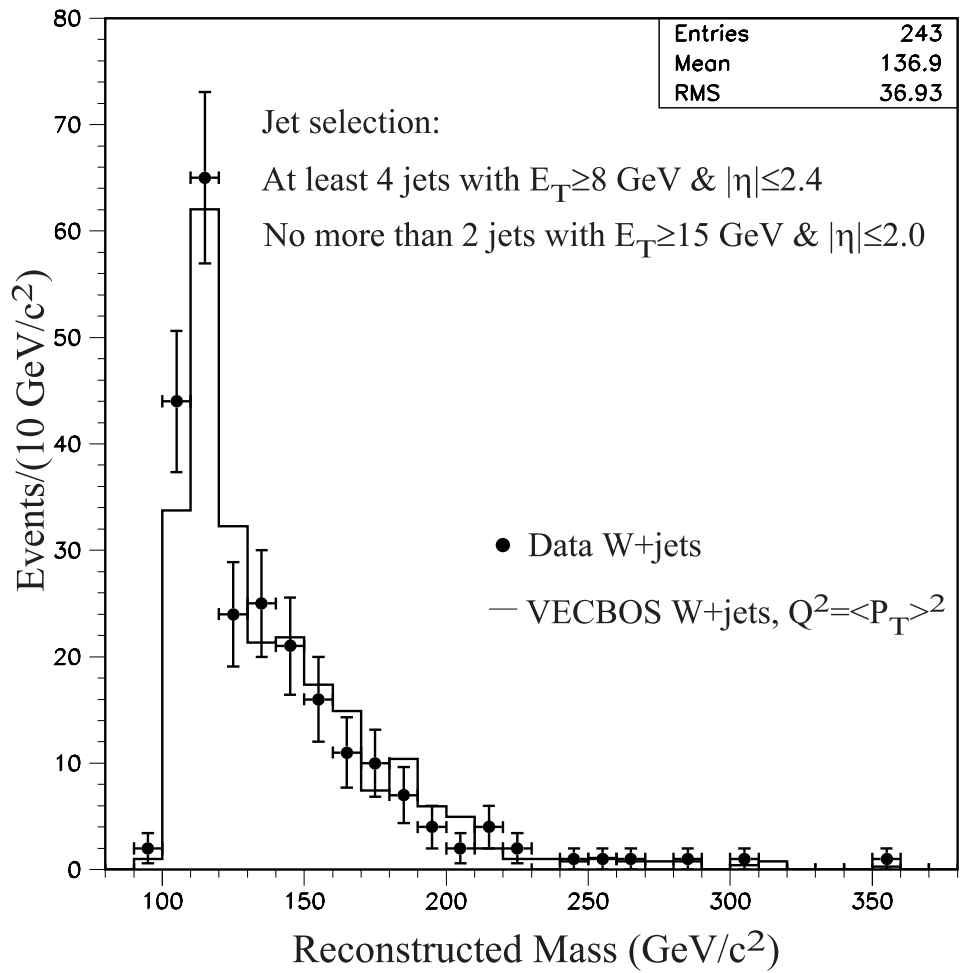}
%\epsffile[0 72 612 720]{$PRD_FIG/loose.eps}
%\end{center}
\caption{Reconstructed-mass distributions for events with an isolated
lepton, \met, and no more than two jets with $\Et>15 \ \gev$. The expected
fraction of $\ttbar$ in this sample is $\approx$0.7\%. The points
are data and the histogram is the distribution from {\small{VECBOS}}.} 
\label{loosewj}
\end{figure}

\begin{figure}[ht]
%\leavevmode
%\begin{center}
\hspace{0.5in}
\epsfysize=6.5in
%\epsffile{plugw.eps}
\epsffile[0 72 612 720]{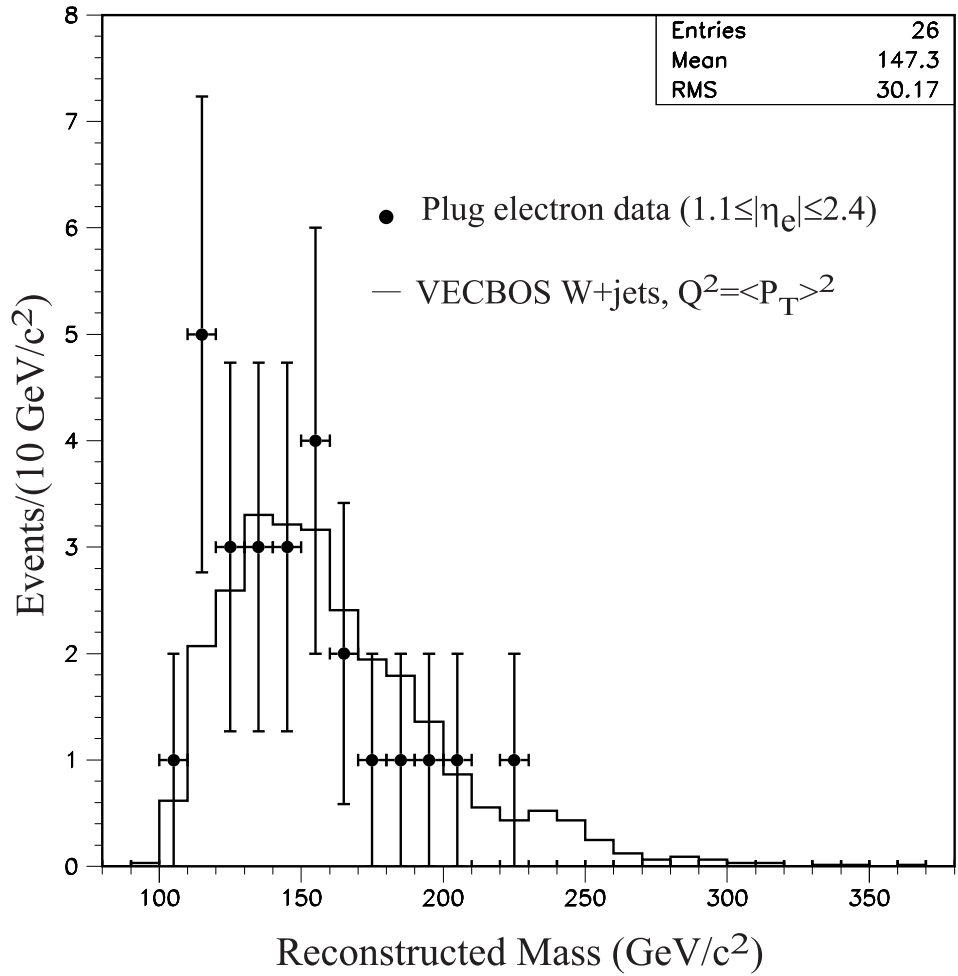}
%\epsffile[0 72 612 720]{$PRD_FIG/plug.eps}
%\end{center}
\caption{Reconstructed-mass distribution for events containing at
least four jets, \met, and a non-central electron (reconstructed
in the region $1.1<|\eta |<2.4$). The expected
fraction of $\ttbar$ in this sample is $\approx$0.2\%. The points
are data and the histogram is the {\small{VECBOS}} distribution.}
\label{plugw}
\end{figure}

The background modeling is checked by
comparing the reconstructed $W$+4 jet mass distributions from some data samples
with the appropriate distributions from the {\small{VECBOS}} simulation. 
The data samples consist of events that fail only one of the top sample
criteria. The samples compared are:
%%
%
%  The reliability of the {\small{VECBOS}} simulation 
%to model the $\ttbar$ background is demonstrated by
%comparing the reconstructed mass distributions from data samples 
%expected to be depleted in $\ttbar$ with corresponding distributions from 
%the {\small{VECBOS}} program. The samples compared are:
\begin{itemize}
\item events failing only the lepton isolation criteria;
\item events having fewer than three jets with $\Et>15~\gev$ and $|\eta |<2$
      (see Section~\ref{s-data});
\item events with a non-central ($1.1<|\eta |<2.4$) primary electron.
%\item No Tag events from the Top Mass Candidate Sample where one of the
%       leading four jets has $\Et<15\ \gev$ or $|\eta |>2$.
\end{itemize}
Each of these samples fails one and only one of the top sample 
criteria. Figure~\ref{noniso} shows the reconstructed-mass spectrum
for candidate $W$+4 jet events in which the primary electron 
is not isolated from jet activity in the event. 
The requirement that the lepton is non-isolated makes it more likely
that the selected data events are from multi-jet or $\bbbar$ production.
The data are compared to the distribution from {\small{VECBOS}}
events which also failed the lepton isolation criteria.
The similarity of the two distributions shows that the non-W/Z
component of the background is well modeled by the
{\small{VECBOS}} simulation. The fraction
of $\ttbar$ events in the data sample is expected to be $\approx$9\%.
A Kolmogorov-Smirnov test applied to these two 
distributions yields a 36\% confidence level for agreement. 
Figure~\ref{loosewj} shows a similar comparison in which 
the events are required to have no more than two jets with $\Et$ greater 
than 15 $\gev$. This sample has an estimated $\ttbar$ contribution
of about 0.7\%. A Kolmogorov-Smirnov test applied to these two 
distributions yields a 45\% confidence level for agreement. 
Figure~\ref{plugw} compares samples of events in which the primary 
electron was reconstructed in the PEM ($1.1<|\eta |<2.4$). 
We expect little or no dependence of the reconstructed mass on the 
$\eta$ value of the primary electron, as evidenced by the similarity
between this {\small{VECBOS}} distribution and the one in Fig.~\ref{vecstd}.
This sample is estimated to have a $\ttbar$ fraction of 0.2\%. A 
Kolmogorov-Smirnov 
test applied to these two distributions yields a 33\% confidence level 
for agreement.
%
%Lastly, we compare in Fig.~\ref{ev75} 
%the mass distribution of the 75 event sample from the original 
%151 event Top Candidate Sample which are not used in the 
%Optimized 
%mass analysis with expectation from $\ttbar$ and background. These events 
%either lack a $b$-tagged jet among the
%leading four jets, or have the fourth highest $\Et$ jet with 
%$\Et<15\ \gev$ or $|\eta |>2$. This sample is estimated to consist of 
%90$\pm$5\% background events. The modeled shape is constructed using an
%admixture of signal ({\small{HERWIG}} \mtop=175 $\gevcc$) and 
%background (\it{VECBOS}) events in a one-to-nine ratio.
%
 
   We expect the events in these three data samples to be
predominantly from the same sources as described in Section~\ref{s-bgcalc},
but in different proportions. In all three cases the {\small{VECBOS}}
simulation agrees with the reconstructed-mass distribution in the data.
Therefore we assume that the {\small{VECBOS}} simulation models satisfactorily
the reconstructed-mass distribution of the background events in the 
mass subsamples.

%\section{Alternate techniques}
%
%  Other analyses which have different selection criteria and/or modified 
%formulations of the $\chi^2$ have been performed. The analyses are
%aimed at improving the probability for choosing the correct combination.
%The first of these analyses, the $\mLik^{\star\star}$ analysis, uses two 
%additional terms in the $\chi^2$ to aid in choosing the correct combination.
%The second analysis uses a looser definition for $b$-jet tagging to
%increase the number of double $b$-tagged events. Values of the top
%quark mass from these two analyses are
%consistent with the results presented in this report
%and are summarized in Appendix~\ref{s-other}. 

%\begin{figure}[ht]
%\leavevmode
%\begin{center}
%\hspace{0.75in}
%\epsfysize=6.0in
%\epsffile{ev75.ps}
%\end{center}
%\caption{Reconstructed mass distribution for the 75 events which pass the
%Top Mass Candidate selection criteria, but are rejected by the
%Optimized 
%mass analysis. Such events are characterized by 
%having the fourth largest $\Et$ jet with $\Et<15\ \gev$ or $|\eta |>2$, or have
%no $b$-tagged jets among the leading four jets.
%The points show the data distribution, and the shaded regions
%show the model predictions for 90\% background ({\small{VECBOS}}), 
%and an admixture of 90\% background and 10\% $\ttbar$ ({\small{HERWIG}}
%\mtop=175 $\gevcc$). A Kolmogorov-Smirnov test applied to these two 
%distributions yields a 47\% confidence level for agreement.}
%\label{ev75}
%\end{figure}

%% file: likelihood.tex
\chapter{DESCRIPTION OF THE LIKELIHOOD PROCEDURE}
\label{s-like}

A likelihood procedure is used to extract a measurement of the top quark
mass from the reconstructed-mass distributions of the data samples and the 
$\ttbar$ signal and background models, along with the constraint on the 
background fractions.  This section describes the likelihood and discusses
its validation with the help of simulated experiments.

%
%For each
%mass subsample, the reconstructed-mass distributions from $\ttbar$ 
%events are modeled over a wide range of input top quark masses,
%ranging from 120 to 220 GeV/$c^2$.
%The distribution of reconstructed masses 
%for background events are described by a single distribution for 
%each mass subsample. 
%
%The fraction of background in each mass subsample
%(see Section~\ref{s-bgopt}) is constrained to the expected number 
%within the estimated uncertainty.

\section{Parametrization of the reconstructed-mass distributions}
\label{s-par}
We use the {\small{HERWIG}} Monte Carlo plus detector simulation to 
model the shape of the reconstructed-mass distribution in $\ttbar$ events. 
Event samples are generated at several different values of 
the top quark mass ranging from 120 to 220 $\gevcc$. The 
{\small{VECBOS}} Monte Carlo program is used to model
the shape of the background distribution. The $\ttbar$ and
background samples are processed using the CDF simulation,
and the same analysis is applied to them as to the data sample.
Histograms of reconstructed masses from
these samples are referred to as templates. 

Since we generated templates for only a finite number of input
top quark masses, extraction of a measured top quark mass from the
data will require an interpolation. This can be achieved in one of 
two ways. The first consists of fitting the data separately at each input top
quark mass value to a combination of signal (at the given mass)
and background. The maximum likelihood is then interpolated from the
resulting likelihood values. The second method requires that the templates 
themselves be interpolated as a function of input top quark mass. 
The signal templates then become a function of both the input top
quark mass and the reconstructed mass. The background templates are
only functions of the reconstructed mass. The
likelihood can then be defined as a smooth function of both input
mass and reconstructed mass, and no further interpolation
is needed. Previous publications ~\cite{cdf-evidence,cdf-discovery},
with lower statistics, have used the first method of interpolation.
However, the second method employs a single
%results in lower systematic uncertainties from both the 
interpolation process and uses optimally the finite Monte Carlo statistics in 
the templates. We have adopted the latter method for this analysis.

\subsection{Signal parametrization}
\begin{figure}[ht]
%\leavevmode
%\begin{center}
\hspace{0.5in}
\epsfysize=6.5in
%\epsffile{herwig_templ_6_svx_sing.ps}
\epsffile{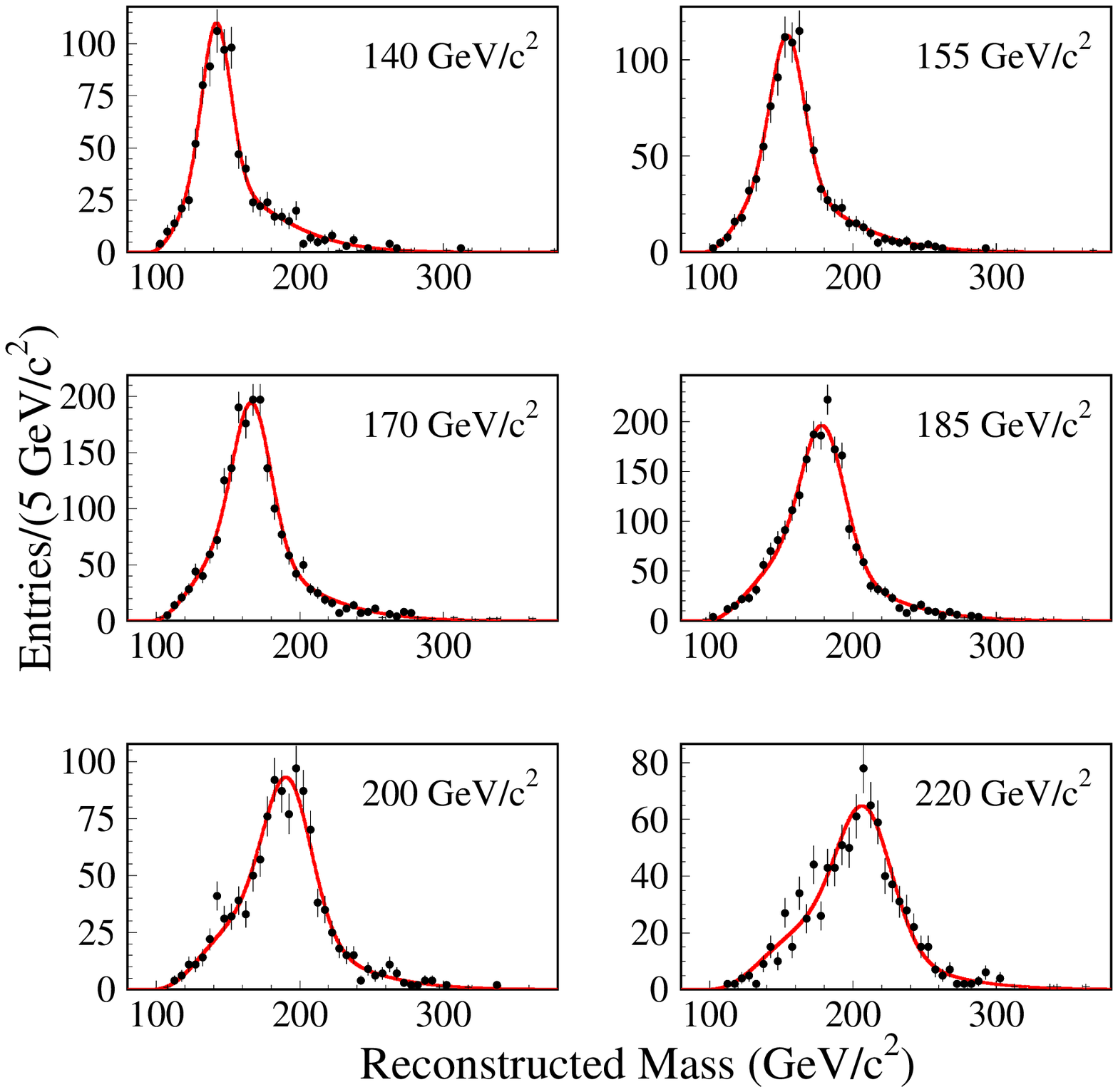}
%\end{center}
\caption{Reconstructed-mass distributions from simulated $\ttbar$
events for several input values for the top quark mass used in the
simulation.  The overlayed curves are predictions from the
parametrization of templates at 18 different top mass values. 
The distributions shown are for
the subsample corresponding to events with exactly one SVX-tagged
jet.}
\label{herwig_templ_svx_sing}
\end{figure}

\begin{figure}[ht]
%\leavevmode
%\begin{center}
\hspace{0.5in}
\epsfysize=6.5in
\epsffile{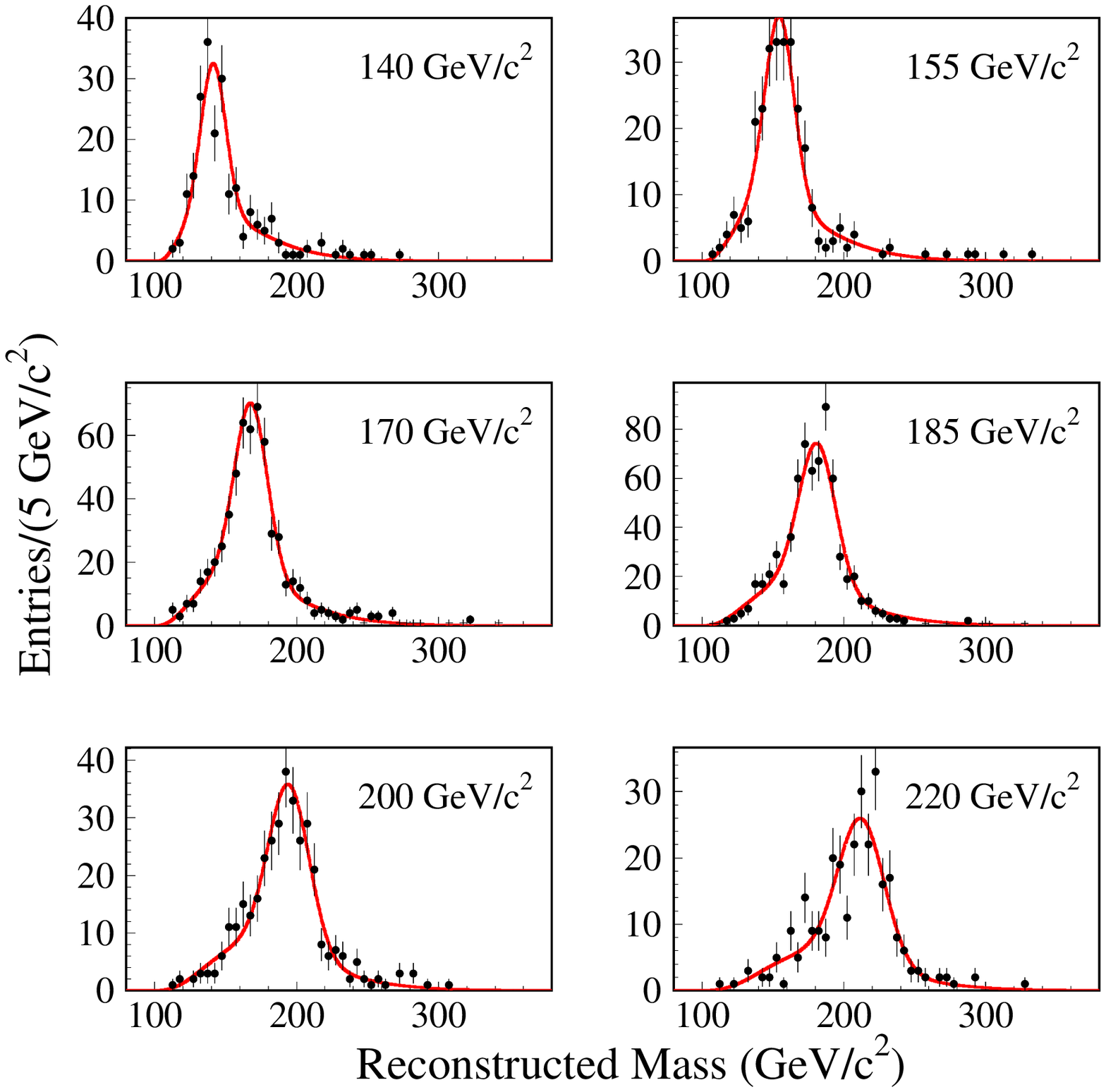}
%\epsffile{$PRD_FIG/herwig_templ_6_svx_doub.ps}
%\end{center}
\caption{Reconstructed-mass distributions from simulated $\ttbar$
events for several input values for the top quark mass used in the
simulation. The overlayed curves are predictions from the
parametrization of templates at 18 different top mass values.  
The distributions shown are for
the subsample corresponding to events with exactly two SVX-tagged
jets.}
\label{herwig_templ_svx_doub}
\end{figure}

\begin{figure}[ht]
%\leavevmode
%\begin{center}
\hspace{0.5in}
\epsfysize=6.5in
\epsffile{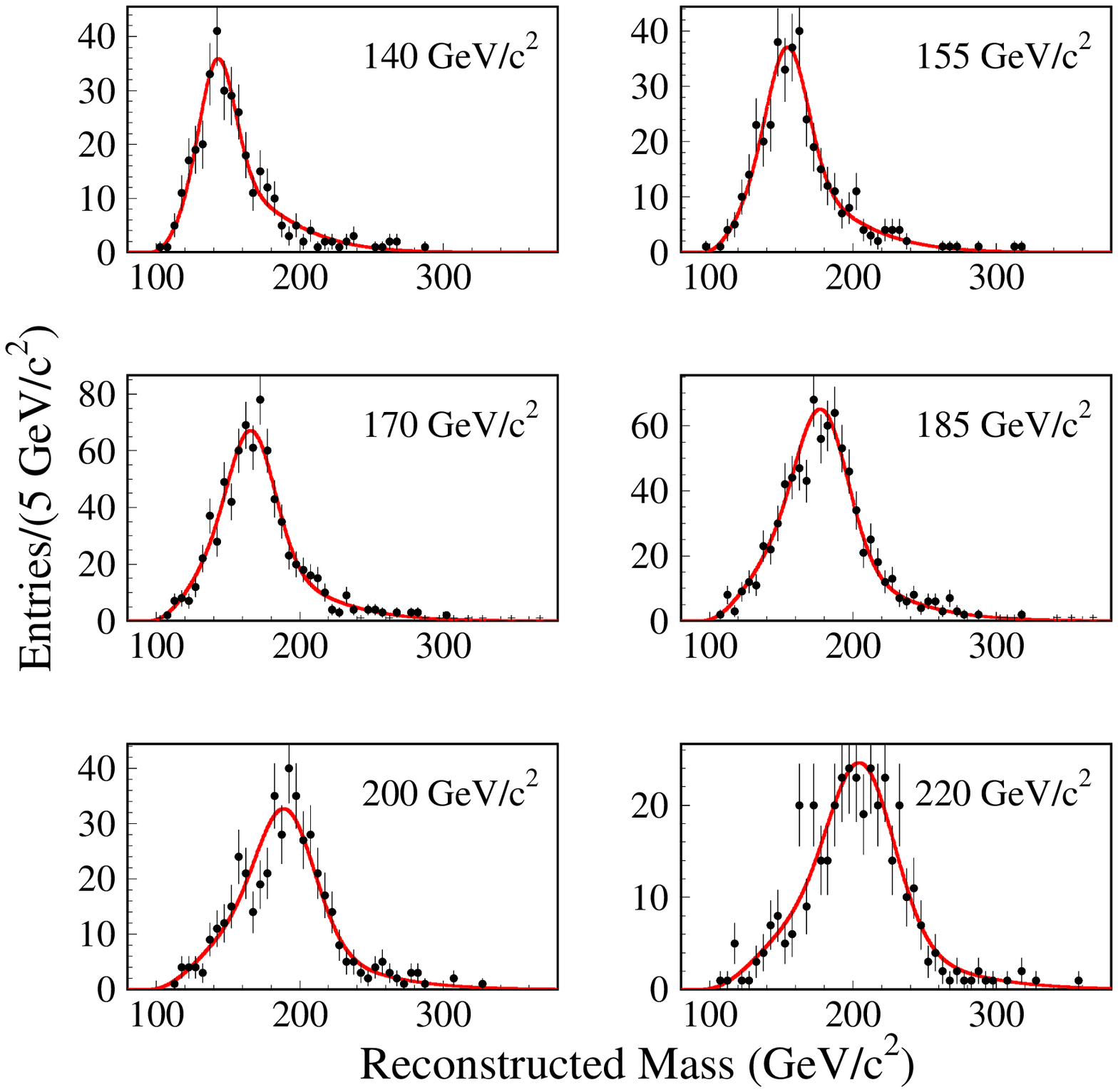}
%\epsffile{$PRD_FIG/herwig_templ_6_slt.ps}
%\end{center}
\caption{Reconstructed-mass distributions from simulated $\ttbar$
events for several input values for the top quark mass used in the
simulation. The overlayed curves are predictions from the
parametrization of templates at 18 different top mass values.   
The distributions shown are for
the subsample corresponding to events with one or more SLT-tagged
jets and no SVX-tagged jets.}
\label{herwig_templ_slt}
\end{figure}

\begin{figure}[ht]
%\leavevmode
%\begin{center}
\hspace{0.5in}
\epsfysize=6.5in
\epsffile{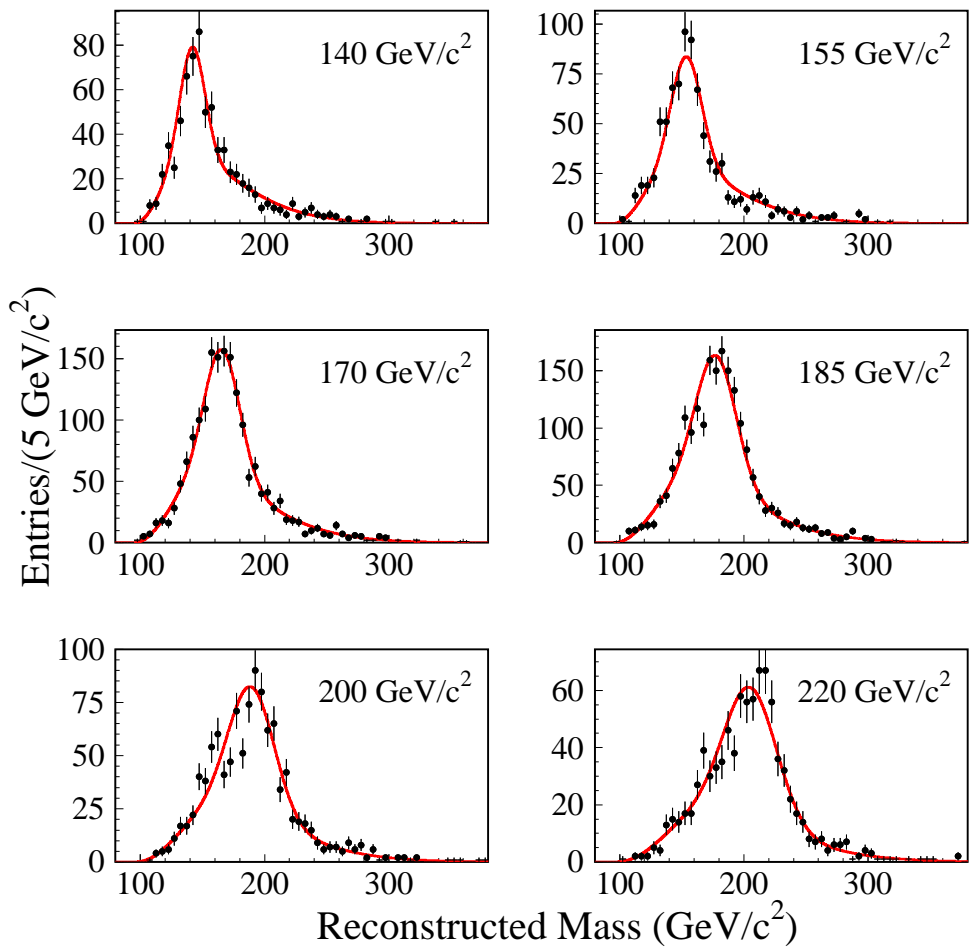}
%\epsffile{$PRD_FIG/herwig_templ_6_not15.ps}
%\end{center}
\caption{Reconstructed-mass distributions from simulated $\ttbar$
events for several input values for the top quark mass used in the
simulation. The overlayed curves are predictions from the
parametrization of templates at 18 different top mass values.   
The distributions shown are for
the subsample corresponding to events with no SVX-tagged or SLT-tagged
jets, and the fourth jet having $E_T>15$ GeV and $|\eta |<2$.}
\label{herwig_templ_not15}
\end{figure}

A single function, $f_s$, is used to model the distribution of reconstructed 
top masses for $\ttbar$ events for any given value of the input top quark mass
between 120 and 220 $\gevcc$:
\begin{equation}
f_s(M_{t},P_k) = N\,\left[P_6\, f_1 (M_{t},P_{1,2,3})\,+\,
                          (1-P_6)\,f_2(M_{t},P_{4,5})\right],
\label{eq:signal_shape}
\end{equation}
\noindent where:
\begin{eqnarray}
f_1(M_{t},P_{1,2,3}) & = & \frac{P_3^{1+P_2}}{\Gamma(1+P_2)}\,
                           (M_{t}-P_1)^{P_2}\, e^{-P_3 (M_{t}-P_1)}, \\
f_2(M_{t},P_{4,5})   & = & \frac{1}{\sqrt{2\pi} P_5}\, e^{-\frac{1}{2}
                           \left(\frac{M_{t}-P_4}{P_5}\right)^2}, \\
P_i                  & = & \alpha_i + \alpha_{i+6}\,M_{top}.
\label{eq:signal_shape_in}
\end{eqnarray}
For each mass subsample, six pairs of parameters $(\alpha_{i},\alpha_{i+6})$
are needed to describe how the distribution of reconstructed mass ($M_{t}$) 
evolves with the input top quark mass (\mtop).  For example, $\alpha_{4}$ and 
$\alpha_{10}$ ($\alpha_{5}$ and $\alpha_{11}$) describe how the mean (width) 
of the Gaussian portion of the reconstructed-mass distribution changes with 
\mtop.  The parameter values and their covariance matrix are obtained by a 
chisquare fit to the templates~\cite{bettelli}.  
Six of the 18 templates for the SVX Single sample are shown in 
Fig~\ref{herwig_templ_svx_sing} together with the predictions obtained
from the fit parameter values. 
Figs~\ref{herwig_templ_svx_doub}--\ref{herwig_templ_not15} show the
same six templates 
for the SVX Double, SLT, and No Tag subsamples respectively. 
The fit chisquares per degree of freedom ($dof$) are:
1.17 for 555 $dof$, 1.07 for 335 $dof$, 0.96 for 454 $dof$, and 1.36 for 589 
$dof$, for these four subsamples, respectively. 

%\begin{figure}[ht]
%\leavevmode
%\begin{center}
%\hspace{0.75in}
%\epsfysize=6.0in
%\epsffile{norm_resid.ps}
%\end{center}
%\caption{Normalized residuals of the bin contents to the parameterization
%for the SVX Single subsample. The normalized residuals are the difference
%between the discrete bin contents and the fitted curve divided by the
%statistical uncertainty. The four figures correspond to the the templates
%using input top quark mass values of 130, 150, 175, and 200 $\gevcc$.}
%\label{norm_resid}
%\end{figure}

%\begin{figure}[ht]
%\leavevmode
%\begin{center}
%\hspace{0.75in}
%\epsfysize=6.0in
%\epsffile{mean_sigma_pullssvx.ps}
%\end{center}
%\caption{The mean value (top) and width (bottom) of the distributions
%of normalized residuals
%between the parameterized and binned mass distributions for
%simulated $\ttbar$ events having exactly 1 SVX tag.}
%\label{f-intp2}
%\end{figure}

%\begin{figure}[ht]
%\leavevmode
%\begin{center}
%\hspace{0.75in}
%\epsfysize=6.0in
%\epsffile{sigma_pullssvx1.ps}
%\end{center}
%\caption{
%Standard deviations ($\sigma$) of the pull distribution, defined in the text,
%between the parameterized and observed reconstructed-mass distributions
%for simulated $\ttbar$ events having exactly 1 SVX tag.
%}
%\label{f-intp3}
%\end{figure}

\subsection{Background parametrization}
\label{s-backshape}
\begin{figure}[ht]
%\leavevmode
%\begin{center}
\hspace{0.5in}
\epsfysize=6.5in
\epsffile{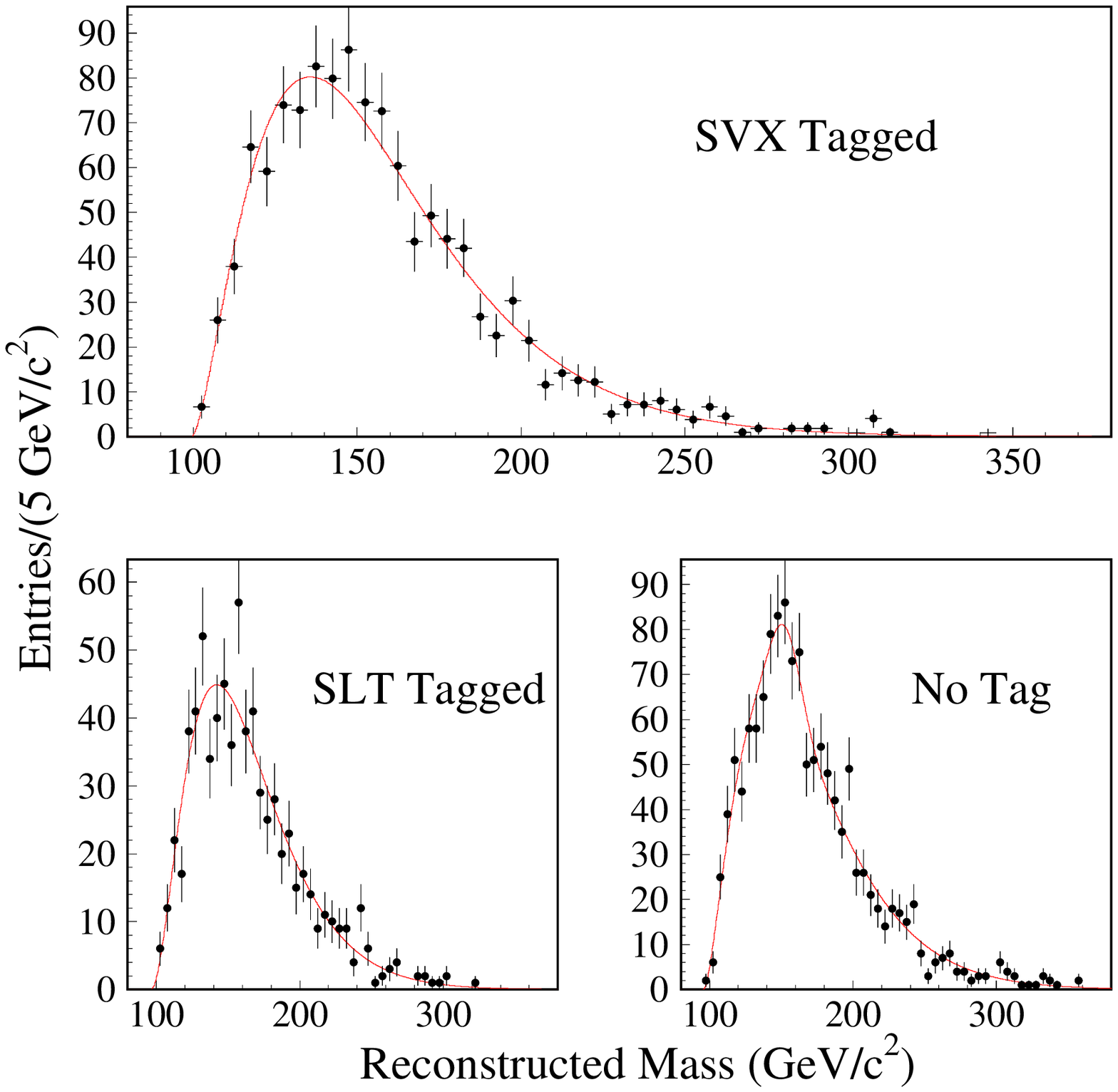}
%\epsffile{$PRD_FIG/vecbos_templ.ps}
%\end{center}
\caption{Reconstructed-mass distribution for $W$+jets events 
generated with the {\small{VECBOS}} Monte Carlo simulation. The
smooth curves are the parametrizations of the reconstructed-mass
distributions. The distributions are for SVX tagged,
SLT tagged, and No Tag events used in the mass analysis.}
\label{vecbos_templ}
\end{figure}

The fitting of the distribution of reconstructed masses from {\small{VECBOS}}
is performed in a similar fashion to the signal templates, but with fewer 
parameters and no dependence on \mtop. For the tagged subsamples,
the background distribution shape can be described by $f_1$,
whereas the No Tag subsample requires the 
additional freedom of $f_2$ to adequately describe its shape. 
Figure~\ref{vecbos_templ} shows the parametrizations of the background 
distributions for the SVX tagged, SLT (no SVX) tagged, and No Tag
events. Because
of limited statistics and low probability for obtaining
two SVX tagged jets in the {\small{VECBOS}} Monte Carlo simulation,
we assume the same background shape for SVX Double and SVX Single 
tag events. The mass measurement is insensitive to this assumption 
because the
expected background fraction for double tag events is only 4\%.
In Section~\ref{vecbos-comp} we compared distributions from top-depleted 
data samples with analogous {\small{VECBOS}} distributions to show 
that the {\small{VECBOS}} Monte Carlo simulation models the shape of 
the $\ttbar$ backgrounds quite well.

\section{Definition of the mass likelihood}
\label{m-like}
The value of \mtop for each mass subsample is
obtained from a maximum likelihood procedure applied to the 
observed mass distribution.  
The procedure allows the template fit parameters and the background fraction
$x_b$ to vary about their central values within their respective 
uncertainties. 
%Because the template fit 
%parameters have relatively small uncertainties, their fitted values
%are nearly identical to the input values. 
%In addition the
%background constraint does not allow the fitted number of background to
%deviate far from the expected values. 
The only parameter which is entirely unconstrained in the mass likelihood fit 
is \mtop. The reconstructed masses for the events in each 
of the four mass subsamples are tabulated in Appendix A.
Since the subsamples are statistically independent, the probability of 
observing these four sets of masses can be expressed as a product of
four individual likelihood functions, one for each subsample.  These four
likelihoods have the same form:
\begin{equation}
{\cal L} = {\cal L}_{shape} \times {\cal L}_{background} \times  
           {\cal L}_{param},
\label{eq:likfun}
\end{equation}
\noindent where:
\begin{eqnarray}
{\cal L}_{shape}      & = & \prod_{i=1}^{N_{\rm events}} 
                            \left[(1-x_b) f_s (M_i, M_{top},\vec{\alpha}) + 
                            x_b f_b (M_i,\vec{\beta})\right], \\
{\cal{L}}_{background}& = & P(x_b), \\
{\cal{L}}_{param}     & = & \exp\left\{-\frac{1}{2}\left[
                            (\vec{\alpha}-\vec{\alpha_{0}})^{T}\, U^{-1}\,
                            (\vec{\alpha}-\vec{\alpha_{0}})\right.\right.
                            \nonumber\\
                      &   & \hspace*{3em}\mbox{}+\left.\left.
                            (\vec{\beta}-\vec{\beta_{0}})^{T}\, V^{-1}\,
                            (\vec{\beta}-\vec{\beta_{0}})\right]
                            \rule{0mm}{3.1ex}\right\}.
\label{eq:likfun2}
\end{eqnarray}
The likelihood ${\cal L}_{shape}$ is the joint probability density for
a sample of $N_{\rm events}$ reconstructed masses ${M_i}$ to come from a parent
distribution with background fraction $x_b$ and signal fraction $(1-x_{b})$. 
The background likelihood ${\cal L}_{background}$, discussed in 
Section~\ref{s-bgopt}, constrains the fraction of background events to the 
expected value within its uncertainties (see Fig.~\ref{backg}). The expected 
background fraction and number of background events are related via 
$N_b = x_b\times N_{obs}$, where $N_{obs}$ is the number of observed events 
for that subsample. 
The vectors $\vec\alpha$ and $\vec\beta$ determine the shapes of the
signal ($f_s$) and background ($f_b$) distributions. They are constrained by
${\cal{L}}_{param}$ to agree with the nominal values, $\vec\alpha_0$ and 
$\vec\beta_0$, via their covariance matrices $U$ and $V$ respectively.
The inclusion of ${\cal{L}}_{param}$ in the likelihood definition is due
to the finite statistics of the Monte Carlo samples used
to determine $f_s$ and $f_b$. Furthermore, by parametrizing
the signal probability $f_s$ as a continuous function of $M_{top}$, 
the likelihood is inherently a continuous function of $M_{top}$ as well. 

To extract the top quark mass for each subsample, we minimize $-\log{\cal L}$ 
with respect to $M_{top}$, $x_b$, $\vec\alpha$ and $\vec\beta$. 
The statistical uncertainty on $M_{top}$ is taken as the change in $M_{top}$ 
which results in a 0.5 unit increase in $-\log{\cal L}$ along the line
on which  $-\log{\cal L}$ is minimized with respect to variations in all
the other fit parameters.  
%When calculating this
%statistical uncertainty, $-\log{\cal L}$ is kept minimized with 
%respect to all the other fit parameters. 
The statistical uncertainty has contributions not only from the
finite statistics in the data sample, but also from 
the uncertainty in the expected background and the finite
statistics in the mass templates. However, the latter two contributions
account for less than 1\% of the total statistical uncertainty.
The top quark mass and its statistical uncertainty for the four subsamples
combined are extracted in the same way as above from the product of the
four subsample likelihoods.

\section{Tests of the likelihood procedure on simulated experiments}
\label{like-perf}

The performance of the likelihood scheme was tested using simulated events 
from Monte Carlo programs. We performed a large number of simulated 
experiments, each consisting of four subsamples with the same numbers of 
events ($N_{obs}^{i}, i=1,\ldots,4$) as observed in the four data subsamples.
Each experiment subsample contained $N_b^i$ background events and 
$N_s^i=(N^i_{obs}-N_b^i)$ $\ttbar$ events, where $N_b^i$ is a binomial
fluctuation of the expected background.  The $\sum_i N_s^i$
and $\sum_i N_b^i$ distinct mass values for each simulated experiment were 
chosen at random from the discrete templates for signal and 
background events (Figs~\ref{herwig_templ_svx_sing}--\ref{vecbos_templ}). 
The four sets of masses
%of 5 SVX Double Tag, 15 SVX Single Tag, 14 SLT (no SVX) Tag, 
%and 42 No Tag events 
were fit using the same likelihood procedure that was used to fit the 
data sample. Each simulated experiment yielded a fitted top quark mass,
a statistical uncertainty and a maximum likelihood value. 
The self-consistency of the likelihood procedure was tested by comparing
these returned values with expectations.

\begin{figure}[ht]
%\leavevmode
%\begin{center}
\hspace{0.5in}
\epsfysize=6.5in
\epsffile{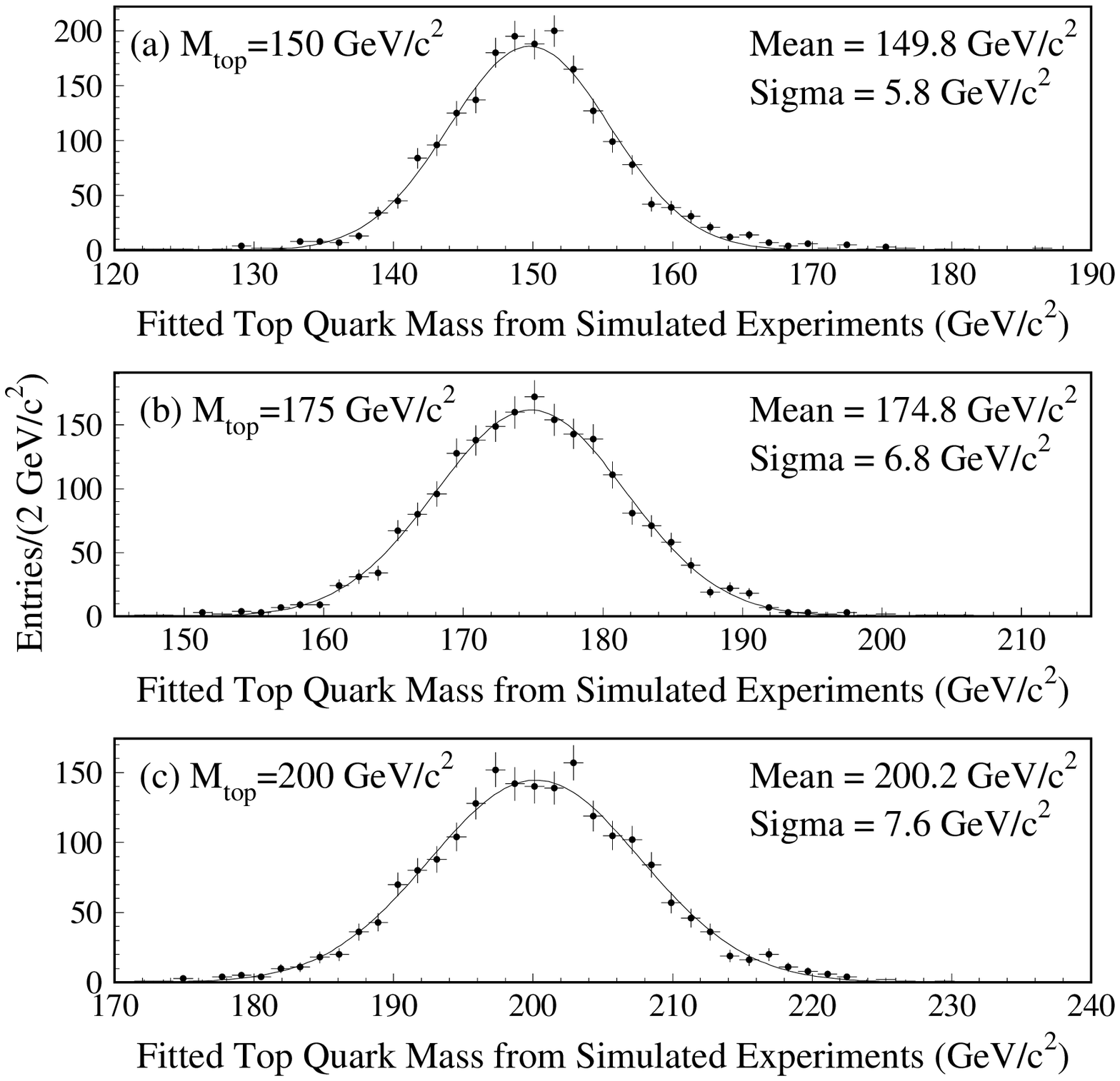}
%\epsffile{$PRD_FIG/pe_mass.ps}
%\end{center}
\caption{The value of the top quark mass returned by the fitter 
for input top quark masses of 150, 175, and 200 GeV/$c^2$. 
Each simulated experiment contains an admixture of signal and 
background events as described in the text.}
\label{pe_mass}
\end{figure}

%    Several checks were made on the consistency of the procedure
%using these simulated experiments. In simulated experiments we can
%choose the input value of the top quark mass, by simply 
%picking events at random from the appropriate mass template (see  
%Figure~\ref{herwig_templ_svx_sing}, for example). Several samples of simulated
%experiments were generated, each at a different input value for the
%top quark mass. 
%A prerequisite for the likelihood procedure is that the mean value 
%of the distribution of returned top masses should be equal (within 
%statistical uncertainty) to the input value used to generate the 
%simulated experiments. 
  Figure~\ref{pe_mass} shows the distribution of returned masses from the 
likelihood fit for input top quark masses of 150, 175, and 200 $\gevcc$.
The curves are fits to Gaussians, and have central values 
of 149.8, 174.8, and 200.2 $\gevcc$, and $\sigma$ of 5.8, 6.8, and 
7.6 $\gevcc$.
In each case the mean of the distribution is consistent with the input value,
which demonstrates that the procedure introduces little or no bias into
the top quark mass measurement.
%
%parameterization of the templates does
%not introduce any systematic shifts in the fitted top quark mass.
The $\sigma$ of the distributions reflects the expected statistical 
uncertainty on the top quark mass measurement for experiments which have 
the same expected background and $b$-tag composition as our 
Run 1 data sample. Based on the fitted $\sigma$'s one expects to achieve
a statistical uncertainty on \mtop of $\approx$4\%.

%Figure~\ref{output_input} 
%shows the median value of the distribution of returned top masses for 
%several values of the input top quark mass ranging from 120 to 220 GeV/c$^2$. 
%The likelihood procedure is observed to return the correct 
%value for a wide range of input top quark masses. 

\begin{figure}[ht]
%\leavevmode
%\begin{center}
\hspace{0.5in}
\epsfysize=6.5in
\epsffile{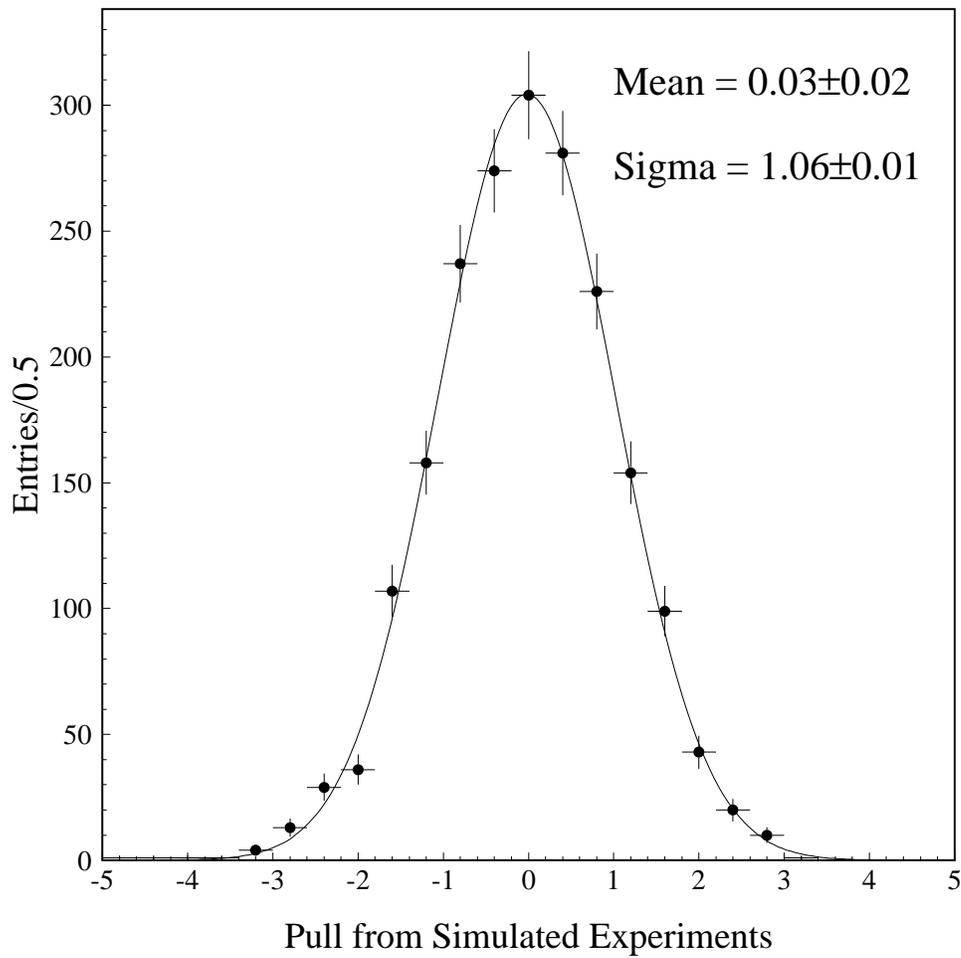}
%\epsffile{$PRD_FIG/pe_pull.ps}
%\end{center}
\caption{The distribution of pulls between the returned value of the
top quark mass and the input value of 175 GeV/$c^2$. Each simulated 
experiment contains an admixture of signal and background events as described 
in the text.}
\label{pe_pull}
\end{figure}

\begin{figure}[ht]
%\leavevmode
%\begin{center}
\hspace{0.5in}
\epsfysize=6.5in
\epsffile{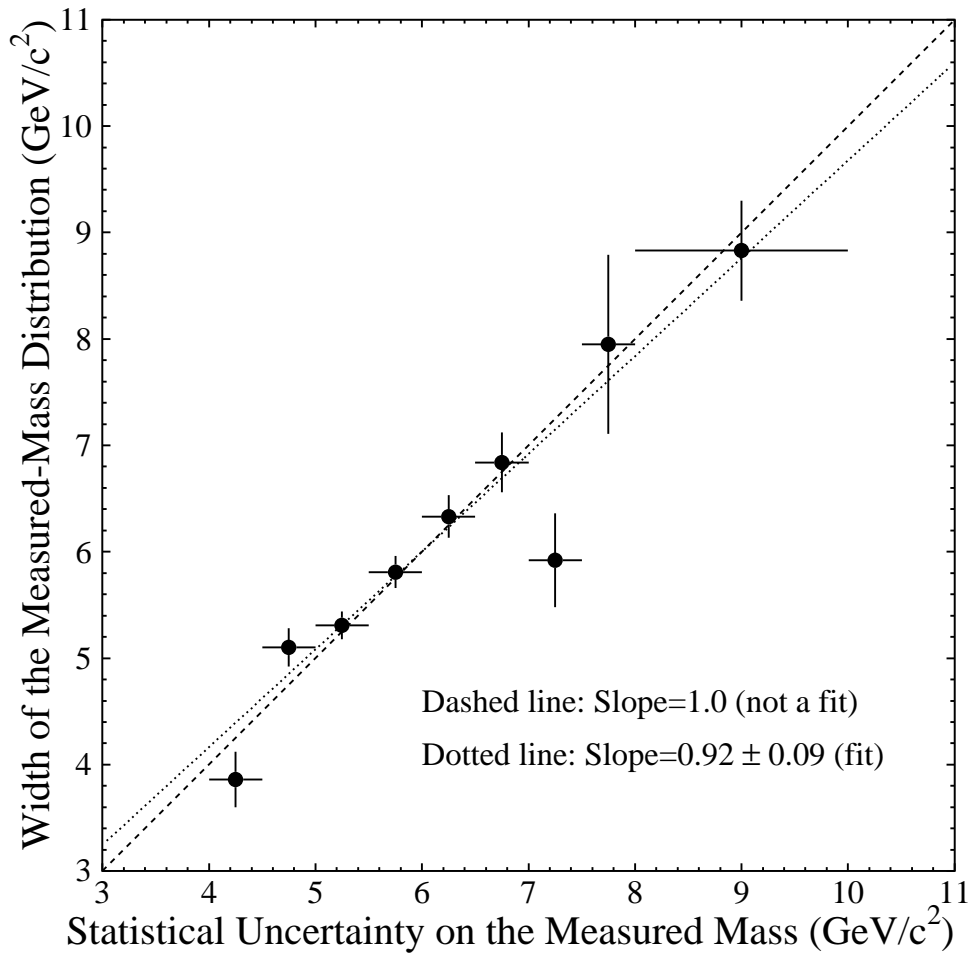}
%\epsffile{$PRD_FIG/masswidth.ps}
%\end{center}
\caption{The Gaussian width of the mass distribution as a function
of the statistical uncertainty returned by the likelihood. Each simulated 
experiment contains an admixture of signal and background events as described 
in the text. The dashed line, which has a slope of 1.0, is not a fit. The
dotted line is a fit to the points, and has a slope of 0.92$\pm$0.09.}
\label{masswidth}
\end{figure}

The statistical uncertainty returned by the likelihood procedure
should reflect the deviation of the returned top quark mass from the
input value. The pull, defined by 
\begin{equation}
 {\rm Pull}\; =\; \frac{M_{exp}-M_{input}}{\sigma^{stat}_{M}},
\label{eq:pull_def}
\end{equation}
\noindent is used to check the consistency between the measured deviation 
on the top quark mass and the estimated statistical uncertainty. In the
above expression, $M_{exp}$ is the fitted top mass value returned by the 
likelihood, $M_{input}$ is the input value used to generate the (simulated) 
experiment, and $\sigma^{stat}_{M}$ is the statistical uncertainty on $M_{exp}$
returned by the fitter. Figure~\ref{pe_pull} shows the pull distribution
for the simulated experiments generated for $M_{top}=175 \ \gevcc$. 
The width is close to unity, which indicates that the statistical uncertainty
returned by the fitter accurately reflects the deviation of the fitted value
from the input value. Alternately, in Figure~\ref{masswidth}, we take slices 
in $\sigma^{stat}_{M}$, and evaluate the width of the corresponding
($M_{exp}-M_{input}$) distribution. The points have a slope of 0.92$\pm$0.09,
which supports using the statistical uncertainty returned by the fitter
as a measure of the statistical uncertainty for a given experiment. 

%\begin{figure}[ht]
%\leavevmode
%\begin{center}
%\hspace{0.75in}
%\epsfysize=6.0in
%\epsffile{output_input.ps}
%\end{center}
%\caption{The median value of the distribution of top quark masses 
%returned by the fitter as a function of the input mass value. Each 
%simulated experiment contains an admixture of signal and background 
%events as described in the text.}
%\label{output_input}
%\end{figure}

%% file: results.tex
\chapter{RESULTS}
\label{like-data}

Having tested the mass likelihood procedure on simulated experiments, 
we now apply it to the data sample. Two separate fits are performed. 
The first is the standard mass likelihood fit defined in Section~\ref{m-like}.
The second is the mass likelihood fit with the background fraction constraint
removed. In this case, the background fraction is determined only from the 
shapes of the reconstructed-mass distributions for signal and background.
After presenting these results, we check the consistency of the data with 
Monte Carlo expectation for $\ttbar$ plus background events in the
expected proportion.

\section{Fits to data}

\begin{figure}[ht]
%\leavevmode
%\begin{center}
\hspace{0.5in}
\epsfysize=6.5in
\epsffile{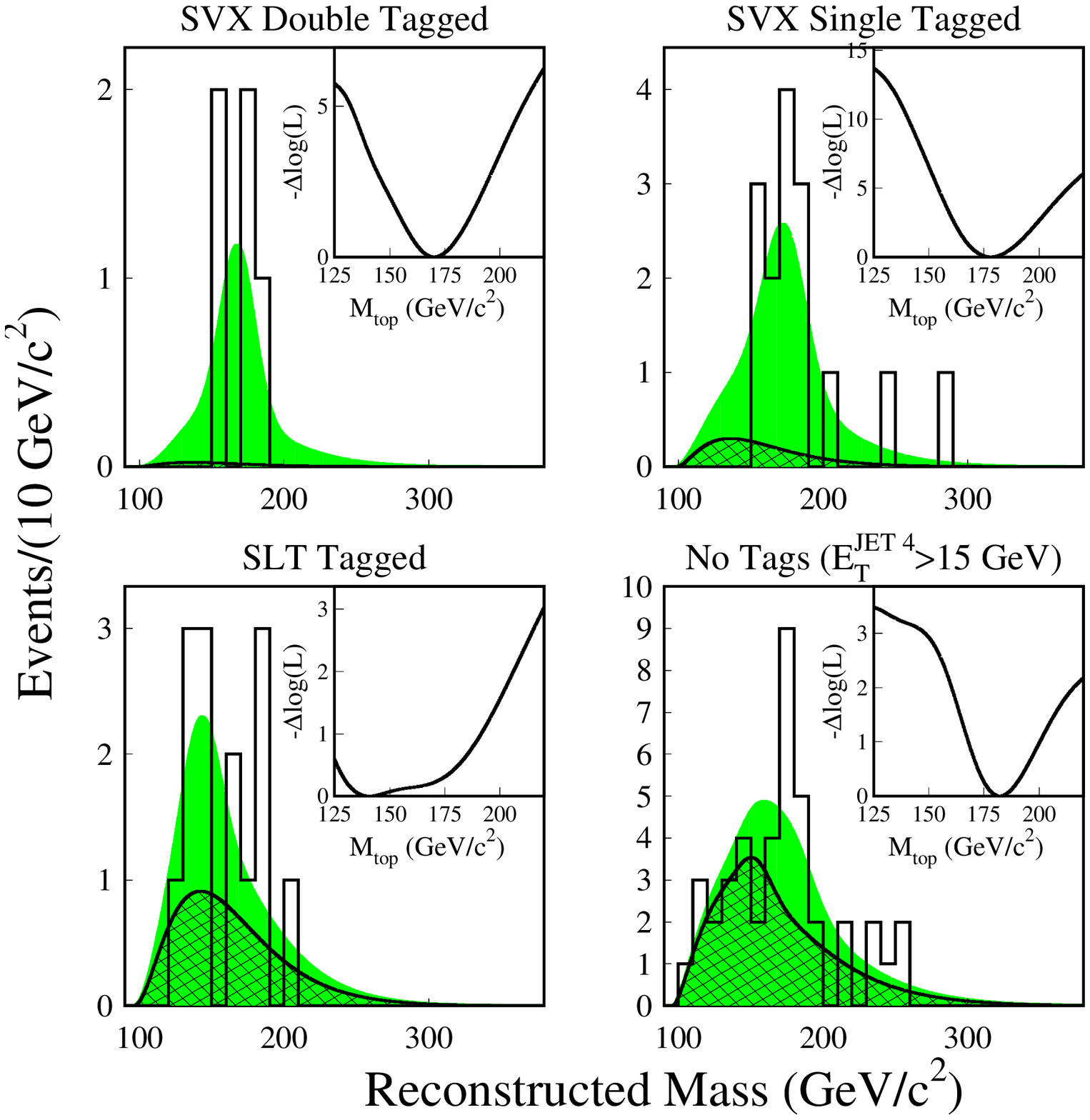}
%\epsffile{$PRD_FIG/opt4_err_mod.ps}
%\end{center}
\caption{Results of applying the likelihood procedure to the four subsamples.
The figure show the data (histogram), fitted background (shaded hatched region),
and fitted signal (shaded non-hatched region). The insets show the shapes of 
$-\log{\cal L}$ versus top mass, from which we extract the fitted top quark
mass and its statistical uncertainty.}
\label{opt4_err_mod}
\end{figure}

\begin{figure}[ht]
%\leavevmode
%\begin{center}
\hspace{0.5in}
\epsfysize=6.5in
\epsffile{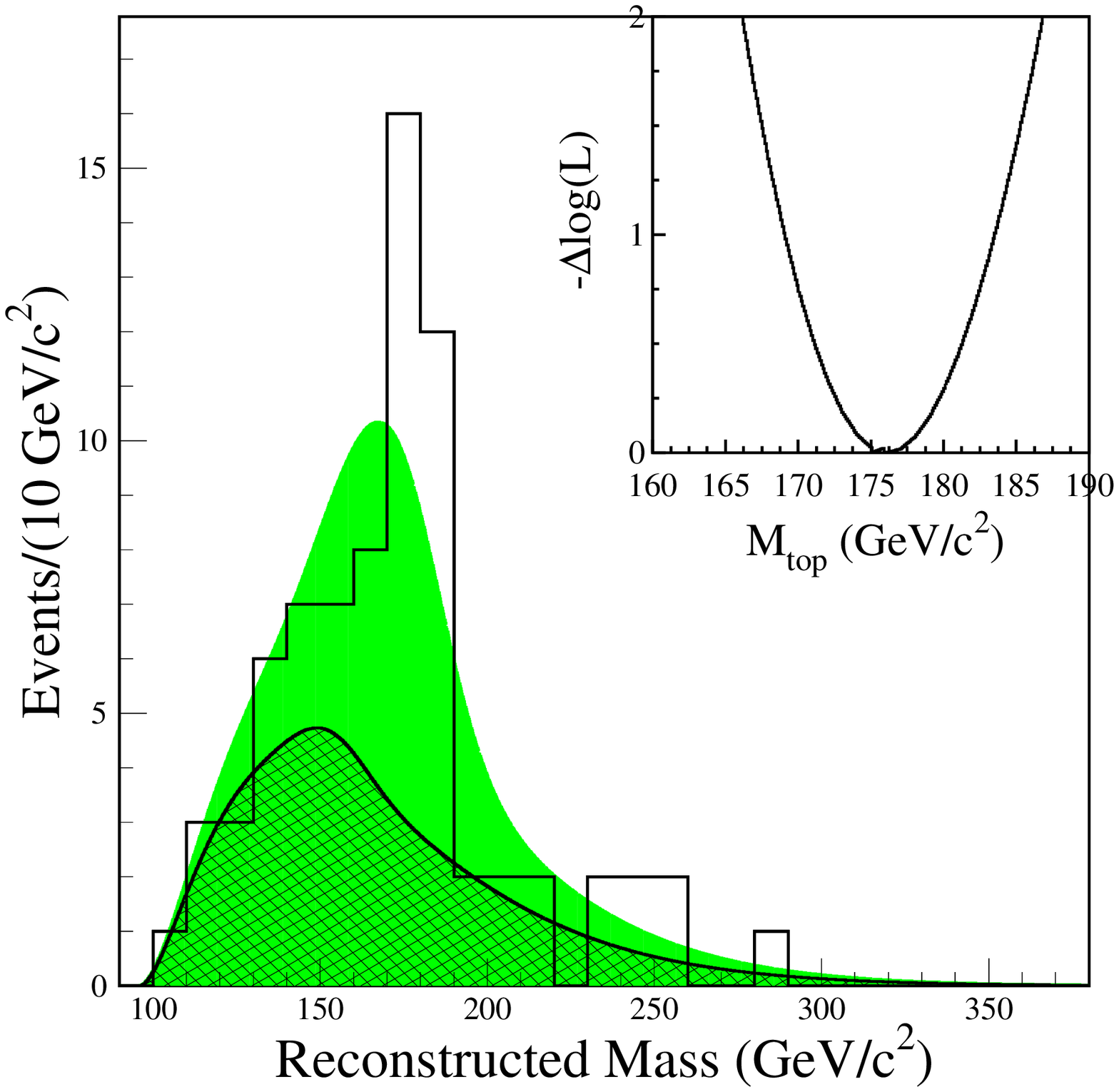}
%\epsffile{$PRD_FIG/opt1_err_mod.ps}
%\end{center}
\caption{Result of applying the likelihood procedure to the combined 
subsamples.  The figure shows the data (histogram), fitted background (shaded 
hatched region), and fitted signal (shaded non-hatched region). The inset 
shows the shape of $-\log{\cal L}$ versus top mass, from which we extract 
the best estimate of the top quark mass and its statistical uncertainty.}
\label{opt1_err_mod}
\end{figure}

The mass likelihood procedure is applied to the events in the four mass 
subsamples.  The measured values for the top quark mass for each subsample 
and the combined results are presented in Table~\ref{fit-results}. As 
discussed in Section~\ref{m-like}, the statistical uncertainties contain
contributions from both the statistics in the data, the uncertainties in the 
expected background, and the uncertainties in the template fit parameters. 
The latter two contribute less than 1\% to the total statistical uncertainty. 
Table~\ref{fit-results} also shows the fitted background fractions, which are 
constrained to the expected values via the background likelihoods in 
Fig.~\ref{backg}. The mass fits for the four mass subsamples are statistically
consistent with one another and are shown in Figure~\ref{opt4_err_mod}.
For each subsample, the background shape has been normalized to the fitted 
number of background events via $N_b^i = x_b^{fit}\times N_{obs}^i$, and the 
signal plus background has been normalized to the number of data events 
($N_{obs}^i$).  The combined fit to all four subsamples is shown in 
Figure~\ref{opt1_err_mod}. 

To investigate the impact of the background constraining term on the fitted 
top quark mass, we also performed mass likelihood fits with the
constraint on the fraction of background
removed. In this case, the shape of the mass distribution 
determines the background fraction. The results of the mass fits are presented
in Table~\ref{fit-results-uncon}. Several observations can be made from a 
comparison of these unconstrained mass fits with the constrained ones in 
Table~\ref{fit-results}.  First, the tagged subsamples fit to zero background,
although with large uncertainties, while the No Tag subsample yields a similar
background content whether the background is free to float or not.  Secondly, 
the masses show little sensitivity to removal of the background constraint.
In general, one would expect the removal of the constraint to result in an 
increased statistical uncertainty since information is being removed 
from the likelihood fit.  For all subsamples however, the uncertainty in the 
mass decreases when the background constraint is removed. This is because the 
fitted number of signal events becomes larger.  

Since the background rates in the four mass subsamples are correlated,
it is not correct to allow their background fractions to float relative
to one another. On the other hand, it is reasonable to
investigate whether the background constraint is affecting the
top quark mass measurement. The results in Tables~\ref{fit-results}
and ~\ref{fit-results-uncon} indicate very little sensitivity to the
background constraint.

%%%%%%%%%%%%%%%%%%%
% Summary Table of All Results
%%%%%%%%%%%%%%%%%%%

%\begin{table}[ht]
%\begin{center}
%\begin{tabular}{|c|c|c|c|}
%\hline
%Data Sample & Number of Events & $x_b^{fit}$ & Top Mass (GeV/c$^2$) \\ 
%\hline\hline
%SVX Double                &  5  & $0.05^{+0.04}_{-0.02}$ & $170.1^{+9.5}_{-9.0}$ \\ \hline
%SVX Single                & 15  &  $0.13\pm 0.04$ & $178.0^{+8.3}_{-7.5}$\\ \hline
%SLT                      & 14  &  $0.40\pm0.08$ & $142.1^{+33}_{-14}$ \\ \hline
%No Tags                   & 42  & $0.55\pm0.13$ & $180.8^{+9.7}_{-8.5}$ \\ 
%\hline\hline
%All subsamples  &  76  & 0.41$\pm$0.07 & $175.9\pm 4.8$ \\ \hline
%\end{tabular}
%\end{center}
%\caption{Results of applying the mass likelihood procedure to the
%four subsamples and for all subsamples combined. The background
%fractions are constrained to their expected values via the curves 
%in Fig.~\ref{backg}.
%For the four subsamples, we show the fitted background fraction and
%the fitted top mass. Also shown is the final mass value obtained when
%combining all four subsamples. The combined background fraction is
%the average of the $x_b$ fit results weighted by the number of events
%in the subsamples.}
%\label{fit-results}
%\end{table}

\begin{table}[ht]
\begin{center}
\begin{tabular}{cccc}
\hline\hline
\rule{0mm}{5mm}
Data sample & Number of events &   $x_b^{fit}$   & Top mass (GeV/$c^2$)     \\
\hline
\rule{0mm}{5mm}
SVX Double      &  5   &  $0.03\pm 0.02$         &  $170.0^{+9.4}_{-8.9}$   \\
\rule{0mm}{5mm}
SVX Single      & 15   &  $0.15^{+0.05}_{-0.04}$ &  $178.0^{+8.5}_{-7.6}$   \\
\rule{0mm}{5mm}
SLT             & 14   &  $0.53\pm0.09$          &  $140.6^{+40.5}_{-14.6}$ \\
\rule{0mm}{5mm}
No Tags         & 42   &  $0.69^{+0.09}_{-0.10}$ &  $182.1^{+11.7}_{-9.9}$  \\
\hline
\rule{0mm}{5mm}
All subsamples  &  76  &  $0.51\pm 0.06$         &  $176.1^{+5.2}_{-5.0}$   \\
\hline\hline
\end{tabular}
\end{center}
\caption{Results of applying the mass likelihood procedure to the
four subsamples and for all subsamples combined. The background
fractions are constrained to their expected values via the curves 
in Fig.~\protect\ref{backg}.
For the four subsamples, we show the fitted background fraction and
the fitted top mass. Also shown is the final mass value obtained when
combining all four subsamples. The combined background fraction is
the average of the $x_b$ fit results weighted by the number of events
in the subsamples.}
\label{fit-results}
\end{table}

\begin{table}[ht]
\begin{center}
\begin{tabular}{cccc}
\hline\hline
\rule{0mm}{5mm}
Data sample & Number of events & $x_b^{fit}$  & Top mass (GeV/$c^2$)     \\
\hline
\rule{0mm}{5mm}
SVX Double      &   5  &  $0.0^{+0.6}_{-0.0}$ &  $169.9^{+9.2}_{-8.7}$   \\
\rule{0mm}{5mm}
SVX Single      &  15  &  $0.0^{+.1}_{-0.0}$  &  $177.6^{+7.8}_{-7.1}$   \\
\rule{0mm}{5mm}
SLT             &  14  &  $0.0^{+0.8}_{-0.0}$ &  $146.2^{+26}_{-16}$     \\
\rule{0mm}{5mm}
No Tags         &  42  &  $0.53\pm0.22$       &  $180.8^{+10.1}_{-8.3}$  \\ 
\hline
\rule{0mm}{5mm}
All subsamples  &  76  &  $0.29\pm0.20$       &  $176.2\pm 4.8$          \\
\hline\hline
\end{tabular}
\end{center}
\caption{Results of applying the mass likelihood procedure to the
four subsamples and for all subsamples combined. The background
fractions are free parameters in the mass likelihood fit.
For the four subsamples, we show the fitted background fraction and
the fitted top mass. Also shown is the final mass value obtained when
combining all four subsamples. The combined background fraction is
the average of the $x_b$ fit results weighted by the number of events
in the subsamples.}
\label{fit-results-uncon}
\end{table}

\section{Comparison of data to expectations}

Up to this point we have assumed that our data sample is a
mixture of standard model $\ttbar$ signal plus background. Using
simulated experiments (with \mtop=175 $\gevcc$), we quantify the probability 
that our data sample is consistent with a mixture of $\ttbar$ plus background 
with the background fractions given in Table~\ref{fit-results}.

\begin{figure}[ht]
%\leavevmode
%\begin{center}
\hspace{0.5in}
\epsfysize=6.5in
\epsffile{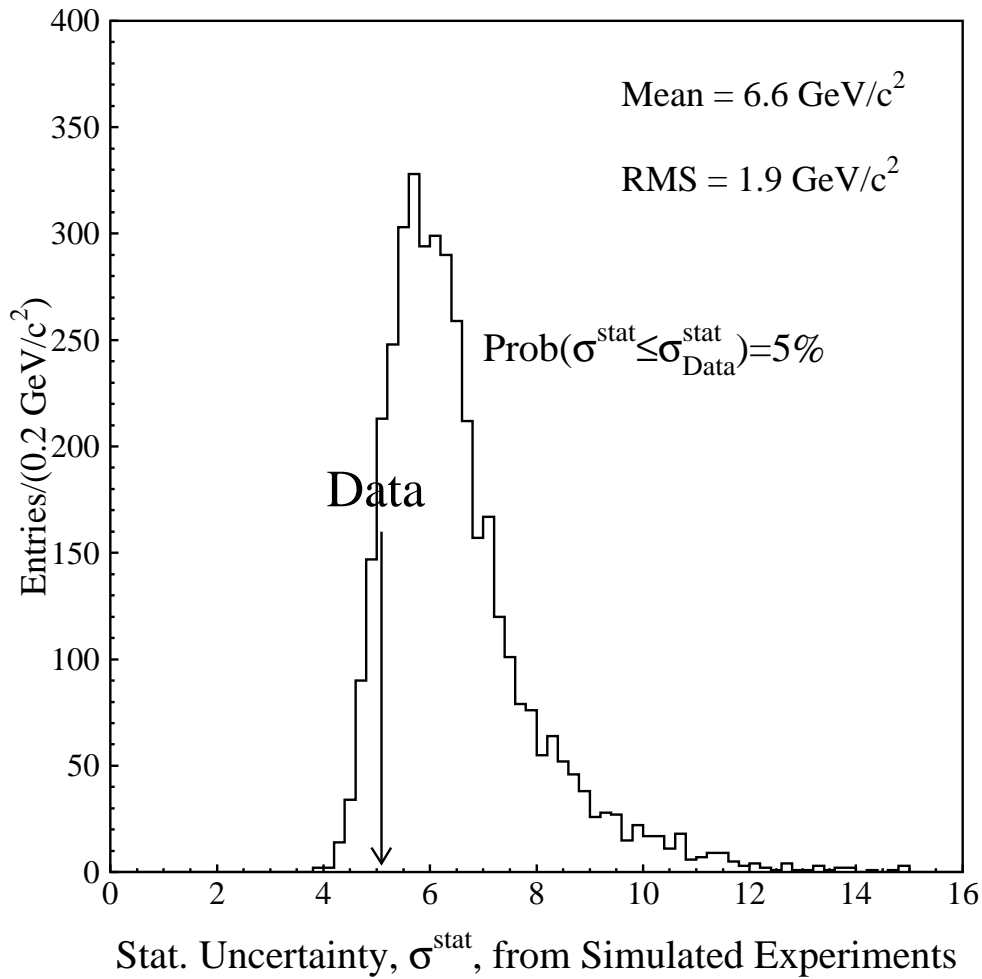}
%\epsffile{$PRD_FIG/pe_error.ps}
%\end{center}
\caption{Distribution of statistical uncertainties from
simulated experiments of $\ttbar$ (\mtop=175 $\gevcc$) plus background. 
Each simulated experiment contains a mixture of signal and background events 
as described in the text.  Also shown is the statistical uncertainty obtained
from our data sample.  The probability for obtaining a smaller uncertainty in
the simulated experiments is 5\%.
\label{pe_error}}
\end{figure}

We first check that the statistical uncertainty obtained from the
data sample (5.1 $\gevcc$) is reasonable.
Figure~\ref{pe_error} shows the distribution of statistical
uncertainties from simulated experiments along with the value we obtain
for our data sample. We find that 5\% of simulated experiments
yield a statistical uncertainty equal to or smaller than the value 
from our data sample. While this number is small, it is reasonable.

\begin{figure}[ht]
%\leavevmode
%\begin{center}
\hspace{0.5in}
\epsfysize=6.5in
\epsffile{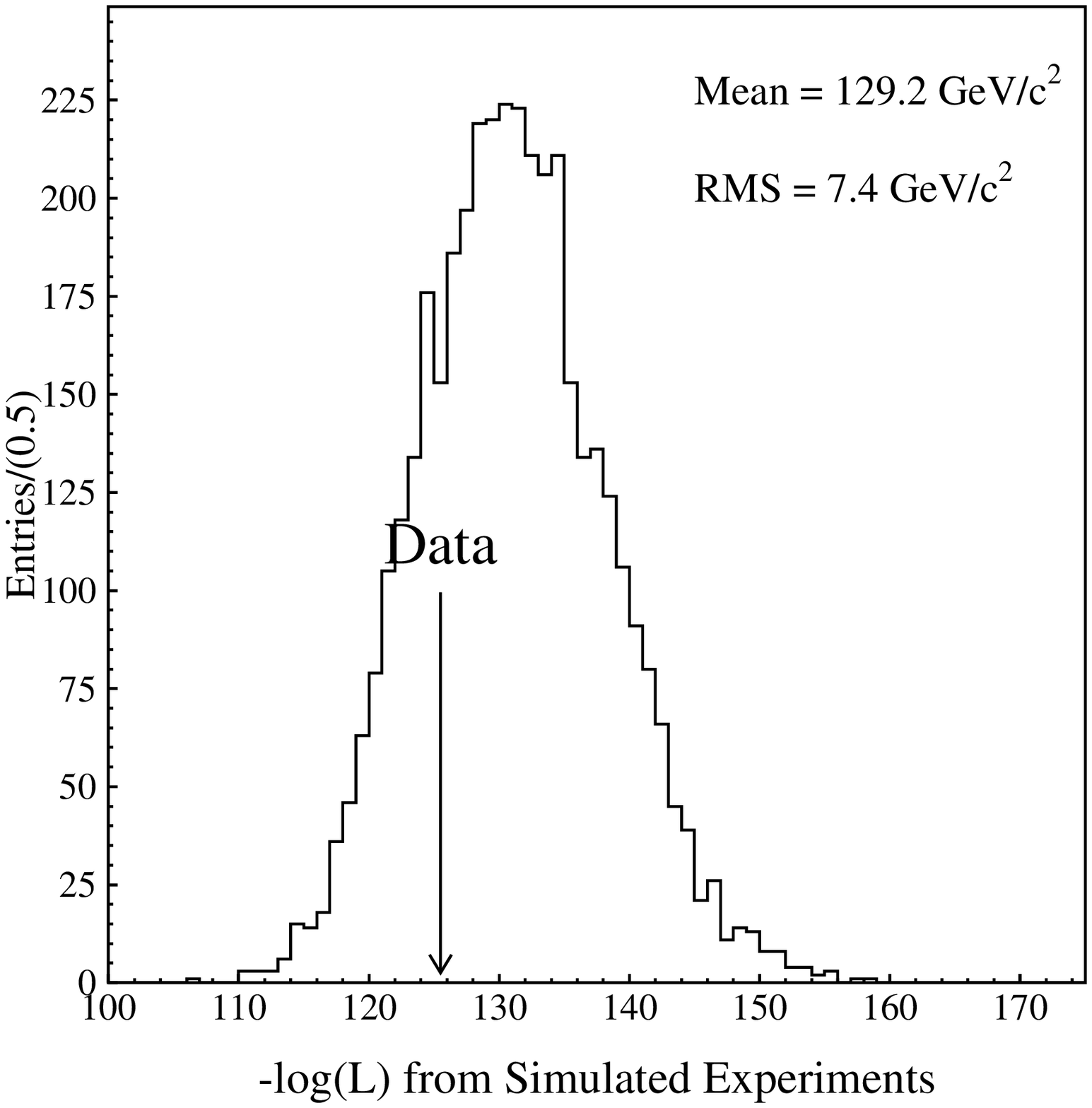}
%\epsffile{$PRD_FIG/pe_lik.ps}
%\end{center}
\caption{Distribution of the minimum value of $-\log{\cal L}$ from 
simulated experiments of $\ttbar$ (\mtop=175 $\gevcc$) plus background. 
Each simulated experiment contains a mixture of signal and background events 
as described in the text.  Also shown is the minimum $-\log{\cal L}$ value
obtained from our data sample.  The probability for obtaining a larger value 
of $-\log{\cal L}$ in the simulated experiments is 79\%.
\label{pe_lik}}
\end{figure}

A further check is provided by comparing the minimum of the negative
log-likelihood obtained in the data sample with the values obtained
from a large sample of simulated experiments (Fig.~\ref{pe_lik}). 
A value of the negative log-likelihood larger than expected from simulated 
experiments might indicate that either the reconstructed-mass
distribution is not well modeled or that the background fractions
in the sample are not properly estimated. The distribution shows that the 
value obtained from the data is quite consistent with standard model $\ttbar$ 
plus background, as evidenced by the 79\% probability of obtaining a value of 
$-\log{\cal L}$ larger than the one seen in the data. 
%We have also compared
%the cumulative distribution of the reconstructed masses in data
%with the expectation from $\ttbar$ plus background (using the
%fitted background fractions). The comparison, shown in Figure~\ref{cumdist}, 
%gives a confidence level of 64\%, indicating that the distribution
%of reconstructed masses in data is well described by the model. 
%
%In Section~\ref{like-perf}, it was demonstrated that the uncertainty on the
%top mass returned from the likelihood procedure accurately reflects
%the statistical precision of the event sample. With this in mind,
%it is clear that the value of statistical uncertainty returned by
%the likelihood procedure for our data sample is an appropriate estimator 
%for the precision of our top quark mass measurement.
%

%\begin{figure}[ht]
%%\leavevmode
%\begin{center}
%\hspace{0.5in}
%\epsfysize=6.5in
%\epsffile{cumdist.ps}
%\end{center}
%\caption{The cumulative distribution of the reconstructed masses for the
%combined four subsamples. The solid line is the distribution for the 76
%data events, and the dashed line is the sum of top and background, weighted
%by the appropriate background fractions. A Kolmogorov-Smirnov test gives
%a confidence level of 64\% that the data fits the model of $\ttbar$ plus
%background.}
%\label{cumdist}
%\end{figure}

%The same likelihood technique used to combine the results from differently
%tagged subsamples can also be used to combine results from
%other decay modes. CDF has reported respective 
%results~\cite{dilepton, hadronic}
%using the decay channels $\tllx$ and $\thad$ of $186 \pm 12 \gevcc $ and
%$159 \pm 25 \gevcc $.  Figure~\ref{all} shows the likelihood distribution
%for these and for the optimized lepton+jet analysis and the cumulative
%likelihood for all three channels.  The minimum of $M_{TOP} = 176.0 \pm 4.1 
% \gevcc $ comprises the single best estimate of the top quark mass.  In 
%addition the the statistical uncertainty, there are also systematic sources
%of uncertainty and they form the content of the following section. 
%Table~\ref{t-amass} shows the best 
%mass estimates from each decay channel and the combined estimate.

\section{Results from $b$-tagged events}

\begin{figure}[ht]
%\leavevmode
%\begin{center}
\hspace{0.5in}
\epsfysize=6.5in
\epsffile{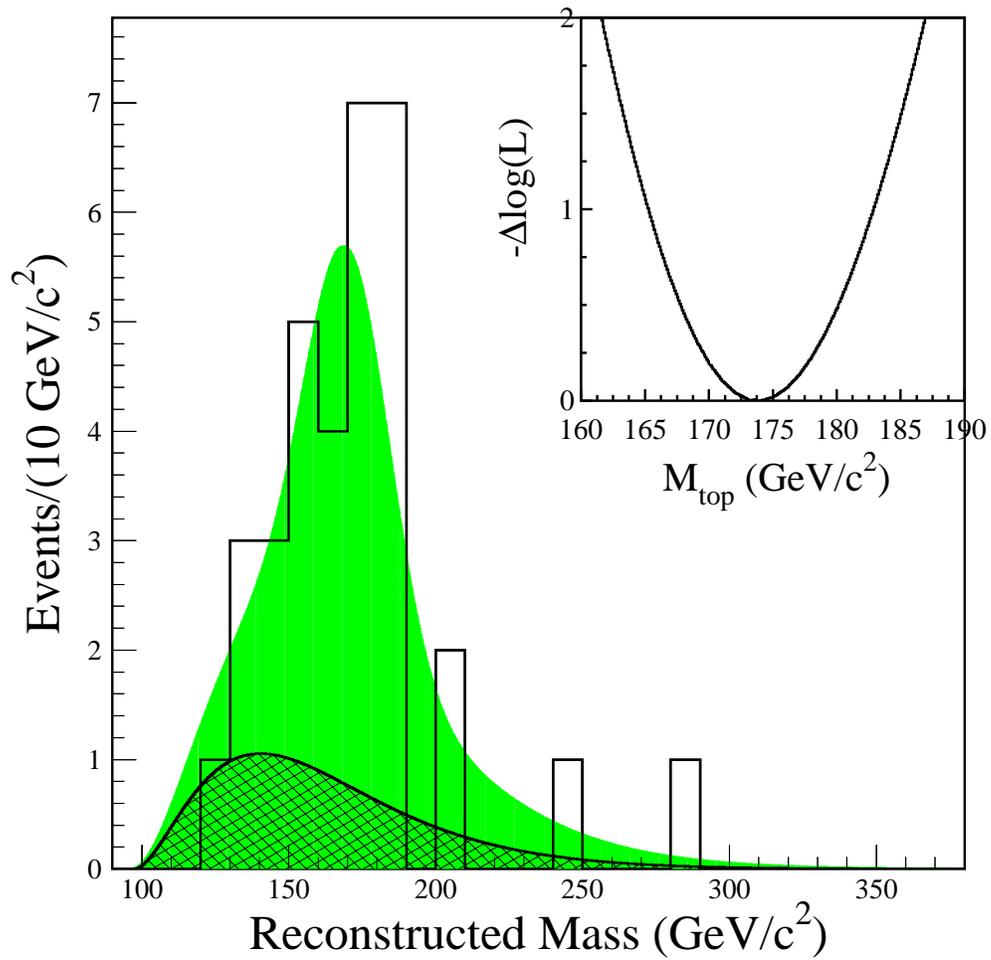}
%\epsffile{$PRD_FIG/std_err.ps}
%\end{center}
\caption{Result of applying the likelihood procedure to the 34 $b$-tagged 
events, treated as a single sample.  The figure shows the data
(histogram), fitted background (shaded 
hatched region), and fitted signal (shaded non-hatched region). The inset 
shows the shape of $-\log{\cal L}$ versus top mass, from which we extract 
the top quark mass and its statistical uncertainty.}
\label{std_err}
\end{figure}

In previous publications~\cite{cdf-evidence,cdf-discovery}, the
top quark mass was measured using only events containing SVX and/or
SLT tagged jets among the leading four jets as a single sample
(7 events in Ref~\cite{cdf-evidence}, 19 events in
Ref~\cite{cdf-discovery}). The final sample of 34 $b$-tagged events
has been analyzed as part of our four subsample fit using 
the likelihood method described in Sections~\ref{s-algor} 
and \ref{s-like}.
The 34 tagged events may be treated as three subsamples 
or they may be fit
as a single 34 event sample~\cite{eddy}. The results of fitting the 34 
$b$-tagged events as a single sample are shown in Fig.~\ref{std_err}.
The likelihood fit yields a top quark mass of 173.7$\pm$6.2(stat.) $\gevcc$ 
with a top fraction of 75\%. Treating the 34 $b$-tagged events
as three subsamples, we obtain a top quark mass of 174.0$\pm$5.7(stat.)
$\gevcc$. The 8\% improvement is consistent with expectations from
simulated experiments (see Section~\ref{s-optimize}).

\section{Alternate techniques}

  Other analyses which have different selection criteria and/or modified 
formulations of the $\chi^2$ have been performed. The analyses are
aimed at improving the probability for choosing the correct combination.
The first of these analyses, the $\mLik^{\star\star}$ analysis, uses two 
additional terms in the $\chi^2$ to aid in choosing the correct combination.
The second analysis uses a looser definition for $b$-jet tagging to
increase the number of double $b$-tagged events. Values of the top
quark mass from these two analyses are
consistent with the results presented in this report
and are summarized in Appendix~\ref{s-other}.

%% file: chapter_9.tex
\chapter{SYSTEMATIC UNCERTAINTIES}
%\newpage
%\section {SYSTEMATIC UNCERTAINTIES}
\label{s-sys}

%\section{Overview}
%\label{s-sysovr}

% The measured value of the top quark mass is subject to 
%uncertainties related to the measurements of the observables
%in the event, e.g. the jets and lepton momenta, and the Monte Carlo 
%models used to extract the top quark mass. 
%%Systematic uncertainties 
%%from these sources generally result from  modeling of the 
%%data sample in the Monte Carlo simulations. 
%In this section we describe the evaluation of these systematic 
%uncertainties and the corresponding uncertainty on the 
%top quark mass.

 The general procedure for estimating the systematic uncertainty on
$M_{top}$ from a given source is handled similarly for all sources.
For a given source of uncertainty, we make a change in the input
value, and evaluate the impact on the measured top quark mass. The change
is either a one standard deviation (1$\sigma$) uncertainty on the
variable in question,
or a change in an input assumption.
The change in the top quark mass is evaluated using
simulated experiments (see Section~\ref{like-perf}).
We perform a large number of simulated experiments with (a)
the nominal input value of the variable or the standard assumption, 
and (b) a ``1$\sigma$'' shift in the variable value 
or the changed assumption. The reconstructed-mass 
distribution from each simulated experiment is fit using the same
likelihood procedure as used on the data sample, thus obtaining
a measured top quark mass. The likelihood procedure
includes the same templates as used with the data.
The systematic uncertainty is defined as the difference in the 
median top quark mass between the two ensembles (a) and (b).
The distribution of reconstructed top quark 
masses from simulated experiments in which all inputs are
set to their nominal values was shown in Fig.~\ref{pe_mass}.

\section{Uncertainties from the energy scale}
\label{s-jtetsys}

  The event reconstruction algorithm
varies the measured momenta of the jets, leptons,
and unclustered energy to fit the kinematics of the hypothesized 
$\ttbar$ decay. The energy scale for electrons and the momentum scale for
muons are known to a precision of 0.14\% and 0.065\%
respectively~\cite{wmass}. This  uncertainty 
has a negligible effect on
the uncertainty in the top quark mass measurement.
The uncertainty on the quantity $X_T$, the transverse energy beyond the
partons associated with the $\ttbar$ event, has been discussed in Section
~\ref{s-uclus}. This uncertainty is large, but large variations of $X_T$ do not
%Section~\ref{s-uclus}, and because it is ascribed a large 
%uncertainty (100\%) in the kinematic mass fit, it does 
have a significant impact on the mass reconstruction. The $\met$ is
evaluated through the measurements of the charged lepton,
the jets, and the unclustered energy, and is therefore not
an independent measurement (see Section~\ref{s-uclus}). 
To avoid correlations it is not used 
as a measurement in the fit, but is used to estimate a starting value for
the transverse momentum of the neutrino. Thus, the energy scale uncertainty in
the measured top quark mass is dominated by the
uncertainty in the measurement of the jet momenta.

%
%\subsection{Uncertainty on top quark mass from the jet 
%$\Et$ uncertainty}

\begin{figure}[htbp]
\epsfysize=6.5in
\hspace{0.5in}
\epsffile[0 72 612 720]{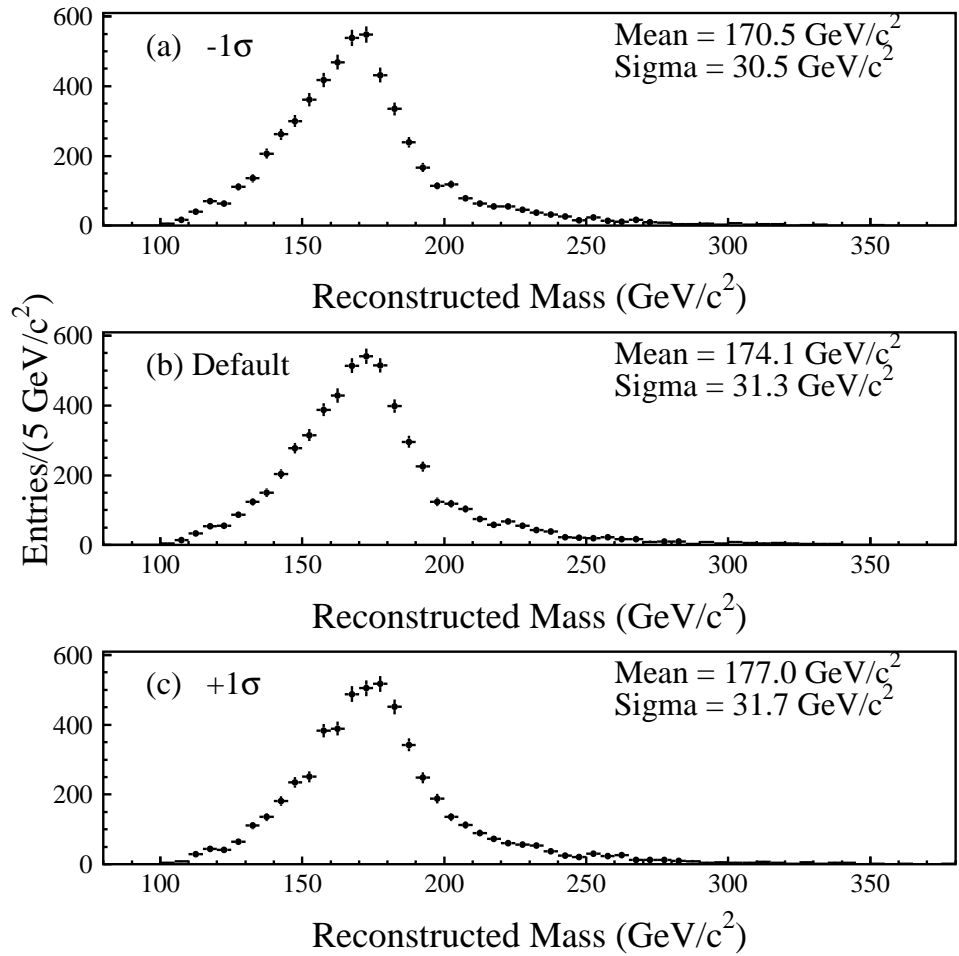}
%\epsffile[0 72 612 720]{$PRD_FIG/jetet_sys.ps}
\caption{Distribution of reconstructed masses for SVX Single tagged
  events from the {\small{HERWIG}} $\ttbar$
Monte Carlo simulation using an input top mass 175 $\gevcc$ for (a) 
a $-1\sigma$
shift in the jet $\Pt$ scale, (b) no shift in the jet $\Pt$ scale, and (c)
a +1$\sigma$ shift in the jet $\Pt$ scale. These distributions are used 
as inputs to generate the samples of simulated experiments described
in the text.}
\label{jetet_sys}
\end{figure}

\begin{figure}[htpb]
\epsfysize=6.5in
\hspace{0.5in}
\epsffile[0 72 612 720]{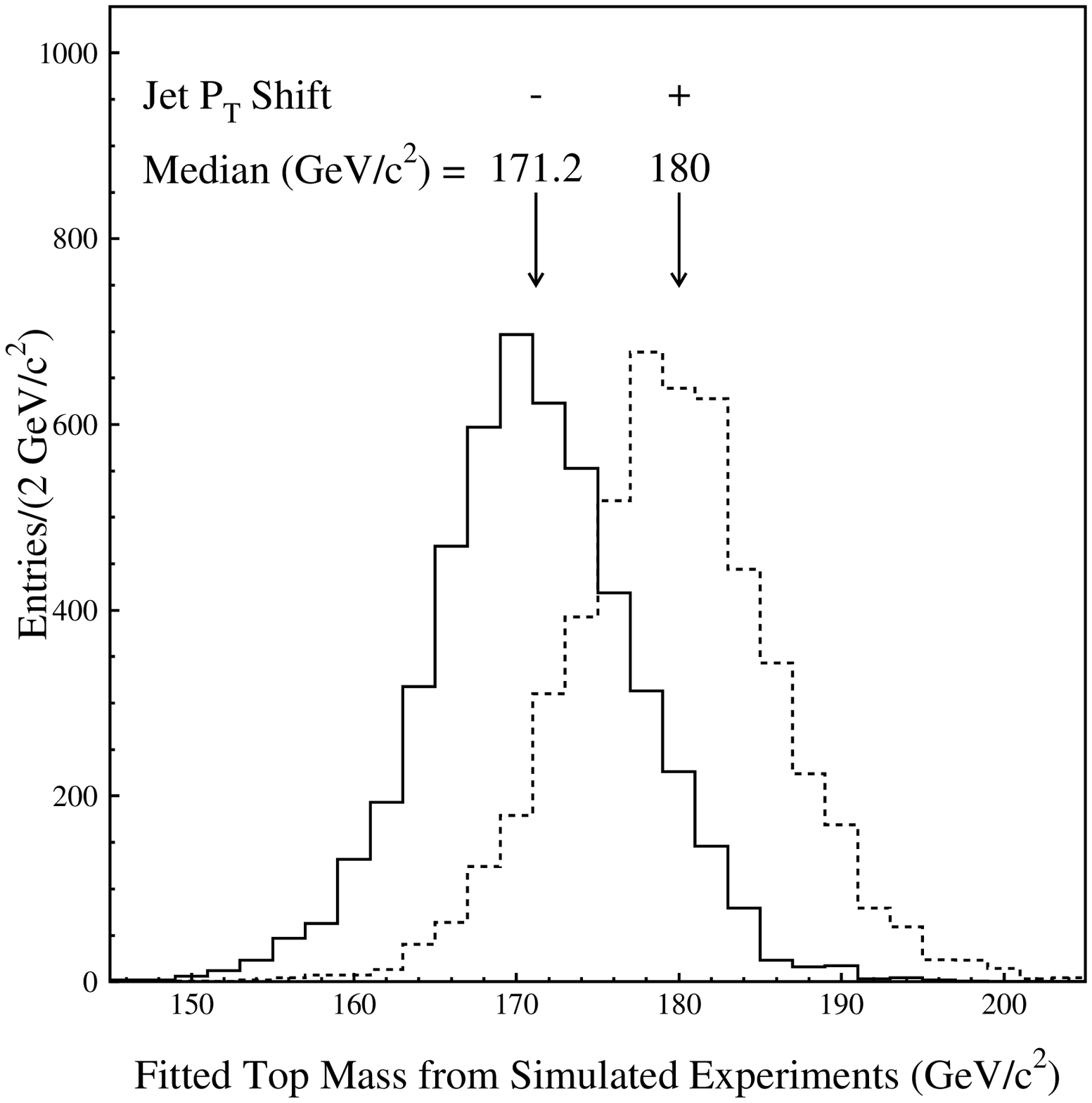}
%\epsffile[0 72 612 720]{$PRD_FIG/jetet_pseudo.ps}
\caption{Distribution of fitted top masses returned from the
likelihood procedure for simulated experiments. The solid histogram
shows the distribution when the jet $\Pt$ scale is shifted down 
by $-1\sigma$, and the dashed histogram
shows the results when the
jet $\Pt$ scale is shifted up by +1$\sigma$. The median top masses
for each are indicated on the figure, from which we obtain a
systematic uncertainty of 4.4 $\gevcc$.}
\label{jetet_pseudo}
\end{figure}

   The total uncertainty in the jet $\Pt$ scale is taken
as the quadrature sum of all uncertainties discussed in
Section~\ref{s-jetcor}. We apply +1$\sigma$ and $-1\sigma$
shifts
to the jet momenta in $\ttbar$ signal and background events,
and measure the effect on the measurement of the top quark mass.
For the SVX Single subsample, 
the distributions of reconstructed masses for 
$-1\sigma$ and +1$\sigma$ shifts in the $\Pt$ scale are shown in 
Figs.~\ref{jetet_sys}(a) and (c) respectively.
These distributions may be compared to Fig.~\ref{jetet_sys}(b)
which shows the distribution obtained from the default momentum scale.
As expected, a clear shift in the reconstructed-mass spectrum is 
observed. We generate analogous distributions for the other three mass 
subsamples and for the background mass distribution.
To obtain the systematic uncertainty, we generate
two large samples of simulated experiments.
In the first sample, we choose the reconstructed masses for $\ttbar$ events
at random from distributions like the one in Fig.~\ref{jetet_sys}(a). In the
second sample, we use distributions like the one in Fig.~\ref{jetet_sys}(c).
The simulated experiments in each
of these samples are fit using the standard templates and the likelihood 
technique described in Section~\ref{s-like}. The median
top mass from the simulated experiments in the two samples
differ because of the applied jet $\Pt$ scale shifts. The
distribution of reconstructed top masses from the two 
(jet $\Pt$ shifted) samples are displayed in Fig.~\ref{jetet_pseudo}.
We take half the difference between the medians of
the $-1\sigma$ and +1$\sigma$ distributions (from Fig.~\ref{jetet_pseudo})
as the uncertainty on the top quark mass measurement due to the $\Pt$
scale uncertainty. Using this prescription, we obtain a top mass
uncertainty of $\pm 4.4 \ \gevcc$ from the jet $\Pt$ scale.
%Similar results are obtained if we compare simulated experiments using
%the default jet $\Pt$ scale with templates in which the jet $\Pt$
%scale is shifted.

\section{Initial and final state hard radiation}
\label{gluon}

QCD radiation that produces jets can originate
from the outgoing (final state) partons,
the incoming (initial state) partons, or
from interference among the two.  The interference
effect is expected to be small~\cite{lynorr1} and is
not considered here.

 The effects of initial state radiation (ISR) and final
state radiation (FSR) on the measurement
of \mtop are studied using the {\small PYTHIA} program
since it allows the two effects to be studied in isolation
from one another. The approach used to evaluate the 
systematic uncertainty due to ISR is to compare the median 
mass from simulated experiments using the standard
{\small PYTHIA} settings to the median mass from simulated experiments
with ISR turned off. The median mass from simulated experiments
for the no-ISR {\small PYTHIA} sample is found to be lower than 
that of the standard {\small PYTHIA} sample by 2.6 $\gevcc$. 
The uncertainty is taken to be one half of the shift in median mass 
between the standard {\small PYTHIA} simulation and the no-ISR 
{\small PYTHIA} simulation, which is 1.3 $\gevcc$. 
We assume that the shift is symmetric with the amount of
ISR and therefore the uncertainty is $\pm$1.3 $\gevcc$. 

Extracting the effects due to final state radiation is a more subtle 
exercise because {\small PYTHIA}, like {\small HERWIG}, describes
jet formation through a parton shower.
% which ascribes the entirety of a jet to final state radiation.  
The effects
of modeling the softer components on the measurement of \mtop
are described by the studies of soft gluon radiation 
(see Section~\ref{s-fijetcor}).  
In this discussion, we are therefore referring to the ``harder'' 
component of FSR, which leads to extra jets in the
final state. To isolate the effect of FSR, we use a
sample of {\small{PYTHIA}} events which have ISR turned 
off. We select a subsample of these events that have
exactly four jets ( either four high-$\Et$ jets, $\et>15\ \gev$ 
and $|\eta |<2.0$, or three high-$\Et$ jets plus one 
with $\et>8\ \gev$ and $|\eta |<2.4$),
all of which can be uniquely matched to the partons from
$\ttbar$ decay (within a cone of 0.4). Using the
procedure of simulated experiments, we take the systematic
uncertainty to be half the difference between
the no-ISR {\small{PYTHIA}} events with no restriction on the
number of jets and the subsample of events with exactly four jets
uniquely matched to the partons from
$\ttbar$ decay. We assume that this difference is approximately
independent of the amount of ISR present. The median top mass from the 
no-ISR sample
with exactly four jets is found to be larger than the default 
no-ISR sample by 4.4 $\gevcc$. The systematic uncertainty is taken 
to be one half of the difference in the median top masses between 
the two samples, or 2.2 $\gevcc$. As with ISR, we assume that 
the shift in top mass is symmetric with respect to the amount of FSR,
so the systematic uncertainty is $\pm2.2\;\gevcc$.

\section{Background mass distribution}

%  As described previously, we rely on the {\small{VECBOS}}
%Monte Carlo simulation to model the shape of the backgrounds
%(which are primarily $W$+multi-jets events). Published results
%\cite{vecbos_w} indicate that the properties of jets in
%$W$-boson events are in general agreement with the predictions
%from the {\small{VECBOS}} Monte Carlo simulation. 

In generating the default background distributions with the {\small{VECBOS}}
program,  we used the $W$+3 partons matrix elements and chose a scale of 
$Q^{2}=\langle\Pt\rangle^{2}$. 
This $Q^2$ scale is not only used in {\small{VECBOS}} for the computations
of the matrix elements and the evaluation of the parton distribution functions,
but it is also used in the evolution of the parton showers to limit the
$\Pt$ of additional jets~\cite{dglap}. As a result, the shape of the 
reconstructed-mass distribution is sensitive to the choice 
of scale. The systematic 
uncertainty from using the assumed background shape was evaluated by
changing the $Q^2$ scale from $\langle\Pt\rangle^{2}$ to $M_{W}^{2}$. 
%These background shapes are compared in Fig.~\ref{vecbos_bg}. 
Simulated experiments
using $\ttbar$ and the modified background shape ($Q^{2}=M_{W}^{2}$) were 
fit to the default signal and background probability distributions as
described in Section~\ref{s-like}. The median mass from these simulated 
experiments was found to differ by 1.3 $\gevcc$ from simulated experiments 
using the default background shape. The systematic uncertainty from
the background shape modeling is taken to be this difference symmetrized, or
$\pm 1.3\;\gevcc$.

%\begin{figure}[htbp]
%\leavevmode
%\begin{center}
%\hspace{0.75in}
%\epsfysize=6.0in
%\epsffile{vecbos_bg.eps}
%\end{center}
%\caption{Fitted mass distribution for different values of the $Q^2$ parameter
%for simulated $W$+multi-jet events from the {\small{VECBOS}} Monte Carlo program.}
%\label{vecbos_bg}
%\end{figure}

%\begin{figure}[ht]
%\leavevmode
%\begin{center}
%\epsfysize=6.0in
%\epsffile{zjetbg.eps}
%\end{center}
%\caption{Fitted mass distribution for different values of the parameter
%for simulated Z+multi-jet events from the {\small{VECBOS}} Monte Carlo program and
%from candidate Z+multi-jet events observed in the data.}
%\label{f-zbg}
%\end{figure}

\section{$b$-Tagging }
\label{btagbias}

 A systematic uncertainty in the top mass measurement may arise from an
uncertainty in the {\small{SVX}} and {\small{SLT}} tagging efficiencies. For 
{\small{SVX}} tagging, the primary uncertainty comes from the 
possible $\Et$ dependence of the {\small{SVX}} tagging efficiency
which may differ from the simulation. The {\small{SVX}} tagging efficiency 
in data relative to the simulation is parametrized
as a function of the jet $\Et$, and is nearly flat. 
%with an average value of 0.87$\pm$0.07. 
The systematic uncertainty is
evaluated by assuming the largest possible $\Et$ dependence of this ratio 
given the available statistics. 
%This results in a ratio which decreases from
%$\approx$0.92 at $\Et=15 \ \gev$ to $\approx$0.76 at 120 GeV.
Comparison of results obtained using a flat ratio to those obtained with 
a sloped one gives a mass shift of only 0.1 $\gevcc$

  The jet $\Et$ dependence of the {\small{SLT}} tagging efficiency
is better known than in the {\small{SVX}} case. However, a systematic
uncertainty does arise from the uncertainty in the expected ratio of
true to fake {\small{SLT}} tags in $\ttbar$ events. To estimate the
sensitivity of our top quark mass measurement to this ratio, we generate
Monte Carlo $\ttbar$ events in which all {\small{SLT}} tags are either 
(a) true tags, or (b) fake tags. We then produce two large samples
of simulated experiments, each having the same number of observed events as
in our four mass subsamples and including the appropriate background 
contributions. 
The two samples of simulated experiments consist 
of either events all from set (a) or all from set (b). The median top quark
mass values from the two samples of simulated experiments differ by
0.8 $\gevcc$. We take half of this difference, or 0.4 $\gevcc$, as the
corresponding systematic uncertainty in the top quark mass.
Combining the systematic uncertainties from {\small{SVX}} and 
{\small{SLT}} tagging, we find an overall systematic uncertainty
of $\pm 0.4\;\gevcc$.

%have been studied in great detail using
%large $\psi\to\mu\mu$ and $\gamma\to e^+e^-$ samples and the
%statistical uncertainties are negligible. The largest uncertainty 
%for SLT tagged jets is the exact ratio of real-to-fake SLT tags
%in $\ttbar$ events. The real-to-fake ratio in $\ttbar$ events is
%estimated to be uncertain at the level of $\approx$20\%. To evaluate
%the uncertainty, simulated experiments are generated which have 
%100\% real SLT tags, and another sample which has 100\% fake SLT
%tags. We take half the difference in the median mass from the
%simulated experiments as an estimate of the systematic uncertainty.
%We find a 2.2 $\gevcc$ mass shift which only effects the SLT tagged
%subsample. When combining all four subsamples, the overall effect
%is to produce a systematic uncertainty from b-tagging of $\pm 0.4\gevcc$.

\section{Parton distribution functions}
\label{partdissys}

All of the Monte Carlo samples used to measure the top mass were generated
with the MRSD0$^{\prime}$~\cite{mrsd0} set of parton distribution functions
(PDF).  This was the preferred PDF at the time the samples were generated.  
Newer distribution functions now exist, in particular ones which fit CDF's 
inclusive jet cross section.  One such PDF, CTEQ4L~\cite{cteq4l}, provides a 
higher gluon content at lower momentum fraction than MRSD0$^{\prime}$.
We take the shift in the median top mass between samples generated with the 
two PDF's as the relevant uncertainty. We therefore assign a systematic 
uncertainty of $\pm 0.3\;\gevcc$ in the top quark mass from this source.

\section{Monte Carlo generators}
\label{mcgensys}

The effect of using different Monte Carlo generators has also been
studied. Previously, this was evaluated from the difference between
the {\small{HERWIG}} and {\small{ISAJET}} simulations. 
%Most of the difference between these two generators can be attributed to
Because of the evidence that independent fragmentation does not
reproduce some aspects of the data (energy flow 
around and between jets~\cite{color_coh,andy}) we will not use ISAJET
here. We evaluate the
systematic uncertainty from the choice of Monte Carlo generators
via the mass shift between the {\small{HERWIG}} and {\small{PYTHIA}} 
simulations. This gives a systematic uncertainty of $\pm 0.1\;\gevcc$.
%Since this uncertainty is negligible, it is detailed here for 
%completeness but not included in the final list of systematic
%uncertainties.
 
\section{Summary of systematic uncertainties}
\label{sumsyst}

The relevant systematic uncertainties studied for the top mass
measurement are listed in Table~\ref{t-allsys}.  Combining all of these
effects in quadrature gives a total systematic uncertainty of 
$\pm 5.3\;\gevcc$, or $\pm 3\%$ of 176.1 $\gevcc$.

%%%%%%%%%%%%%%%%%%%%%%%%%%%%%%%%%%%%%%%%%%%%%%%%
\begin{table}[ht]
\begin{center}
\begin{tabular}{lc}
\hline\hline
Source                            &  Uncertainty ($\gevcc$) \\ 
\hline
Jet energy measurement            &       4.4               \\
Initial and final state radiation &       2.6               \\
Shape of background spectrum      &       1.3               \\
$b$-Tagging                       &       0.4               \\
Parton distribution functions     &       0.3               \\
Monte Carlo generators            &       0.1               \\
\hline
Total                             &       5.3               \\
\hline\hline
\end{tabular}
\end{center}
\caption{Systematic uncertainties on the measurement of the top 
quark mass for this analysis.}
\label{t-allsys}
\end{table}

%% file: ch10.tex
%% This is CHAPTER10.TEX. 
%% Extracted from test_s10.tex that checked that TEX things seemed OK.

\chapter{COMBINED TOP QUARK MASS}
\label{s-combmass}

The most precise measurement of the top quark mass in any
single decay channel is obtained with events in the lepton+jets
topology. The analysis of such events leads to a mass of 
${\rm 176.1\pm 5.1 (stat.)\pm 5.3 (syst.)\;\gevcc}$.  Measurements in 
the all-hadronic~\cite{all-hadron} and dilepton~\cite{lbl-mass} 
decay topologies have also been made and can be combined with the 
lepton+jets result to reduce the overall uncertainty.  Here we make 
some brief remarks on these analyses, and describe how the three 
measurements were combined.

\section{All-hadronic topology}

The top quark mass measurement in the all-hadronic topology used a sample of 
136 events that satisfied several selection criteria, including the 
requirement of six or more jets, at least one of which was tagged as a $b$
by the SVX. The estimated background in the sample was $108\pm 9$ events.
The method for extracting a top mass was similar to the one used for the 
lepton+jets topology, and included a kinematic fit to each event and a 
likelihood fit to the resulting reconstructed-mass distribution.  The results 
of the likelihood fit yielded a measured top quark mass of
${\rm 186\pm 10 (stat.)\pm 12 (syst.)\;\gevcc}$~\cite{all-hadron}. 
A reevaluation of the systematic uncertainty on this measurement has led to 
a more accurate estimate of 5.7 $\gevcc$~\cite{lbl-mass}.  
Appendix~\ref{dilep-appendix} describes the details of this reevaluation.
  
\section{Dilepton topology}

The dilepton topology includes $\ttbar$ events in which the $W^+$ and $W^-$ 
bosons each decay into an $e\nu$ or $\mu\nu$ final state. The presence of 
two neutrinos, which are not observed in our detector, prevents a 
straightforward event-by-event kinematic fit to the $\ttbar$ decay hypothesis.
Therefore, we have measured the top quark mass from dilepton events using a 
weighting method~\cite{d0-dilepton,kondo,varnes}. In this method the 
vector sum of the neutrino 
transverse momenta, as predicted after making certain assumptions, is compared
to the observed missing transverse momentum~\cite{lbl-mass}.  
From a sample of eight events with an estimated background of $1.3\pm 0.3$ 
events we obtain a mass of 
${\rm 167.4\pm 10.3 (stat.)\pm 4.8 (syst.)\;\gevcc}$.
A brief description of the method, and some additional information 
not reported in Ref.~\cite{lbl-mass} is given in 
Appendix~\ref{dilep-appendix}.

\section{Combining the measurements}

Each of the three top quark mass measurements is associated with a statistical
and systematic uncertainty. The statistical uncertainties are uncorrelated,
since the samples are statistically independent. However, the systematic
uncertainties are correlated, and these correlations must be included when 
combining the results.

The systematic uncertainties in the measurements from each decay
topology~\cite{lbl-mass} are assigned to one of five independent categories:
\begin{enumerate}
 \item jet energy scale,
 \item signal model (ISR, FSR, PDF, $b$-tagging),
 \item Monte Carlo generator,
 \item background model,
 \item Monte Carlo statistics.
\end{enumerate}
The assignment of the systematic uncertainties for each of the three 
mass analyses to these categories is shown in Table~\ref{comb-sys}.
In the lepton+jets measurement, the statistical uncertainty in the
Monte Carlo simulation is included in the global statistical uncertainty.
\begin{table}[ht]
\begin{center}
\begin{tabular}{cccc}
\hline\hline
                    & \multicolumn{3}{c}{Systematic uncertainty ($\gevcc$)}\\
\cline{2-4}
Systematic category & Lepton+jets & All-hadronic & Dilepton \\
\hline
Jet energy scale    &   4.4       &    5.0       &   3.8    \\
Signal model        &   2.6       &    1.8       &   2.8    \\
M.C. generators     &   0.1       &    0.8       &   0.6    \\
Background model    &   1.3       &    1.7       &   0.3    \\
M.C. statistics     &    -        &    0.6       &   0.7    \\
\hline
Total               &   5.3       &    5.7       &   4.8    \\
\hline\hline
\end{tabular}
\end{center}
\caption{Systematic uncertainties for each of the three mass analyses 
grouped into the five categories. Also shown is
the total systematic uncertainty for each analysis.}
\label{comb-sys}
\end{table}

For each of the five categories, the systematic uncertainties in each of 
the three measurements are assumed to be either uncorrelated or 100\% 
correlated.  The jet energy scale uncertainty is taken to be
100\% correlated since all three analyses use the same detector and
the same jet clustering algorithm.  The systematic uncertainties 
coming from the signal model and the Monte Carlo generator are also 
assumed to be 100\% correlated since all three analyses use the 
{\small HERWIG} Monte Carlo generator to simulate $\ttbar$ events. 
The uncertainties in the background shape are assumed to be uncorrelated
because the background processes for each analysis are different.
The correlation coefficients between the three pairs of analyses
are given in Table~\ref{corr-coeff}.
\begin{table}[ht]
\begin{center}
\begin{tabular}{cccc}
\hline\hline
                    & \multicolumn{3}{c}{Correlation coefficients} \\ 
\cline{2-4}
Systematic category &   LJ/AH     &    LJ/LL     &   AH/LL \\
\hline
Jet energy scale    &   1.0       &    1.0       &   1.0    \\
Signal model        &   1.0       &    1.0       &   1.0    \\
M.C. generators     &   1.0       &    1.0       &   1.0    \\
Background model    &   0.0       &    0.0       &   0.0    \\
M.C. statistics     &   0.0       &    0.0       &   0.0    \\
\hline\hline
\end{tabular}
\end{center}
\caption{Correlation coefficients between the three mass analyses 
for the five categories of systematic uncertainty. Here, LJ signifies
the lepton+jets analysis, AH the all-hadronic analysis, and LL the dilepton
analysis.}
\label{corr-coeff}
\end{table}

The inputs into the calculation for combining the mass measurements are 
the three top quark mass measurements cited in this section, their 
statistical uncertainties, and the systematic uncertainties and their 
correlations as listed in Tables~\ref{comb-sys} and \ref{corr-coeff}. 
The calculation uses a generalized chisquare method with full covariance 
matrix (see for example~\cite{PDG}), and yields:
\begin{equation}
              m_{t} = 176.1 \pm 6.6\;\gevcc.
\end{equation}
In the calculation, the central value can be written as the weighted
sum of the three input central values. The weights, which depend on the 
statistical and systematic uncertainties and the correlations, are found
to be 0.65 (lepton+jets), 0.19 (dilepton), and 0.16 (all-hadronic).
If we define a statistical uncertainty on the combined result as the 
quadrature sum of the weighted individual statistical uncertainties, 
that combined statistical uncertainty is $\pm 4.2\;\gevcc$.
The combined systematic uncertainty, defined as the quadrature difference 
between the total and statistical uncertainties, is then $\pm 5.1\;\gevcc$.

%% file: conclusion2.tex
\chapter {SUMMARY}
\label{s-conclusion}

  The first evidence of the production of top quark pairs in $\ppbar$
collisions was reported by CDF in 1994~\cite{cdf-evidence}.
From a sample of seven candidate lepton+jets events with an expected background
of 1.4 events, the top quark mass was measured to 
be 174$\pm$10(stat)$_{-12}^{+13}$(syst) $\gevcc$. 
Since that time, both CDF and D0 have analyzed their full Run 1 data
samples and have published their measurements of the $\ttbar$ 
production cross section~\cite{xsec,d0xsec} and top quark 
mass~\cite{d0-mass-prl,cdf-mass-prl}.  Indirect measurements of the top quark 
mass using data from LEP and SLC have been made~\cite{lep2-topmass}, and are 
consistent with the direct measurements, although with a substantially larger 
uncertainty.

This report has described in detail the best single channel
measurement of the top quark mass. A letter on the measurement
is already published~\cite{cdf-mass-prl}. The likelihood method uses
parametrized templates, which results in a continuous likelihood
shape as a function of the top quark mass from which the top quark
mass and statistical uncertainty are evaluated.
The statistical precision of
the top quark mass measurement has benefitted from a larger data sample
than earlier measurements~\cite{cdf-evidence,cdf-discovery}, and through
subdivision of the data sample into non-overlapping subsamples according 
to the $b$-tagging information. 
Systematic uncertainties have been considerably
reduced, primarily through a better understanding of the jet energy 
measurements which resulted in smaller uncertainties on jet $\Pt$. 
%uncertainties~\cite{cdf-evidence,cdf-discovery}. 

From the 106 \pb~ Run 1 
data sample, we measure the top quark mass in the lepton+jets topology
to be 176.1$\pm$5.1(stat.)$\pm$5.3(syst.) $\gevcc$. Measurements of the 
top quark mass in the all-hadronic~\cite{all-hadron} and 
dilepton~\cite{lbl-mass} decay topologies are consistent with this 
measurement. Combination of the three measurements from CDF gives a top 
quark mass of 176.1$\pm$6.6 $\gevcc$. The D0 collaboration has also
published results on the top quark mass measurement in the lepton+jets 
and dilepton channels, from which they obtain 
a combined top quark mass of 172.1$\pm$7.1 $\gevcc$~\cite{newd0}.
The measurements of the top quark mass from the CDF and D0 experiments are 
consistent with each other, therefore,
their Run 1 measurements have been combined  to obtain a top quark mass
at the Tevatron of 174.3$\pm$ 5.1 $\gevcc$~\cite{tev-mass}. This
measurement represents the most precise measurement of any of the
quark masses.

%  Measurements of the top quark mass have also been published by the
%D0 Collaboration~\cite{newd0}. They measure a top quark mass of
%173.3 $\pm$ 5.6(stat.) $\pm$ 5.5(syst.) $\gevcc$ in the lepton+jets topology.
%This result is combined with the measurement in the dilepton topology

% A number of alternate top mass analyses (see Appendix B) were pursued during 
%the course of the analysis described in this report. While these analyses
%were aimed at improving the resolution on the top quark mass measurement,
%the analysis presented in the body of this report ultimately provided the 
%lowest total uncertainty.

     We thank the Fermilab staff and the technical staffs of the
participating institutions for their vital contributions.  This work was
supported by the U.S. Department of Energy and National Science Foundation;
the Italian Istituto Nazionale di Fisica Nucleare; the Ministry of Education,
Science, Sports and Culture of Japan; the Natural Sciences and Engineering 
Research Council of Canada; the National Science Council of the Republic of 
China; the Swiss National Science Foundation; the A. P. Sloan Foundation; the
Bundesministerium fuer Bildung und Forschung, Germany; and the Korea Science 
and Engineering Foundation.

%% file: appendix.tex
\appendix

\chapter {EVENTS IN THE MASS ANALYSIS}

%\newpage

The individual reconstructed masses of all events in the four subsamples are
listed in Tables~\ref{t-opt-double} through~\ref{t-opt-notag}.

\begin{table}[ht]
 \begin{center}
 \begin{tabular}{llc}
  \hline\hline
      &       & Mass \\
  Run & Event &     ($\gevcc$) \\
\hline
40758  &   44414  &  175.3\\
67824  &  281883  &  170.1\\
65581  &  322592  &  152.7\\
67971  &   55023  &  183.5\\
68464  &  547303  &  151.1\\
\hline\hline
\end{tabular}
\end{center}
\caption{List of events which are in the SVX Double subsample.
Shown are the run and event numbers
and the reconstructed top mass for the solution having the lowest $\chi^2$.}
\label{t-opt-double} 
%\end{table}

%\begin{table}[ht]
 \begin{center}
 \begin{tabular}{llc}
  \hline\hline
      &       &        Mass      \\       
  Run & Event &      ($\gevcc$)  \\     
\hline
43096&47223  &  288.6 \\ 
45610&139604  &  180.0 \\
45879$^{\star}$&123158  &  180.1 \\
59698$^{\star}$&31639  &  187.4 \\ 
63247&65096  &  161.0 \\ 
63641&3054  &  173.3 \\ 
68006&44672  &  243.4 \\ 
64901&569801  &  156.3\\ 
69683$^{\star}$&135095 &  163.2\\
56911$^{\star}$&114159  &  156.7\\
67515&298909  &  174.6\\
 68312$^{\star}$&821014  &  202.4\\
68739&425355  &  170.9\\
69781$^{\star}$&266905  &  182.8\\
56669&21631  &  152.1\\
\hline\hline
\end{tabular}
\end{center}
\caption{List of events which are in the SVX Single subsample.
Shown are the run and event numbers
and the reconstructed top mass for the solution having the lowest $\chi^2$.
Events labelled with a $^{\star}$ have both SVX and SLT tagged jets.}
\label{t-opt-single} 
\end{table}

\begin{table}[ht]
 \begin{center}
 \begin{tabular}{llc}
  \hline\hline
      &        &         Mass    \\
 Run  &  Event &      ($\gevcc$) \\
\hline
45705  &   54765  &  186.3\\
45880  &   31838  &  130.4\\
43351  &  266423  &  162.4\\
66368  &   91765  &  137.9\\
66500  &  421896  &  173.0\\
67879  &   30394  &  141.1\\
69005  &  181134  &  129.6\\
58908  &   41102  &  138.6\\
60998  &  423792  &  162.0\\
61334  &   57897  &  183.1\\
64721$^{\star}$  &  229200  &  181.0\\
65298$^{\star}$  &  747402  &  149.4\\
65648  &  203840  &  203.2\\
67515  &  616477  &  149.9\\
\hline\hline
\end{tabular}
\end{center}
\caption{List of events which are in the SLT subsample.
Shown are the run and event numbers and the 
reconstructed top mass for the solution having the lowest $\chi^2$.
Events labelled with a $^{\star}$ have two SLT-tagged jets.}
\label{t-opt-slt} 
\end{table}

\begin{table}[ht]
 \begin{center}
 \begin{tabular}{llccllc}
  \cline{1-3}\cline{5-7}\\[-4.2mm]
  \cline{1-3}\cline{5-7}
      &        & Mass       &&     &        & Mass \\
  Run & Event  & ($\gevcc$) && Run & Event  & ($\gevcc$)\\
\cline{1-3}\cline{5-7}
46492  &   57501  &  179.2 && 58696  &   83095  &  137.6\\
41301  &   45902  &  175.7 && 59948  &  105232  &  115.4\\
43421  &   65648  &  147.8 && 60634  &  350037  &  151.2\\
47757  &  262594  &  219.6 && 61167  &  332223  &  167.3\\
45757  &   30003  &  173.0 && 63265  &    5385  &  255.2\\
45144  &  107403  &  189.2 && 64041  &  473567  &  247.5\\
60656  &   96710  &  180.3 && 64997  &   78806  &  192.0\\
60746  &  121257  &  180.1 && 65179  &  215794  &  195.7\\
61511  &   75858  &  113.0 && 67391  &   50780  &  184.9\\
62981  &   85084  &  125.0 && 67757  &  631972  &  172.0\\
64861  &  121618  &  178.8 && 68144  &  100373  &  178.3\\
64934  &  400688  &  215.4 && 68231  &   78554  &  177.7\\
66046  &  507038  &  164.2 && 68374  &  312573  &  139.1\\
66207  &   12039  &  154.4 && 68553  &  707057  &  130.4\\
66315  &  365275  &  230.3 && 68570  &  897728  &  142.6\\
67862  &  631243  &  114.2 && 68593  &   88427  &  144.0\\
68006  &  176291  &  120.9 && 69519  &  430034  &  160.0\\
68939  &  352425  &  173.1 && 70000  &   26023  &  161.1\\
69520  &  307639  &  235.2 && 57438  &   71994  &  253.1\\
70578  &  351956  &  143.0 && 64901  &  505659  &  108.1\\
70986  &  227609  &  176.2 && 67397  &  105755  &  190.0\\
\cline{1-3}\cline{5-7}\\[-4.2mm]
\cline{1-3}\cline{5-7}
\end{tabular}
\end{center}
\caption{List of events which are in the No Tag subsample.
Shown are the run and event numbers, 
and the reconstructed top mass for the solution having the lowest $\chi^2$.}
\label{t-opt-notag} 
\end{table}

%% file: other.tex
\chapter {ALTERNATE MASS ANALYSES}
\label{s-other}

   A number of alternate mass analyses have been performed using
the Run 1 data sample. We discuss two alternate analyses which are aimed
at improving the statistical and/or systematic uncertainty
on the top quark mass measurement using some subsample of events. 
Another goal is to check our
default technique by employing complementary strategies by: 
(i) using more event information associated with
$b$ tagging and jet charge, and (ii) reducing the sample to the 
most complete events, ie. those where we have two b-tagged
jets. The first of these techniques
includes additional terms in the likelihood function, which improves 
the probability for choosing the correct jet-to-parton configuration
at the expense of reduced statistics. The second technique uses three
$b$-tagging algorithms to explore a subsample of the data set that consists 
of events with two $b$-tagged jets among the leading four jets.
Neither of these two techniques is found to yield a more precise 
measurement than the mass analysis described in the body of this
report. In this appendix, we briefly describe these two mass
analyses.

%   In Section~\ref{s-fperform}, we showed that the fraction
%of events reconstructed in the lepton+jets topology which
%have all four jets assigned correctly to the $\ttbar$ decay 
%partons was relatively small. 
%In particular,
%only 21\% of events with no tags and 28\% (25\%) of events with
%one SVX (SLT) tag have all four jets assigned correctly 
%to the $\ttbar$ decay partons. The fraction of correct
%assignments increases to nearly 50\% when two of the jets
%are SVX-tagged. It should be noted that the fraction of correct
%assignments is limited by the detector acceptance. 
%About 70\% of SVX Double tagged events have the four highest
%$\Et$ jets matching to the $\ttbar$ daughter partons, whereas

%this number is $\approx$55\% for the other three subsamples
%(see Table~\ref{mc_ev_frac}). 
%Two analyses have been developed which 
%are aimed at improving the fraction of correct jet-to-parton
%assignments. These two analyses are described in this
%section of the report.

\section{The $\mLik^{\star\star}$ fitting technique}
\label{s-lstar}

The $\mLik^{\star\star}$ technique~\cite{lanzoni} aims at improving
the fraction 
of correct jet-to-parton assignments by combining three independent 
sources of event information into a single parameter. These
sources are:
\begin{itemize}
\item  $\Ki$ for $\ttbar$-like kinematics as described in 
       Section~\ref{s-algor};
\item  probability for the jets assigned as $b$ jets to originate from $b$ 
       quarks, and the two jets assigned to the hadronic W-decay to originate 
       from light quarks.  The probability is evaluated using the jet 
       probability (JPB) algorithm~\cite{cdf-evidence,JETPROB1,jetprob2};
\item  probability to observe a given jet charge~\cite{JETCHARGE} for $b$ and 
       $\bbar$ quarks in $\ttbar$ events.
\end{itemize} 
  
\subsection {Definition of $\mLik^{\star\star}$}

The JPB algorithm evaluates for each charged track in a jet the
probability that it comes from the primary vertex. For each jet the
track probabilities are combined into an overall probability (JPB)
that the jet is consistent with the zero lifetime hypothesis.
Due to the long lifetime
of $b$ hadrons, the JPB distribution for $b$ quark
jets exhibits a strong peak near zero. Non-$b$ jets in $\ttbar$
events are produced either through the decays of W-bosons to $(u,d)$ 
and $(c,s)$ quark pairs, or production of gluon jets from initial or
final state radiation. With the exception of the charmed quarks,
the non-$b$ jets exhibit a flat JPB distribution. The
charm quark jets produce a small peak near zero which can be ignored
given its relative size. Unless otherwise noted, charm quark jets
are understood to be included in the ``non-$b$'' quark distribution of 
JPB.

We incorporate the JPB variable into the $\chi^2$ definition by introducing 
the following selection function:
\begin{equation}
 \mLik^{\star}  =  {\Ki} - {\rm 2} \cdot \ln\left[{\mPr}({\rm JPB_{1}}) 
                                  \cdot {\mPr}({\rm JPB_{2}}) 
                                  \cdot {\mPr}({\rm JPB_{3}})
                                  \cdot {\mPr}({\rm JPB_{4}})\right].
\label{eq:likjpb}
\end{equation}
The $\chi^2$ is the same as the one defined in Section~\ref{s-algor},
and $\mPr$(${\rm JPB_{i}}$) is the probability density for the $i^{th}$-jet 
assignment (i=1,...4). The $\mPr$ functions in $\mLik^{\star}$   
depend only on jet type, since one function is appropriate for  
both $b$ and $\bbar$ jets, and another for non-$b$ quarks.
While the $\chi^2$ value is in general different for each of the 
24 combinations, only six distinct values occur for the second term
in $\mLik^{\star}$. Groups of four, corresponding to the interchange 
of the $b$ and $\bbar$ quarks (and the two neutrino $\Pz$ solutions), 
have the same contribution from this second term.

We used  the {\small HERWIG} Monte Carlo and the full CDF detector simulation  
to generate  the ($b$ and $\bbar$) and non-($b$ and $\bbar$)~\cite{JETPROB2}
probability density distributions. We only considered events
in which the leading four jets corresponded to the four 
primary partons from $\ttbar$ decay, which limits us to 56\% of
the sample.  Of this subset, we found that the largest
fraction of correct assignments based on selecting combinations
with minimum $\mLik^{\star}$ was 48\%, which was obtained with a 
%
%The size of the jet clustering cone
%and the minimum track $\Pt$ were chosen to maximize the probability for
%correctly assigning both $b$-tagged jets to the $b$ partons. Only events in
%which the leading four jets corresponded to the four primary partons from the
%$\ttbar$ decay were considered. The simulation predicts this occurs in 
%56\% of events. The largest fraction of correct assignments  
%was 48\% (i.e., about half of the 56\%) which was obtained for a 
jet clustering cone size of 0.4 and a minimal track $\Pt$ of 1.0 $\gevc$.

To incorporate additional information pertaining to the charge of the $b$ and
$\bbar$ jets,
%Because the $\mLik^{\star}$  selection function is insensitive to swapping 
%of the $b$ and $\bbar$, 
we define a new selection function,
\begin{equation}
\mLik^{\star \star} = \mLik^{\star} -{\rm 2}\cdot\ln\left[{\mCPr}(Q_{\rm b}) 
                      \cdot {\mCPr}(Q_{\rm \bbar})\right],
\label{eq:ldstar}
\end{equation}
\noindent where ${\mCPr}(Q_{jet})$ is the jet charge  probability density. 
The jet charge is defined as in Ref.~\cite{JETCHARGE}:
\begin{equation}
Q_{jet} \; =\; \frac {\sum_{i=1}^{n_{trk}} q_{i} \cdot       
             {|{\vec{p_{i}} \cdot  {\vec{e}}}|}^k} 
                     {\sum_{i=1}^{n_{trk}} 
             {|{\vec{p_{i}} \cdot  {\vec{e}}}|}^k},
\label{eq:qjet}
\end{equation}
\noindent where $\vec{e}$ is the unit vector along the jet axis, $q_{i}$ and 
$\vec{p_{i}}$ are the charge and momentum of the $i^{th}$ track, and the sum
extends over all $n_{trk}$ charged particles in a fixed cone around the jet. 
To determine optimal choices for the cone size and the weighting factor $k$, 
we varied the jet cone size from 0.35 to 1.0 and $k$ from 0.4 to 1.2,
and compared the significance of separation between the 
$b$ and $\bbar$ ${\mCPr}(Q_{jet})$ distributions.
The results were relatively insensitive to the exact values of these
parameters. Since we found no strong dependence on these parameters, we 
chose the same cone size as used to calculate the JPB 
probability and for simplicity selected $k=1$.

\subsection{Event selection and number of expected background events}
%\hspace*{0.8cm} 

  In this analysis we select events with at least one
SVX or SLT tag. All of the standard lepton and jet 
corrections discussed previously in this paper are applied.  
A total of 34 events are accepted which are identical to the 
tagged events shown in Table~\ref{t-sample}. Since the JPB
algorithm uses tracks reconstructed in the silicon vertex detector,
we require that each event has at least one jet with associated SVX tracks.
We also require that the combination with the lowest value of 
$\mLik^{\star\star}$ has a $\chi^2$ (as defined in Section~\ref{s-algor}) 
less than 10. Only solutions in which a $b$-tagged jet is assigned to  
a $b$ parton are considered. We find that 27 of the 34 events pass these 
requirements.

   We take the combination with the lowest $\mLik^{\star\star}$ value
as the most likely decay chain of the $\ttbar$ into the four highest
$\Et$ jets. Monte Carlo studies show that switching from $\Ki$ 
to the $\mLik^{\star\star}$ selection increases the probability of 
making the correct jet-to-parton assignments. The probability of
correctly assigning the four highest $\Et$ jets to the $\ttbar$ daughter
partons increases from 30.5$\pm$0.7\% to 37.3$\pm$0.6\%.
%Due to acceptance losses and kinematic selection cuts, 
This fraction is ``a priori'' limited to a maximum of 56\%, due to
jets from ISR and FSR.

  The number of expected background events for
the 34 tagged events is estimated to be $10.2 \pm 1.5$, which
includes a background of $7.6 \pm 1.3$ for the 14 events with only SLT tags.
This analysis reduces 
the number of  {SLT} tagged events from 14 to 7 (no SVX tagged
events are cut out). Using the method described in Section~\ref{s-bgcalc},
the expected background for the 7-event SLT sample was evaluated 
to be $3.2_{-0.6}^{+0.7}$ events. We therefore
calculate an expected background for the 27-event sample of $5.8_{-0.9}^{+1.1}$
events, which corresponds to a background fraction 
$x_{b}$ =$0.21_{-0.03}^{+0.04}$.

\subsection{Result of the likelihood fit}\label{fit}

\begin{figure}[ht]
\epsfysize=6.5in
\hspace{0.5in}
\epsffile{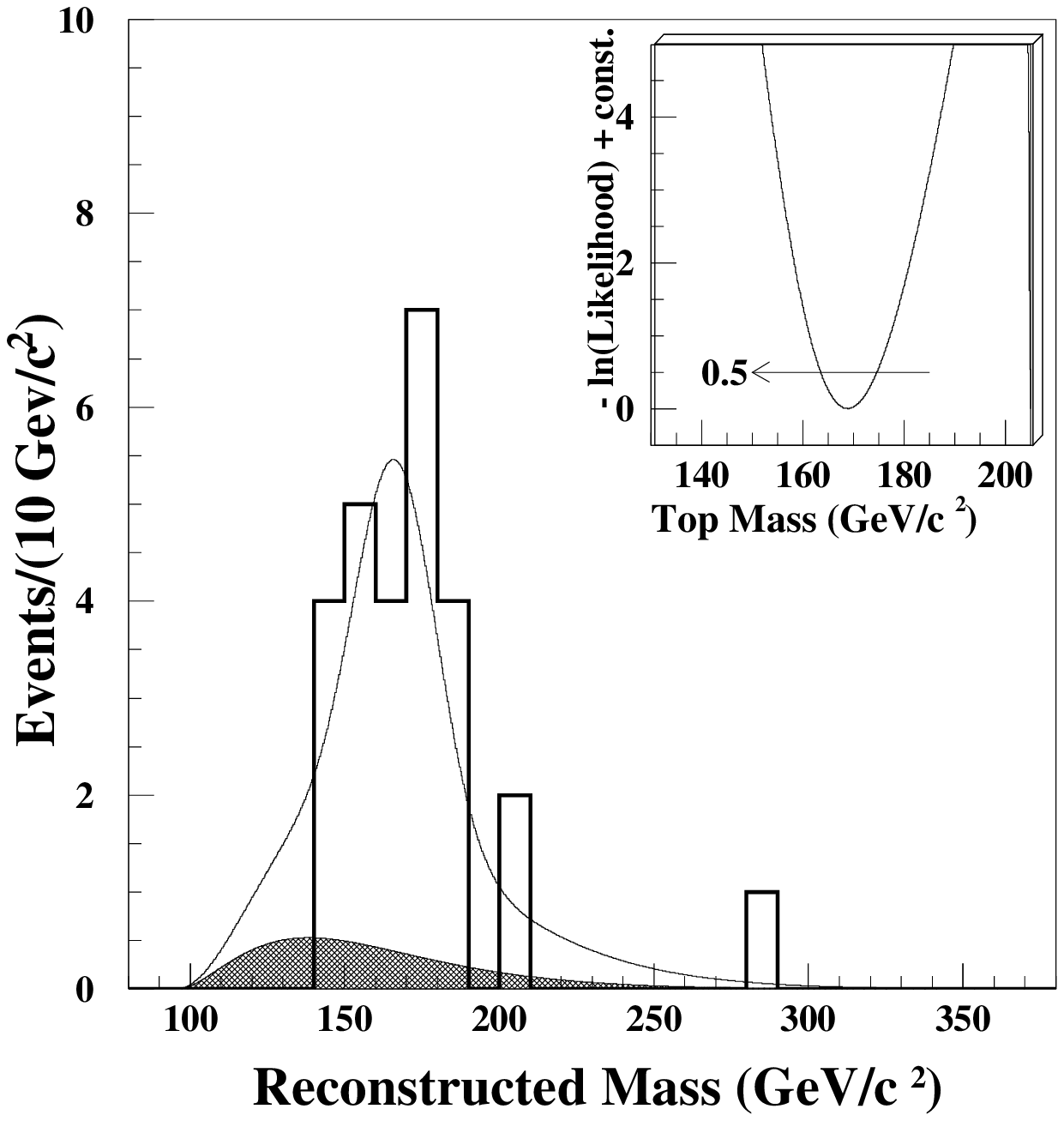}
%\epsffile{$PRD_FIG/datafit_new.eps}
\caption{Results of the $\mLik^{\star \star}$ likelihood fit to 27 $b$-tagged 
events (histogram). The shaded area corresponds to the background returned by 
the fit ($5.2 \pm 1.4$) and the smooth line is the sum of fitted signal 
and background contributions. The inset shows the shape of the 
-log $\mLik^{\star \star}$ versus top mass from which we extract the 
fitted top quark mass and background contribution. }
\label{datafit} 
\end{figure}

  The evaluation of the top mass uses the same techniques described
in Section~\ref{s-like}. The result of the fit is 
shown in Fig.~\ref{datafit}. The histogram represents the reconstructed
mass distribution for the 27 data events. The shaded
area corresponds to the background fraction returned by the fitting procedure,
and the smooth curve shows the sum of the fitted background and signal 
contributions.  The insert displays the likelihood shape with the background 
fraction constrained to $0.21_{-0.03}^{+0.04}$. The  resulting  fit yields:
\begin{equation}
M_{top} \; =\; 170.3^{+5.9}_{-5.4} {\rm (stat.)}\; \gevcc.
% x_{b} \; =\;  0.15^{+0.07}_{-0.05}
\end{equation}

\begin{figure}[ht]
\epsfysize=6.5in
\epsffile{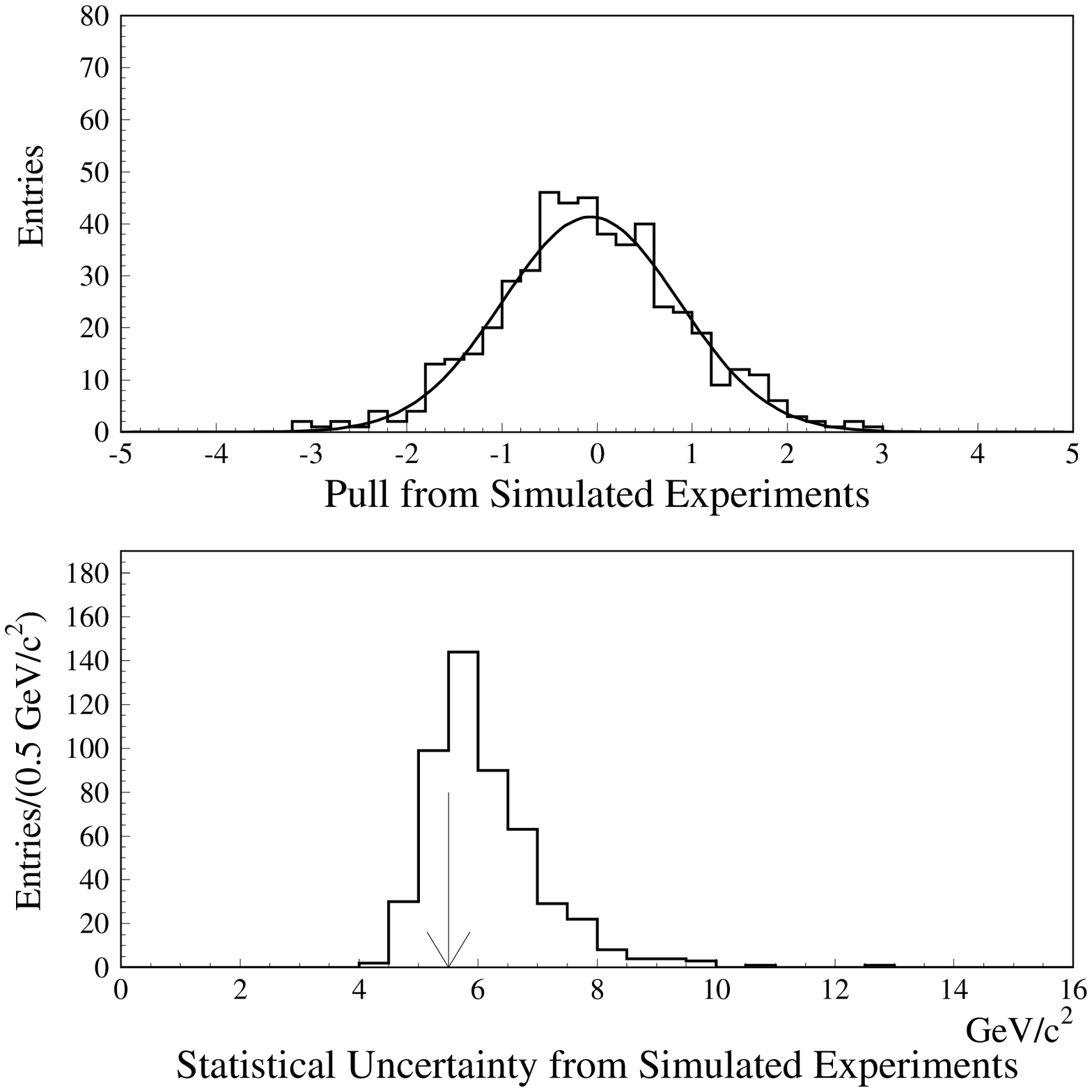}
%\epsffile{$PRD_FIG/pull_prd.eps}
\caption{Results from 500 simulated experiments using the
$\mLik^{\star \star}$ likelihood technique. Each simulated experiment
consists of 27 events, as in the data sample. The upper figure shows 
the pull distribution between
the fitted top mass and the input value (175 $\gevcc$), and
the bottom figure shows the average statistical uncertainty
on the fitted top mass as returned by the likelihood procedure.
The arrow represents the measured value in the data sample.}
\label{pseudo} 
\end{figure}

The soundness of the procedure was tested using simulated
experiments. Figure~\ref{pseudo}(a) shows the pull distributions
for simulated experiments, and Fig.~\ref{pseudo}(b) shows
the average (of the positive and negative) statistical uncertainty 
returned from the likelihood
fit. The arrow indicates the fit result from the data sample. 
We find that 44\% of simulated experiments have a statistical 
uncertainty smaller than measured in the data sample.

Using simulated experiments, we compared the expected statistical 
uncertainty from 34 tagged events using the standard kinematic fit
with 27-event experiments using the $\mLik^{\star \star}$  technique.
The studies indicated that for samples of this size, we could
reduce the top quark mass measurement uncertainty by $\approx 0.5 \ \gevcc$ 
over the standard kinematic $\chi^2$, if we consider the 34 events as
a single sample.

\subsection{Systematic uncertainties}

  The same categories of systematic uncertainties which were
shown in Section~\ref{s-sys} are present in this analysis.
Moreover, we introduce a new systematic uncertainty which
accounts for a possible difference in the tracking efficiency
between data and simulation. This 
uncertainty is introduced because both the JPB algorithm and the
jet charge calculation have some sensitivity to the tracking
efficiency in jets. Using 
simulated experiments, we find an expected uncertainty 
in the top quark mass of 0.9 $\gevcc$ from this source. 
The systematic uncertainties for the $\mLik^{\star \star}$ method 
are summarized  in Table~\ref{systematics}. 
\begin{table}[tbh]
\begin{center}
\begin{tabular}{cc}
\hline\hline
 Systematic              &  Value   \\
 uncertainty             &  ($\gevcc$)  \\   
\hline
  Jet energy  measurement                           &   4.0 \\ 
  Initial and final state radiation                 &   2.7 \\  
  Shape of background spectrum                      &   0.5 \\
  $b$-Tagging                                       &   0.3 \\
  Parton distribution functions                     &   0.6 \\
  Monte Carlo generators                            &   0.8 \\
  Tracking efficiency                               &   0.9 \\
\hline
  Total                                             &   5.1 \\  
\hline\hline
\end{tabular} \\
\caption{Systematic uncertainties for the $ \Lik^{\star \star}$ analysis.}
\label{systematics}
\end{center}
\end{table}

In conclusion, the $\mLik^{\star\star}$ analysis technique has been applied to 
a 27-event subset of the 34 tagged events, and leads to a top quark mass
measurement of $170.3^{+5.9}_{-5.4}$ (stat.) $\pm$ 5.1 (syst.) $\gevcc$.
This value is in good agreement with the results presented in 
Section~\ref{like-data}.

\clearpage
\section{Fitting double $b$-tagged events}
\label{s-ltag}

%   In this mass analysis, we require each event to have 
%two $b$-tagged jets. 
%Studies of candidate
%$\ttbar$ events in which two jets are $b$-tagged have shown 
%that the two non-b tagged jets are consistent with having 
%come from the decay of a W-boson~\cite{wboson}. This
%substantiates the expectation that the $b$-quark jets 
%are tagged with relatively high purity. 
%Requiring two $b$-tagged jets reducing the combinatorics, which in turn increases the fraction 
%of correct jet-to-parton assignments which improves the resolution on
%the top mass.

  This analysis considers only events which contain two 
$b$-tagged jets~\cite{aota}. To increase the acceptance for double $b$-tagged
events beyond what the SVX and SLT algorithms provide, we 
allow one of the $b$ jets to be tagged by the JPB algorithm.
Because the JPB variable
depends upon the impact parameters of the tracks in the jet with
respect to the primary interaction vertex,
this algorithm is correlated with the SVX tagging algorithm. We 
expect little or no correlation between the JPB and SLT tagging algorithms.
We consider a jet $b$-tagged if it has a JPB value less than 5\%.

\begin{figure}[ht]
\epsfysize=6.5in
\hspace{0.5in}
%%\epsffile[0 0 200 600]{dbtag_mass.ps} \vspace{2cm}
\epsffile{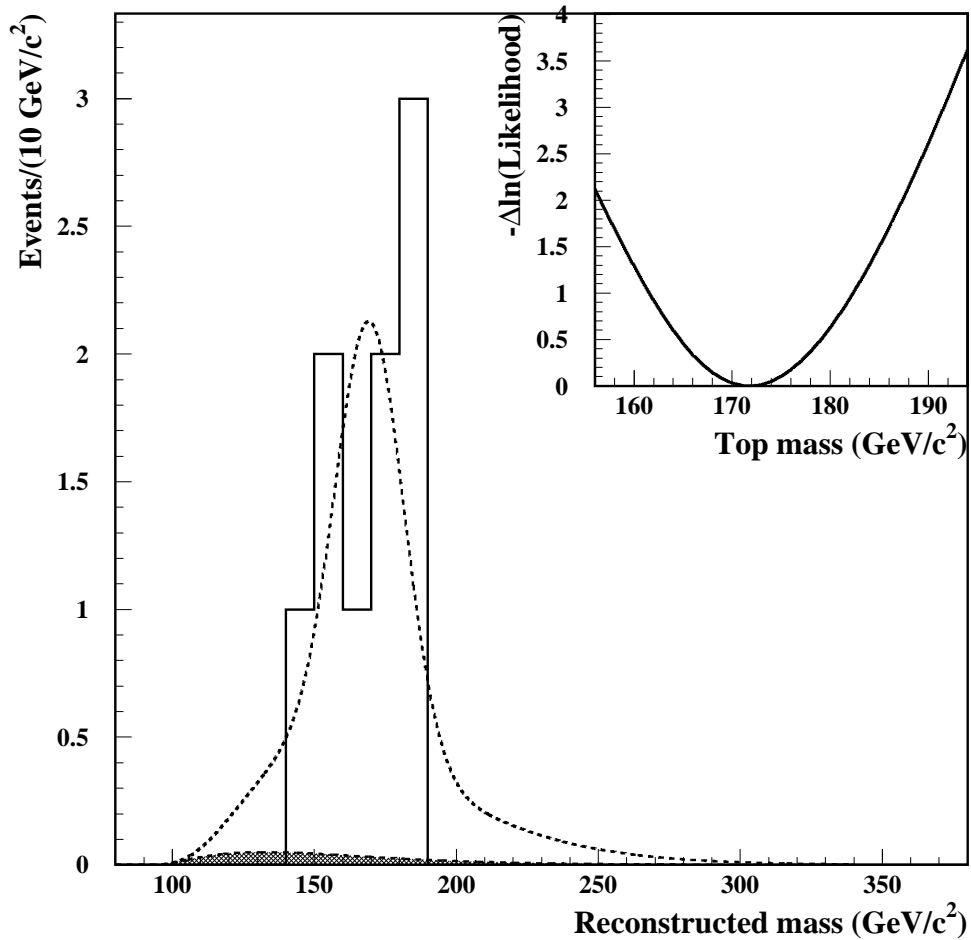} 
%\epsffile{$PRD_FIG/dbtag_mass.ps} 
%\vspace{2cm}
%\vspace{-5cm} \hspace{2cm}
\caption{Distribution of reconstructed mass for the nine data events
and Monte Carlo simulation for the Double b-tag analysis. The background 
distribution (hatched) has been normalized to the expected background of 
0.4 events. The sum of signal+background (dashed line) has been normalized to 
the fitted number of $\ttbar$ and background. The inset shows the 
shape of the likelihood function versus the top quark mass, from 
which we extract the top quark mass to be 171.8$\pm$7.2 $\gevcc$.}
\label{f-dbtmass} 
\end{figure}

   Events are selected using the same selection criteria described
in Section~\ref{s-part}. After we apply analysis cuts 1--7 from 
Section~\ref{s-data}
and require that two jets are tagged by at least one of the three
$b$-tagging algorithms, the data sample consists of 11 events.
Monte Carlo simulations showed that the resolution
on the measured top quark mass can be improved by requiring 
the invariant mass of the two untagged jets to be near the 
$W$-boson mass. A cut of $60<M_{jj}<100 \ \gevcc$ was found to
yield the lowest uncertainty on the measured top quark mass.
Nine of the eleven events are found to survive the $W$ mass cut.
The nine events are a subset of the sample of 34 tagged events. 
The reconstructed top masses of these nine events are listed in 
Table~\ref{t-double} and are shown in Figure~\ref{f-dbtmass}.

  The expected backgrounds are estimated to be 0.22$\pm$0.08 events from
$Wb\bar{b}$ +  $Wc\bar{c}$  processes, 0.05 $\pm$0.02 events from non-$W$
background (e.g. $b\bar{b}$ production), and 0.13 $\pm$0.05 events for 
non-heavy flavor background such as $WW$ and $WZ$ processes. 
The total number of background events is then estimated to be 0.4$\pm$0.1.

    The method for evaluating the top quark mass from this
data sample is the same as the procedure discussed in Section~\ref{s-like}.
The results of the fit are shown in
Fig.~\ref{f-dbtmass}. The figure shows the mass distribution
obtained from data overlayed with the fitted results from the Monte
Carlo simulation. The inset shows the distribution of $-\Delta\log{\cal{L}}$
as a function of the top mass for the nine data events.

   The evaluation of systematic uncertainties are carried out in 
a similar manner to that which was discussed in Section~\ref{s-sys}.
The results are shown in Table~\ref{t-dsys}. The uncertainty due
to background shape is appreciably reduced compared to the 
four subsample analysis because of the smaller background fraction.

%The systematic uncertainty due to the background shape is shown to
%be greatly reduced as compared to the optimize analysis. This is primarily 
%because the background fraction is only about 5\% for this sample, 
%whereas it varies from 5\% for SVX Double Tags to 55\% for No Tag events.
%The systematic uncertainties in the measurement of the top quark mass for 
%events with two $b$-tagged jets is shown in Table~\ref{t-dsys}.

Using the techniques described in this section on the nine
double tagged events, we measure the
top quark mass to be 171.8$\pm$7.2(stat.)$\pm$4.3(syst.) $\gevcc$.
This measurement is consistent with the results
presented in Section~\ref{like-data}. 

\begin{table}[ht]
 \begin{center}
 \begin{tabular}{lcccc}
  \hline\hline
      &        &      & Di-jet mass & Top mass  \\
  Run & Event  & Tags & ($\gevcc$)  & ($\gevcc$) \\
\hline
40758&44414 & SVX + SVX         & 83.9 & 175.4\\
59698&31639 & SVX + (SLT \& JPB)& 79.5 & 187.4 \\
63247&65096 & SVX + JPB         & 81.3 & 161.0 \\
64721&229200& SLT + SLT         & 81.6 & 181.0 \\
65298&747402& SLT + JPB         & 60.0 & 149.4 \\
65581&322592& (SVX \& SLT) + SVX& 66.2 & 152.7 \\ 
67824&281883& (SVX \& SLT) + SVX& 73.3 & 170.1 \\
67971&55023 & SVX + SVX         & 98.1 & 183.5 \\
68464&547303& SVX + SVX         & 87.3 & 151.1 \\
%
%40758&44414 & SVX + SVX         & 83.9 & 175.4$\pm$11.6\\
%59698&31639 & SVX + (SLT \& JPB)& 79.5 & 187.4$\pm$10.0 \\
%63247&65096 & SVX + JPB         & 81.3 & 161.0$\pm$8.2 \\
%64721&229200& SLT + SLT         & 81.6 & 181.0$\pm$14.5 \\
%65298&747402& SLT + JPB         & 60.0 & 149.4$\pm$8.7 \\
%65581&322592& (SVX \& SLT) + SVX& 66.2 & 152.7$\pm$8.9 \\ 
%67824&281883& (SVX \& SLT) + SVX& 73.3 & 170.1$\pm$11.1 \\
%67971&55023 & SVX + SVX         & 98.1 & 183.5$\pm$11.3 \\
%68464&547303& SVX + SVX         & 87.3 & 151.1$\pm$8.4 \\
\hline\hline
\end{tabular}
\end{center}
\caption{List of events used in the double $b$-tagged analysis.
Shown are the run-event numbers, the algorithms which tagged the two
jets, the di-jet mass of the two untagged jets, and the reconstructed
top mass for the solution having the lowest $\chi^2$. If a jet is
tagged by two different algorithms, both tags appear in parentheses. }
\label{t-double} 
\end{table}

\begin{table}[ht]
\begin{center}
\begin{tabular}{lc}
\hline\hline
Systematic uncertainties  & Values (GeV/$c^2$) \\
\hline
Jet energy measurement                  & 4.1 \\
Initial and final state radiation       & 1.1 \\
Shape of background spectrum            & $<$0.1 \\
$b$-Tagging                             & 0.4 \\
Parton distribution functions           & 0.3 \\
\hline
 Total                                  & 4.3 \\
\hline\hline
\end{tabular}
\end{center}
\caption{Summary of systematic uncertainties
 in the top mass measurement from double $b$-tagged events.}
\label{t-dsys} 
\end{table}

%% file: dilep-appendix.tex
\chapter{DETAILS OF THE ALL-HADRONIC AND DILEPTON MASS ANALYSES}
\label{dilep-appendix}

\section{All-hadronic channel}

      A reevaluation of the systematic uncertainties on the measurement 
of the top quark mass in the all-hadronic channel as reported in 
Ref.~\cite{all-hadron} has shown that some of those estimates were overly 
conservative. Since that publication further
studies of the systematic uncertainties have led to better
procedures, which we now apply to all channels. 
The systematic uncertainties which have been revised include: 
initial and final state radiation, fitting procedure, and jet energy scale. 
These revisions are discussed below.

The contribution due to uncertainty in modeling initial
and final state hard radiation was 8.0 $\gevcc$. To evaluate this 
uncertainty, standard {\small{HERWIG}} $\ttbar$ events were compared
to samples which were constructed to have smaller and larger fractions of 
events in which one or more of the final state jets did not match any of 
the daughter quarks from the $\ttbar$ decay. The most evident difference
between the samples was that the width of the reconstructed mass distribution 
broadened as this fraction increased. On the other hand, simulated experiments
showed only a very small shift in the fitted top quark mass. The
systematic uncertainty was evaluated as follows. We generated two samples of
simulated experiments, (a) one using the default {\small{HERWIG}} 
templates, and (b) one using templates which were constructed to have 
90\% of events containing one or more jets that were not matched to the 
daughter quarks from the $\ttbar$ decay. For the default {\small{HERWIG}} 
sample, approximately 60\% of events have one or more jets not matched to a 
quark from the $\ttbar$ decay. In both cases, we evaluated the median and the 
rms width of fitted top quark masses from the simulated experiments.
The systematic uncertainty was taken to be the quadrature difference 
of the widths between samples (a) and (b). This number was then added in 
quadrature with the small shift in the median mass which was observed between 
sample (a) and (b).  Essentially all of the 8.0 $\gevcc$ uncertainty 
was from the increase in the width of the distribution of sample (b). 
%  However, we use the statistical uncertainty as measured in the data, and
%  therefore we are insensitive to the width of the probability distribution.
%  We believe 
%  that this reasoning was incorrect
%  since we only attribute systematic uncertainty to effects which might 
%  cause a shift in the fitted value for the top quark mass. 
%Problems with this method include: 
%(i) only an increase, not a
%decrease, in the amount of hard radiation was considered; (ii) changes
%in the statistical uncertainty returned by the likelihood fit for
%simulated experiments with changed amounts of hard radiation
%were not taken into account; (iii) it is unclear how to relate the
%change in fraction of events with unmatched jets (from 60\% to 90\%
%above) to changes in any physics, and/or Monte Carlo program
%parameters.
Further studies show that the change in width
of the reconstructed mass distribution with increased radiation is reflected
in the statistical uncertainties returned by the fits for simulated
experiments; thus the statistical uncertainty obtained from our
fitting procedure for the data sample already takes into account
this effect.
% (ii) relating the amount of gluon radiation to the 
%fraction of events with non-matching (within $\Delta$ R =0.4) jets 
%was arbitrary as was the choice of the 90% point for the estimation 
%of the uncertainty; neither one of these choices is directly related 
%to changes in gluon radiation parameters in the Monte Carlo. 
A reevaluation, using the same procedure as described in Section~\ref{gluon},
results in a contribution from this source of 1.8 $\gevcc$ \cite{lbl-mass}.
  
Another large source of systematic uncertainty (5.2 $\gevcc$) came from 
the effect of selecting the second-best rather than the best kinematic fit to
each event. A smaller contribution came from considering different
ways of interpolating between likelihood values at discrete top
mass values in order to find the maximum likelihood point. A third
contribution came from the finite Monte Carlo statistics that provided
the expected reconstructed mass distributions at different top mass
values. The first two contributions are no longer identified as
sources of significant systematic uncertainty since they concern the 
robustness of the chosen method.  The contribution from Monte Carlo 
statistics, of 0.3 $\gevcc$, remains.

The jet energy scale uncertainty was determined to be 5.4 $\gevcc$.  Part of 
that (3.7 $\gevcc$), was due to differences in the calorimeter energy scale 
between two versions of the detector simulation.  The source of this 
uncertainty was later corrected.  As a result, the 3.7 $\gevcc$ contribution 
to the uncertainty was eliminated.

A small reorganization of the contributions has occurred, which we mention 
in order to avoid any confusion in a comparison with Ref.~\cite{all-hadron}.
The soft gluon uncertainty (3.0 $\gevcc$) has been moved from the ``gluon
radiation and fragmentation effects'' to the ``jet energy scale'' category.
The Monte Carlo generator uncertainty (0.8 $\gevcc$) has been assigned its
own category.  The result is a new systematic uncertainty of 5.7 $\gevcc$, 
with a breakdown into different contributions as listed in 
Table~\ref{comb-sys}.

\section{Dilepton channel}

The top quark mass measurement in the dilepton channel uses eight observed 
events that pass the standard selection criteria used for
the dilepton channel~\cite{cdf-dilepton,lbl-mass}. 
The criteria require that the leptons have opposite charges, 
that there be at least two jets per event, and
include cuts on the missing transverse energy and the lepton
transverse energies.

This measurement involves two steps: a top mass estimate
is obtained for each event, and then a likelihood fit, which allows
for the presence of background, gives an overall best estimate of the
top quark mass. The second step is similar to that in the lepton+jets
topology, but the first step is appreciably different.

In order to get a mass estimate for an individual event, we determine a 
weight distribution as a function of an assumed top mass, $m_t$. First, we 
assume that the event originates from $\ttbar$ production and decay, that 
the leading two jets are $b$ jets from top decay, and that the leptons 
($e$ or $\mu$) are from associated $W$-boson decays. Next, we assume a 
value for the top mass, $m_t$, assume pseudorapidity values, $\eta_1$ and 
$\eta_2$, for the two neutrinos, and solve for the two neutrino momenta.
In general there are eight solutions because of a quadratic ambiguity
in each neutrino's longitudinal momentum and a choice of pairing 
leptons with jets.  For each solution, we denote as $\MET^p$ the
vector sum of the solution's neutrino transverse 
momenta. Then we assign a weight to each solution according to how
well $\MET^p$ agrees with the event's measured missing transverse
energy, $\MET^m$, as follows:
%%%%%***
\begin{equation}
g(m_t,\eta_1,\eta_2) = 
       \exp{\left( -\frac{(\MET^p_x - \MET^m_x)^2} {2\sigma^2} \right)}
         \times 
       \exp{\left( -\frac{(\MET^p_y - \MET^m_y)^2} {2\sigma^2} \right)}
  \label{eq:g_mt_eta}
\end{equation}
%%%%%***
where $\sigma$ is the resolution in each component (x and y) of the
measured unclustered transverse energy (see below).
The experimental resolution in jets and leptons is taken into account by 
sampling the measured quantities many times according to their resolutions.
That is, for each set of assumed $m_t$, $\eta_1$ and $\eta_2$ values
a weight is calculated many times, and the sum is accumulated.  For each 
assumed $m_t$ value, 100 pairs of $\eta_1$ and $\eta_2$ values are assumed 
in turn, and the summed weights are again summed, to give a final summed 
weight, $f(m_t)$, at any $m_t$ value. The $\eta_1$ and $\eta_2$ values are
drawn independently from a Gaussian distribution with unit width and
centered at 0.0, as predicted by {\small{HERWIG}} Monte Carlo simulations.
Thus all the uncertainties on the  $\MET$ measurement are taken into
account, except for the resolution of the unclustered energy. 
We use $\sigma = 4\sqrt{n}$ GeV, where $n$ is the number of interactions 
in the event and comes from studies of low-luminosity minimum-bias events. 
    
For each event, $m_t$ values in the range 90 to 290 $\gevcc$, in 
2.5 $\gevcc$ steps, were assumed in order to give a $f(m_t)$ distribution.
This distribution is used to determine a top mass estimate, as follows.
The position of the maximum value, $f(m_t)_{max}$, is denoted by
$M_{max}$. The first points on either side of $M_{max}$ that have
$f(m_t) \le f(m_t)_{max}/2$ are denoted by $M_1$ and $M_2$. The average of 
$M_1$ and $M_2$ is taken as the top mass estimate.  

\begin{figure}
\epsfysize=6.5in
\hspace{0.5in}
%\epsffile[0 72 612 720]{dil_obs_wts.ps}
\epsffile[0 72 612 720]{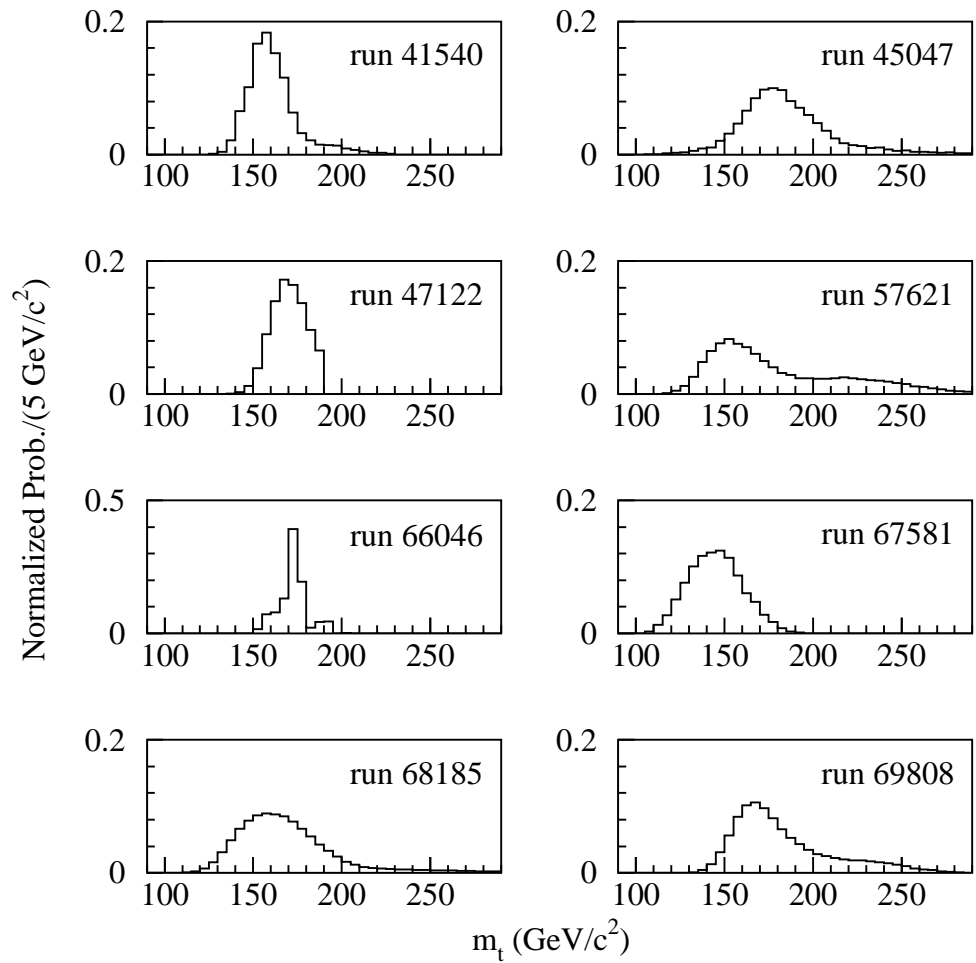}
\caption{Weight distribution $f(m_t)$, normalized to unity, for the eight
observed dilepton events.} 
\label{dilep-evts}
\end{figure}

The $f(m_t)$ distributions, normalized to unity, for the eight events 
are shown in 
Fig.~\ref{dilep-evts}.  The eight events, with their lepton identifications, 
numbers of jets (with uncorrected transverse energy greater than 10 $\gev$ 
and pseudo-rapidity in the range $-2.0$ to $+2.0$ units), and estimated top 
masses are given in Table~\ref{dileps}. 
\begin{table}[ht]
\begin{center}
\begin{tabular}{cccccc}
\hline\hline
Run    &  Event  &   leptons    & $N_{jet}$  &  Top mass  & $\log(P_{ev})$ \\
\hline
41540  &  127085 &  $e^- \mu^+$ &      2     &   158.8    &  0.47    \\
45047  &  104393 &  $e^+ \mu^-$ &      2     &   180.0    &  1.82    \\
47122  &   38382 &  $e^+ \mu^-$ &      2     &   176.3    &  1.40    \\
57621  &   45230 &  $e^+ \mu^-$ &      2     &   156.3    &  2.20    \\
66046  &  380045 &  $e^+ \mu^-$ &      4     &   172.5    & -5.20    \\
67581  &  129896 &  $e^+ \mu^-$ &      2     &   143.8    &  0.44    \\
68185  &  174611 &  $e^+ e^-$   &      2     &   161.3    &  4.10    \\
69808  &  639398 &  $e^- \mu^+$ &      3     &   170.0    &  3.50    \\
\hline\hline
\end{tabular}
\end{center}
\caption{Information on the eight candidate dilepton events used in the
dilepton mass analysis. Shown are: the run and event numbers, the
types of leptons in each event, the number of reconstructed jets 
(with uncorrected $\Pt>10~\gevc$ and $|\eta |<2$), and the top mass 
estimates for each event. Also listed is $\log(P_{ev})$, where $P_{ev}$ 
is the sum of all the weights for the event divided by the number of 
resolution samplings used.}
\label{dileps}
\end{table}

\begin{figure}
\epsfysize=6.5in
\hspace{0.5in}
%\epsffile[0 72 612 720]{dmc_dil_weight.ps}
\epsffile[0 72 612 720]{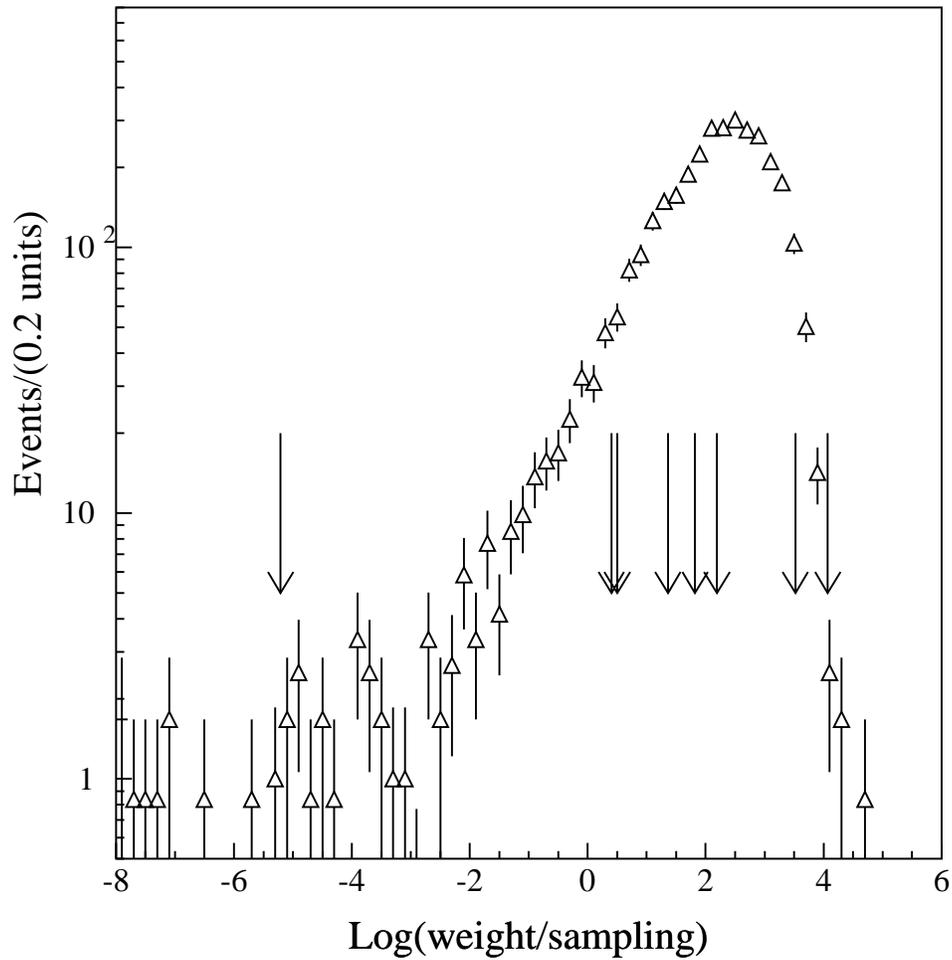}
\caption{Predicted distribution of $\log(P_{ev})$, the total weight sum per
resolution sampling, for the expected $\ttbar$ and background event
mix in the dilepton sample. The arrows indicate the values for the
eight observed events.}
\label{dilep-wts}
\end{figure}

It is useful to define a variable, $P_{ev}$, as the sum of $f(m_t)$ over all 
assumed $m_t$ values, divided by the number of resolution samplings used.
The latter number is 1500 for data and 30 for Monte Carlo events.  This 
variable gives an indication of how easily an event can be fit to the $\ttbar$
decay hypothesis.  The $\log(P_{ev})$ distribution of simulated $\ttbar$ plus 
background events is shown in Figure~\ref{dilep-wts}. The $\ttbar$ events are 
from the {\small HERWIG} simulation with a top quark mass of 175 $\gevcc$.
The $\log(P_{ev})$ values for the eight observed events are listed in
Table~\ref{dileps} and are indicated by arrows in Figure~\ref{dilep-wts}.
The data points all lie within the range spanned by the simulated distribution.
In the simulated events, 0.7\% have $\log(P_{ev}) < -5.2$, the value for the 
lowest data point, so the probability for an eight-event sample to have at 
least one event at $-5.2$ or lower is 5\%.
%
%  None of the 8 events have $P_{ev}$ of zero, so none are totally
%  incompatible with the $\ttbar$ hypothesis (we are assuming that the
%  experimental resolutions we have used are correct). 
%  In fact, all the events have non-zero $f(m_t)$ for all
%  assumed $m_t$ values in the range 150 - 185 $\gevcc$, so none are totally
%  incompatible with the measured top mass.

In Ref.~\cite{lbl-mass} it was noted that the same method could be
applied to events in the lepton+jets topology that had two SVX-tagged jets. 
In such events the two untagged jets (of the four highest $E_T$ jets) are 
assumed to result from $W$-boson decay, and in order to mimic a $W$-boson 
leptonic decay one of those jets is treated as a lepton (electron or muon) 
and the other as a neutrino. In the following we took the jet with lower 
$E_T$ as an unobserved neutrino and recalculated $\MET^m$ for the event.
Then the above dilepton method was applied.

\begin{figure}
\epsfysize=6.5in
\hspace{0.5in}
%\epsffile[0 72 612 720]{dil_2svx.ps}
\epsffile[0 72 612 720]{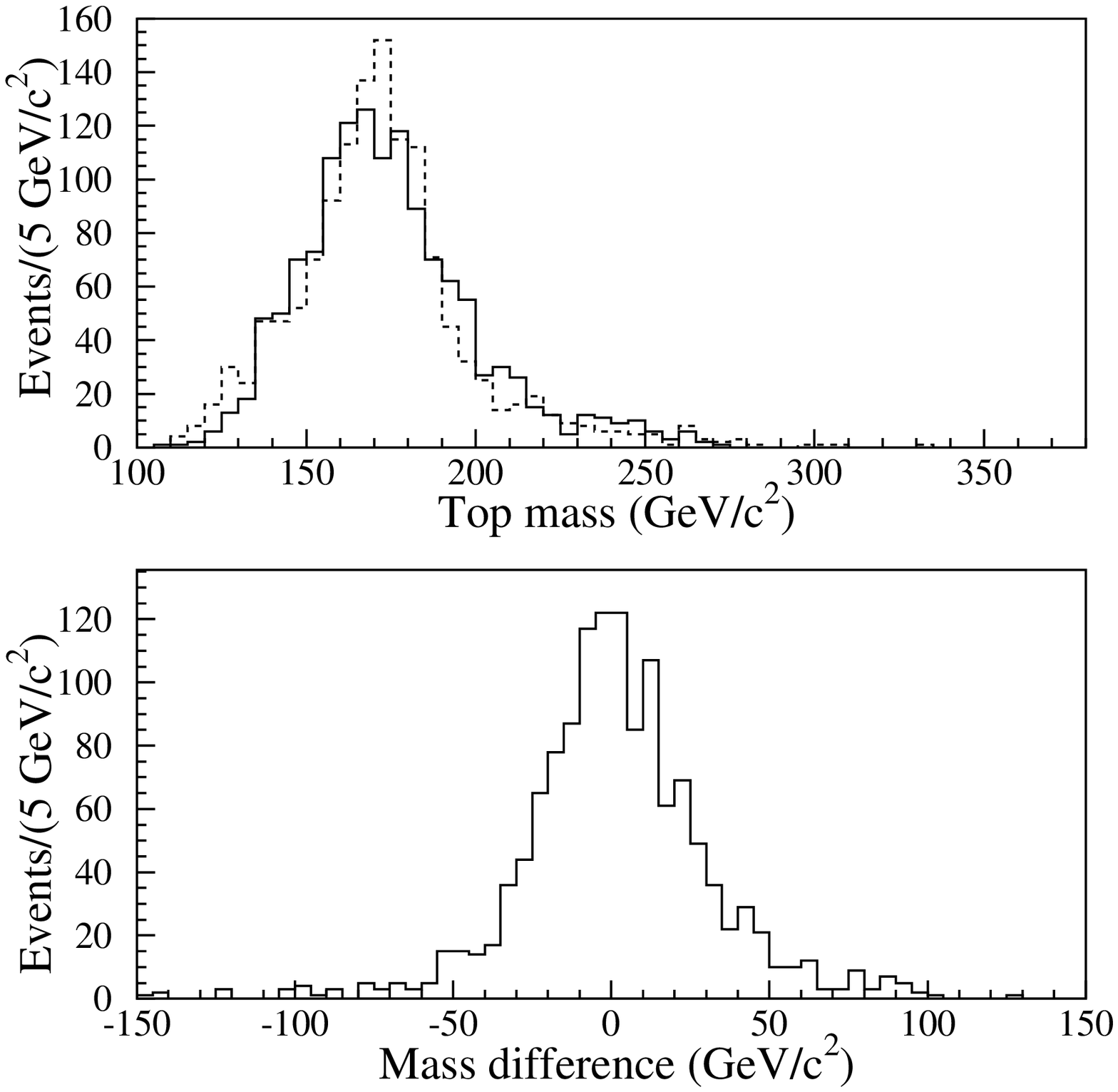}
\caption{(a) Top mass estimates for simulated $\ttbar$ events (top mass
175 $\gevcc$) with two SVX tags using the lepton+jets kinematic fit (dashed) 
and the pseudo-dilepton (solid) methods. (b) The difference per event between 
the top mass estimates from the two methods.}
\label{dilep-2svx}
\end{figure}

The five events in the SVX Double sample were fit with this method. A
top quark mass value of  $181.5 \pm 12.6~ \gevcc$ was obtained. This
value has to be compared with the value shown in Table~\ref{fit-results}
of $170.0^{+9.4}_{-8.9} ~\gevcc$, a difference of 11.5 $\gevcc$. In
order to understand the difference between the two methods a
comparison was made in a Monte Carlo study that used  
a sample of approximately 1300 simulated lepton+jets $\ttbar$ events
with \mtop = 175 $\gevcc$ and with two jets having SVX tags. 
The distribution of the reconstructed mass from
the standard lepton+jets kinematic fit is shown in 
Figure~\ref{dilep-2svx}(a).
Also shown is the top mass estimate per event with the pseudo-dilepton
method described above. The two distributions are similar. The medians
are 170.5 $\gevcc$ and 170.9 $\gevcc$, and the widths are 21.4 $\gevcc$
and 23.4 $\gevcc$, respectively for the kinematic fit and the dilepton
methods. Here the widths are one-half the separation of the $16^{\rm th}$ and
$84^{\rm th}$ percentiles in the distributions.  As expected, the dilepton 
method gives a slightly wider distribution.
In Figure~\ref{dilep-2svx}(b) the mass difference between the two methods is 
plotted for each event.  The width of this distribution is 24.3
$\gevcc$. This shows that the shift of 11.5 $\gevcc$ found for the
five SVX Double events using the two methods is well within expectation.

This study shows that fitting the dilepton events, which are
underconstrained, using the technique described here is just as
valid  and precise as the completely constrained 2-C fit used for the
lepton+jets sample.  In addition, 
if we calculate the statistical correlation between the two 
methods, we obtain a correlation coefficient of 0.36, i.e., fitting
the SVX Double events with this technique could improve the
statistical uncertainty on the mass determination from this channel.

%%  Added Figures

%% file: topmass.bbl
\begin{thebibliography}{99}
\bibitem{etadef} The pseudo-rapidity $\eta$ is defined as 
        $-\log(\tan(\theta /2))$, where $\theta$ is the polar angle measured 
        relative to the positive $z$-axis (taken as the direction of the 
        outgoing proton beam) assuming a $z$-vertex position of zero.
        For the mass fit discussed in Section~\ref{s-algor} we recalculate  
        $\eta$ using the measured  $z$-vertex position for each event.
\bibitem{unclus} The unclustered energy consists of the total energy which is 
        not contained in jets with $\Et>8\ \gev$ and $|\eta |<3.4$.
\bibitem{metdef} The missing $\Et$, $\met$ is defined to be the magnitude of 
        the vector sum of the raw transverse energies in all calorimeter 
        towers with $|\eta |<3.6$.  Only towers which exceed a 
        detector-dependent energy threshold are included in the 
        sum~\cite{cdf-evidence}.
\bibitem{deffid} A fiducial lepton is required to be within a well-understood
        region of the CEM which allows for good lepton identification.
\bibitem{defiso} The isolation, I, is defined as the $\Et$ ($\Pt$) within a 
        cone of $\Delta R<0.4$ centered on the lepton, but excluding the 
        lepton energy, divided by the $\Et$ ($\Pt$) of the lepton.
\bibitem{defraw} Raw jet energies are the observed values after performing 
        the jet clustering, but prior to any corrections. The raw jet energies
        are used to select the data samples and are used in defining the four 
        highest $\Et$ jets.
\bibitem{JETPROB2} Monte Carlo studies showed that the jet probability 
        distributions for light quark and gluon jets were similar, so that a 
        single non-$b$ probability distribution was used.
\end{thebibliography}
